\documentclass[aps,prx,reprint, amsmath, amssymb,superscriptaddress]{revtex4-1}

\usepackage{bm}
\usepackage[retainorgcmds]{IEEEtrantools}
\usepackage{graphicx}
\usepackage{mathrsfs}
\usepackage{amsmath}
\usepackage{amssymb}
\usepackage{color}
\usepackage{amsfonts}
\usepackage{times,txfonts}
\usepackage{nicefrac}
\usepackage[colorlinks=true,linkcolor=blue,urlcolor=blue,citecolor=blue,pdfusetitle]{hyperref}

\usepackage[dvipsnames]{xcolor}
\definecolor{mypink}{rgb}{0.858, 0.188, 0.478}
\definecolor{bondiblue}{rgb}{0.0, 0.58, 0.71}
\definecolor{bleudefrance}{rgb}{0.19, 0.55, 0.91}
\usepackage[authormarkup=none]{changes}
\definechangesauthor[name={Gabriel}, color=red]{g}

\newcommand{\change}[1]{\textcolor{black}{#1}}

\newcommand{\tr}{\text{tr}}
\newcommand{\sech}{\text{sech}}

\newcommand{\sigD}{\Gamma_\text{cl}}
\newcommand{\sigC}{\Gamma_\text{qu}}
\newcommand{\sigB}{\Lambda_\text{cl}}
\newcommand{\sigA}{\Lambda_\text{qu}}

\newcommand{\sigDij}{\gamma_\text{cl}}
\newcommand{\sigCij}{\gamma_\text{qu}}
\newcommand{\sigBij}{\lambda_\text{cl}}
\newcommand{\sigAij}{\lambda_\text{qu}}

\usepackage{array}

\usepackage{pifont}
\newcommand{\cmark}{\ding{51}}%
\newcommand{\xmark}{\ding{55}}%
\newcommand{\lmark}{\fontfamily{pzd}\selectfont\scalebox{2.5}[1.2]{-}}

\begin{document}

\title{Contributions from populations and coherences in non-equilibrium entropy production
} 
\date{\today}
\author{Adalberto D. Varizi}
    \affiliation{Instituto de F\'isica da Universidade de S\~ao Paulo,  05314-970 S\~ao Paulo, Brazil}
    \affiliation{Departamento de F\'isica, Instituto de Ci\^encias Exatas, Universidade Federal de Minas Gerais, 30123-970, Belo Horizonte, Minas Gerais, Brazil}
\author{Mariana A. Cipolla}
    \affiliation{Instituto de F\'isica da Universidade de S\~ao Paulo,  05314-970 S\~ao Paulo, Brazil}    
\author{Mart\'{i} Perarnau-Llobet}
    \affiliation{D\'{e}partement de Physique Appliqu\'{e}e, Universit\'{e} de Gen\`{e}ve, 1211 Geneva, Switzerland}
\author{Raphael C. Drumond}
    \affiliation{Departamento de Matem\'atica, Instituto de Ci\^encias Exatas, Universidade Federal de Minas Gerais, 30123-970, Belo Horizonte, Minas Gerais, Brazil}
\author{Gabriel T. Landi}
    \affiliation{Instituto de F\'isica da Universidade de S\~ao Paulo,  05314-970 S\~ao Paulo, Brazil}

\begin{abstract}
The entropy produced when a quantum system is driven away from equilibrium can be decomposed in two parts, one related with populations and the other with quantum coherences. 
The latter is usually based on the so-called relative entropy of coherence, a widely used quantifier in quantum resource theories.
In this paper we argue that, despite satisfying fluctuation theorems and having a clear resource-theoretic interpretation, this splitting has shortcomings. 
\change{First,  it predicts that at low temperatures the entropy production will always be  dominated by the classical term, irrespective of the quantum nature of the process. 
Second, for infinitesimal quenches,  the radius of convergence diverges exponentially as the temperature decreases, rendering the functions non-analytic.
Motivated by this, we provide here a complementary approach, where the entropy production is split in a way such that the contributions from populations and coherences are written in terms of a thermal state of a specially dephased Hamiltonian. 
}
The physical interpretation of our proposal is discussed in detail.
We also contrast the two approaches  by studying work protocols in a transverse field Ising chain, and a macrospin of varying dimension.

\end{abstract}

\maketitle{}

%
%
\section{\label{sec:int}Introduction and preliminary results}
%
%

Quantum coherence and quantum correlations play a key role in the thermodynamics of microscopic systems \cite{Goold2016,Vinjanampathy2016}. They can be exploited to extract useful work~\cite{Allahverdyan_2004,Scully2007,Korzekwa2016,Manzano2018,Lrch2018,Rodrigues2019,Francica2020}, speed-up energy exchanges~\cite{Hovhannisyan2013,Campaioli2017,campaioli2018quantum,Julia2020},  and   improve heat engines~\cite{Correa2014,Ronagel2014,Brunner2014,Uzdin2015,Manzano2016,hammam2021optimizing}. On a more fundamental level, they alter the possible state transitions in thermodynamic processes~\cite{janzing2006quantum,Lostaglio2015,Cwikli2015},  lead to  new forms of work and heat fluctuations \cite{Allahverdyan2014,Talkner2016,Hofer2017,Bumer2018,Levy2020,Micadei2020}, 
modify  the fluctuation-dissipation relation for work~\cite{Miller2019,Scandi2019,miller2020joint} and may even generate  heat flow reversals~\cite{Lloyd1989,Jennings2010,Jevtic2015a,Micadei2017}.
Understanding the role of coherence in the formulation of the laws of quantum thermodynamics is therefore a major overarching goal in the field, which has been the subject of considerable recent interest.

When a system relaxes to equilibrium, in contact with a heat bath, quantum coherences are known to contribute an  additional term to the entropy production~\cite{Lostaglio2015,Santos2019,Mohammady2020}, which quantifies the amount of irreversibility in the process. 
A similar effect also happens in unitary work protocols~\cite{Francica2019,Varizi2020}.
To be concrete, we focus on the latter and consider a scenario where a system is described by a Hamiltonian $H_t= H(g_t)$, depending on a controllable parameter $g_t$. The system is initially prepared in thermal equilibrium at a temperature $T$, such that its initial state is the thermal state $\rho_0^\text{th} \equiv \rho^\text{th}(g_0) = e^{-\beta H_0}/Z_0$, where $\beta=1/T$ and $Z_0=\tr{e^{-\beta H_0} }$ is the partition function.
At $t=0$, a work protocol $g_t$, that lasts for a total time $\tau$, is applied to the system, driving it out of equilibrium~\cite{Esposito2009,Campisi2011}.
Letting $U$ denote the unitary generated by the drive,  the state of the system after a time $\tau$ will be 
\begin{equation}\label{rho_tau}
    \rho_{\tau} = U \rho_0^\text{th} U^{\dagger}.
\end{equation}
In general,  $\rho_{\tau}$ will be very different from the corresponding equilibrium state
$\rho_\tau^\text{th} = e^{-\beta H_\tau}/Z_\tau$.
This difference is captured by the entropy production (also called non-equilibrium lag in this context)~\cite{Kawai2007,Vaikuntanathan2009,Parrondo2009,Deffner2010},
\begin{equation}\label{EntProd}
    \Sigma = S\big( \rho_{\tau} || \rho_\tau^{\text{th}}\big),
\end{equation}
where $S( \rho || \sigma) = \tr{ \rho (\ln \rho - \ln \sigma)} \geqslant 0$ is the quantum relative entropy.
The non-equilibrium lag is directly proportional to the irreversible work~\cite{Jarzynski1997,Kurchan1998,Talkner2007}, $\Sigma = \beta \big(\langle W \rangle - \Delta F \big)$, where $\langle W \rangle = \tr\big( H_\tau \rho_\tau - H_0 \rho_0^\text{th}\big)$ is the work performed in the process and $\Delta F = F(g_\tau) - F(g_0)$ is the change in equilibrium free energy, $F(g) = \tr\Big\{ H(g) \rho^\text{th}(g)\Big\} - T S(\rho^\text{th}(g))$ (with $S(\rho) = -\tr(\rho \ln \rho)$ being the von Neumann entropy). 
Due to its clear thermodynamic interpretation, $\Sigma$ has been widely used as a quantifier of irreversibility, both theoretically~\cite{Jarzynski1997,Derrida1998,Crooks1998,Kurchan1998,Lebowitz1999,Mukamel2003,Talkner2007,Deffner2010,Guarnieri2018}
and experimentally~\cite{Liphardt2002,Douarche2005,Collin2005,Speck2007,Saira2012,Koski2013,Batalhao2014,An2014,Batalhao2015,Talarico2016,Zhang2018a,Smith2017}.

\begin{figure*}
    \centering
    \includegraphics[width=\textwidth]{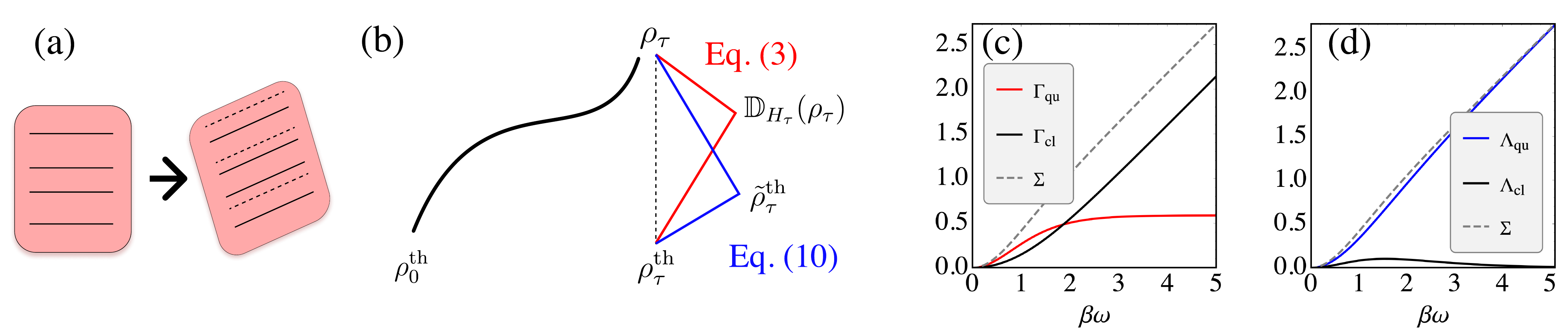}
    \caption{{\bf Difficulties in identifying the classical and quantum contributions to the entropy production.} 
    (a) A work protocol $H(g_t)$ can modify the Hamiltonian in two ways: altering the energy level spacings, which can be viewed as a semi-classical effect, and rotating the energy eigenbasis, which is a fully quantum property with no classical counterpart.
    (b) The entropy production~\eqref{EntProd} compares the final state $\rho_\tau$ with the reference thermal state $\rho_\tau^\text{th}$.
    To understand its classical and quantum contributions, the splitting~\eqref{CD_split} uses an intermediate state $\mathbb{D}_{H_\tau}(\rho_\tau)$. 
    Conversely, in this paper we introduce the splitting~\eqref{AB_split}, which uses the state $\tilde{\rho}_\tau^\text{th}$, in Eq.~\eqref{rho_tilde}. 
    (c) The contributions $\sigC$ and $\sigD$ of Eq.~\eqref{CD_split} in a minimal qubit model, as a function of $\beta \omega$, plotted using Eqs.~\eqref{qubit_Sigma} and~\eqref{qubit_C} with $\theta = 1.1$ (see text for details).  
    (d) The new splitting~\eqref{AB_split} (see also  Eq.~\eqref{qubit_A}), which yields physically more reasonable results at low temperatures.
    }
    \label{fig:abcd}
\end{figure*}

The entropy production $\Sigma$ in Eq. $\eqref{EntProd}$ contains contributions of both a classical and quantum nature. 
This is linked with the fact that the work protocol $g_t$ can modify the Hamiltonian $H(g)$ in two ways. 
On the one hand, it may alter the spacing of the energy levels; 
and, on the other, it may rotate the eigenvectors (Fig.~\ref{fig:abcd}(a)). 
The latter  is directly associated with quantum coherence and to the fact that $[H(g_{t_1}), H(g_{t_2})] \neq 0$, for two different times $t_1, t_2$. 
It therefore has no classical counterpart, and  corresponds to a fundamental feature distinguishing classical and quantum processes.
\change{In general, these two processes will become mixed, and hence}
identifying how each physical process contributes to $\Sigma$ is in general a challenging task.
In the literature, a popular choice is the splitting  put forward 
in~\cite{janzing2006quantum,Lostaglio2015,Santos2019,Francica2019}: 
\begin{equation}\label{CD_split}
    \Sigma = \sigD + \sigC,
\end{equation}
where
\begin{IEEEeqnarray}{rCl}
\label{D}
    \sigD &=& S\big(\mathbb{D}_{H_{\tau}}(\rho_{\tau}) || \rho_\tau^\text{th} \big), \\[0.2cm]
    \sigC &=& S\big( \rho_{\tau} || \mathbb{D}_{H_{\tau}}(\rho_{\tau}) \big) = S\big(\mathbb{D}_{H_{\tau}}(\rho_{\tau})\big) - S(\rho_\tau),
    \label{C}
\end{IEEEeqnarray}
with $\mathbb{D}_{H}(\rho)$ being the super-operator that completely dephases the state $\rho$ in the eigenbasis of  $H$ (explicitly defined below, in Eq.~\eqref{DephOperation}).
The first term, $\sigD$, 
measures the entropic distance between the populations of the actual final state $\rho_\tau$ and those of the reference thermal state $\rho_\tau^\text{th}$, and is generally identified with the classical contribution. 
The term $\sigC$, in turn, is known as the relative entropy of coherence and compares the final state $\rho_{\tau}$ with the dephased state $\mathbb{D}_{H_{\tau}}(\rho_{\tau})$.
It hence captures the contribution from coherences in the energy basis.
By construction, $\sigD$ and $\sigC$ in Eq.~\eqref{CD_split} are both non-negative, which shows that coherences increase the entropy production in the process, as compared to a fully classical (incoherent) scenario.
\change{One should also clarify that, since the changes in populations and coherences are inevitably mixed, the terminology ``classical''~vs.~~``quantum'' is not entirely precise, nor is there a one-to-one relationship between this and the terms ``populations'' and ``coherences''. 
For instance, while $\sigC$ depends only on the basis rotation (coherences), 
$\sigD$ depends on both the changes in energy eigenvalues, as well as the eigenbasis rotation.
Notwithstanding, as we will show, in the case of infinitesimal quenches, these distinctions can be made precise.} 

The splitting~\eqref{CD_split}, first analyzed in \cite{janzing2006quantum}, has been studied in the context of the resource theory of thermodynamics \cite{Lostaglio2015}, relaxation towards equilibrium~\cite{Santos2019,Mohammady2020},thermodynamics of quantum optical systems~\cite{Elouard2020} and work protocols in the absence of a bath~\cite{Francica2019,Varizi2020,Francica2020}.
At the stochastic level, both $\sigC$ and $\sigD$  satisfy individual fluctuation theorems~\cite{Francica2019}, which is a very desirable property. 
Moreover, $\sigD$ has a resource-theoretic interpretation within the resource theory of athermality~\cite{Brandao2013,Horodecki2013}, while $\sigC$ is a natural monotone in the resource theory of coherence~\cite{Baumgratz2014,Streltsov2016a}.
These facts make the splitting~\eqref{CD_split} a valuable tool in understanding the relative contribution of classical and quantum features to non-equilibrium processes. 
However, working with various models, we have observed that this splitting behaves strangely, even in some simple  protocols. More specifically, we identify two main shortcomings.

The first concerns the relative magnitudes of $\sigC$ and $\sigD$:  \change{At low temperatures, $\sigD$ will always be much larger than $\sigC$. 
The reason is purely mathematical:  $\sigC$ is a special kind of relative entropy because it can be expressed as a difference between two von Neumann entropies, as in the second equality of~\eqref{C}. 
As $\beta \to \infty$, $\rho_\tau^\text{th}$ tends to a pure state and hence $S(\rho_\tau)$ tends to zero,
while $S(\mathbb{D}_{H_\tau}(\rho_\tau))\in [0,\ln d]$, where $d$ is the dimension of the Hilbert space. As a consequence, $\sigC$ will always remain finite. 
The term $\sigD$, on the other hand, generally diverges when the support of $\rho_\tau$ is not contained in that of $\rho_\tau^\text{th}$~\cite{Nielsen}, meaning $\sigD$ will grow unbounded when $\beta \to \infty$.
This implies that it is \emph{impossible} to construct a low-temperature process where the quantum term dominates. 
}

\change{More precisely, consider again the two types of drivings depicted in Fig. \ref{fig:abcd} (a):  one that alters the spacing of the energy levels (associated here to a classical process), and one which may rotate eigenvectors (associated here to a quantum process). At strictly zero temperature,  a Gibbs state is invariant under the first class of protocols, and hence we may expect that any entropy production in \eqref{EntProd} has  a quantum origin. However, the opposite identification arises in the splitting \eqref{CD_split}. The reason for this apparent contradiction is rather simple: The splitting \eqref{CD_split} is not characterising whether the driving  generates quantum coherence or not; rather, given a possibly coherent process, it characterises how much the final diagonal and off-diagonal terms contribute to the total entropy production. }



This issue can be neatly illustrated by a minimal qubit model. Consider a qubit which starts at $H_0 = \omega \sigma^z$ and is suddenly quenched ($U = 1$) to $H_\tau = \omega(\sigma^z \cos\theta + \sigma^x \sin\theta)$ (where $\sigma^\alpha$ are Pauli matrices). In this quench the energy levels remain intact and all that happens is that the eigenbasis is rotated by an angle $\theta$. This is thus, by all accounts, a highly quantum process. 
The entropy production~\eqref{EntProd} for this model reads 
\begin{equation}\label{qubit_Sigma}
     \Sigma = 2 t \tanh^{-1}(t) \;\sin^2(\nicefrac{\theta}{2}),
\end{equation}
where $t = \tanh(\beta \omega) \in [0,1]$.
On the other hand, the coherent contribution $\sigC$ in Eq.~\eqref{C}, reads
\begin{IEEEeqnarray}{rCl}
\label{qubit_C}
    \sigC &=& t \tanh^{-1}(t) - t\cos\theta \tanh^{-1}(t\cos\theta)  \IEEEeqnarraynumspace \\[0.2cm]
    &&
    -\frac{1}{2} \ln\Big(1+\sinh^2(\beta\omega)\sin^2\theta
    \Big).
    \nonumber
\end{IEEEeqnarray}
A plot of $\sigC$ and $\sigD = \Sigma - \sigC$ is shown in Fig.~\ref{fig:abcd}(c) as a function of $\beta\omega$, for $\theta = 1.1$. As can be seen, in general both quantities are comparable in magnitude. But, as the temperature goes down ($\beta$ goes up), the classical contribution becomes increasingly larger and eventually dominates. Thus, at very low temperatures, most of $\Sigma$ comes from the population term $\sigD$ and very little from coherences.

\change{The above considerations highlight the fact that splitting the total entropy production \eqref{EntProd} in a classical and quantum contribution may be highly non-trivial, and that different splittings might provide different insights. In particular, we argue that the splitting in Eq. \eqref{CD_split} does not appropriately distinguish coherent from non-coherent drivings (see Fig. \ref{fig:abcd}), but instead characterises how populations and off-diagonal terms contribute to entropy production. In this work, we will propose a new complementary splitting that better incorporates the difference between coherent and non-coherent drivings. }

A second issue with the splitting~\eqref{CD_split} concerns \emph{infinitesimal quenches}. This is a very important scenario, widely studied in the context of critical systems~\cite{Gambassi1106,Dorner2012,Fusco2014a,Goold2018} and quasi-isothermal  processes~\cite{Miller2019,Scandi2019}. 
The idea is to analyze the entropy production perturbatively, for a small instantaneous quench of the work parameter, from $g$ to $g + \delta g$.  
The problem with $\sigC$ and $\sigD$ in this case is that, as will be shown, the parameter $\delta g$  appears multiplied by a factor that increases exponentially with $\beta$. 
Hence, the radius of convergence of $\sigC$ and $\sigD$, in $\delta g$, tends to zero exponentially fast as $\beta\to \infty$.
For $\Sigma$, no such issue arises. 

This is again well illustrated by the qubit example in Eqs.~\eqref{qubit_Sigma} and~\eqref{qubit_C}, where the quench parameter is now the angle~$\theta$. 
We  see that $\Sigma$ in~\eqref{qubit_Sigma} can be readily expanded in powers of $\theta$, for any temperature $\beta$ (or any $t = \tanh(\beta \omega)$).  The same is not true for $\sigC$, however.  The problem is in the third term of Eq.~\eqref{qubit_C}, which is a function of $x = \sinh^2(\beta\omega)\sin^2\theta$.
This quantity appears inside a logarithm, in the form $\ln(1+x)$.  However, a series expansion of $\ln(1+x)$ only converges if $|x|<1$.  And since the prefactor $\sinh^2(\beta\omega)$  grows exponentially with $\beta$, at  low temperatures, extremely small values of $\theta$ are required to validate a series expansion. 

More generally, one can readily show that for $\Sigma$ this issue does not arise.
If we use  $\Sigma = \beta \big(\langle W \rangle - \Delta F \big)$, we find in the case of infinitesimal quenches that
\begin{equation}
    \Sigma = \beta \tr\big\{\Delta H \rho^\text{th}(g_0) \Big\} - \beta \Delta F, 
\end{equation}
where $\Delta H = H(g_0 + \delta g) - H(g_0)$ and $\Delta F = F(g_0 + \delta g) - F(g_0)$. 
A series expansion of $\Sigma$ in $\delta g$ therefore amounts to two things. First, an expansion of $\Delta H$ in powers of $\delta g$, which is entirely independent of $\beta$. 
And second, an expansion of $F(g)$, which is an analytic and generally smooth function (except possibly at a critical point~\cite{Dorner2012}). 
Indeed, if $H(g)$ is linear in $g$, the leading order contribution to the expansion becomes~\cite{Fusco2014a}
\begin{equation}
    \Sigma \simeq  -\frac{1}{2}\beta \delta g^2\frac{\partial^2 F}{\partial g_0^2}, 
\end{equation}
showing that $\Sigma$ is simply proportional to the equilibrium susceptibility, a textbook quantity used throughout equilibrium statistical mechanics.

The above results show that, despite its interesting properties (individual fluctuation theorems and resource-theoretic interpretation), the splitting~\eqref{CD_split} 
\change{is not a fully satisfying splitting of the entropy production into a classical and quantum contribution (in the sense described in Fig. \ref{fig:abcd} (a))}. 
\change{In order to capture the difference between coherent and non-coherent drivings,}  in this paper
we  propose  a different splitting,  which is inspired by the recent results of~\cite{Scandi2019}.  We label it as 
\begin{equation}\label{AB_split}
    \Sigma = \sigB + \sigA. 
\end{equation}
The actual definitions of $\sigA$ and $\sigB$ will be given below in Sec.~\ref{sec:BandA} and a stochastic trajectories formulation will be given in Sec.~\ref{sec:StochTrajectories}.
A comparison in the case of the minimal qubit example is also presented in Fig.~\ref{fig:abcd}(d). 
In this case, using the results of Sec.~\ref{sec:BandA}, one finds the following elegant expression for $\sigA$ (to be contrasted with Eq.~\eqref{qubit_C}):
\begin{equation}\label{qubit_A}
    \sigA = \frac{1}{2} \ln \left(\frac{1-\tanh^2(\beta \omega \cos\theta)}{1-\tanh^2(\beta \omega)}\right).
\end{equation}
As seen in Fig.~\ref{fig:abcd}(d), $\sigA$ and $\sigB$  behave as desired: Since the process is highly coherent, $\sigB$ is very small; and as the temperature goes down, $\sigA$ grows monotonically, showing that cold processes have higher contributions from the coherences.

\begin{figure}
    \centering
    \includegraphics[width=0.45\textwidth]{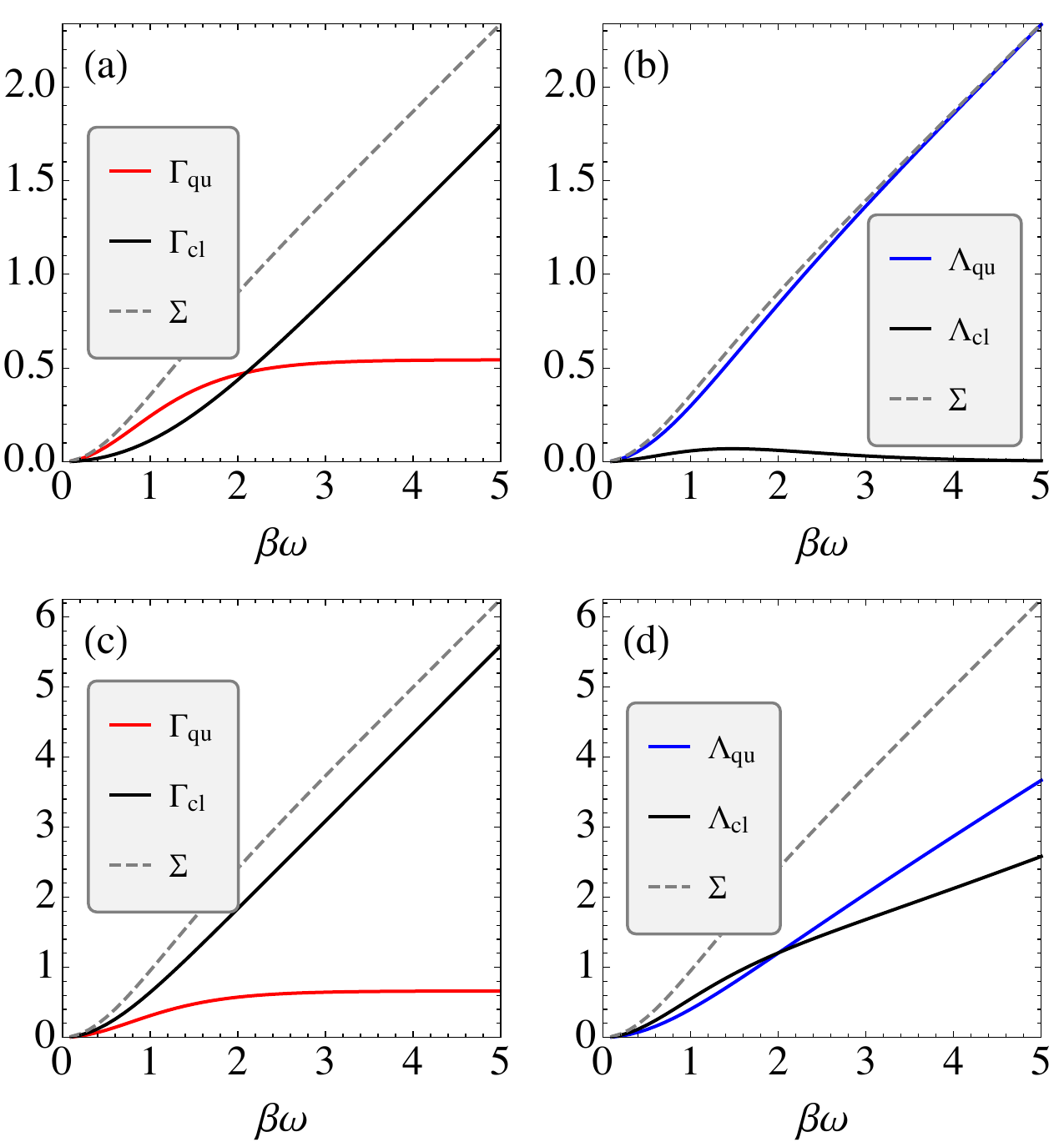}
    \caption{Splitting of the entropy production in a cyclic qubit model, $H_\tau = H_0 = \omega \sigma_z$. The unitary is  generated by an $x$-pulse with a duration $\tau$; that is, $U = e^{-i \tau (H_0 + h_x \sigma_x)}$, where $h_x$ is the pulse intensity. The curves were computed using Eqs.~\eqref{Sigma_free_energy}, \eqref{C_free_energy}, \eqref{D_free_energy}, \eqref{A_free_energy} and~\eqref{B_free_energy}, with $\omega = 1$, $h_x = 1.3$ and two different values of $\tau$: in (a),(b) $\tau = 0.4$ and in (c),(d) $\tau = 1$. 
    }
    \label{fig:qubit2}
\end{figure}

The features discussed in Fig.~\ref{fig:abcd} are not restricted to quenches. To illustrate that we show in Fig.~\ref{fig:qubit2} another qubit example, where  the process is assumed to be cyclic, with $H_\tau = H_0 = \omega \sigma_z$, and the unitary is taken to be generated by an $x$-pulse with a duration $\tau$; that is, $U = e^{-i \tau (H_0 + h_x \sigma_x)}$, where $h_x$ is the pulse intensity. 
Fig.~\ref{fig:qubit2} illustrates the results for $\omega = 1$, $h_x = 1.3$ and two choices of $\tau$: in the upper panels $\tau = 0.4$ and in the lower panels $\tau = 1$. 
The results show that for~\eqref{CD_split} the behavior is always roughly the same, with $\sigD$ always eventually dominating at low temperatures.
Conversely, for the new splitting~\eqref{AB_split} a richer competition is observed. Depending on the parameters we may either have $\sigA$ dominating, or $\sigB$, or both. 

As we will show in this paper, our new  splitting~\eqref{AB_split} \change{more accurately distinguishes which part of the entropy production   is generated by a commuting or non-commuting drive. This provides a complementary approach to the standard splitting in Eq.~\eqref{CD_split}, which instead describes how populations and coherences in the final state contribute to entropy production.
On the other hand, we also note that} $\sigA$ and $\sigB$ do not share some of the nice properties of $\sigC$ and $\sigD$. 
First, $\sigA$ cannot be directly linked with a monotone for coherence or asymmetry~\cite{Streltsov2016a}. 
Second, while $\sigB$ always satisfies an individual fluctuation theorem, $\sigA$ only does so in the case of infinitesimal quenches.
\change{Different properties}  of each splitting are highlighted in Table~\ref{tabABCD}.
We also show that for infinitesimal quenches at high temperatures, both splittings coincide - see Sec.~\ref{ssec:stoch_infinitesimal}.


To illustrate the usefulness of our results,  we analyze our new splitting in two quantum many-body problems. 
Previous works have focused on the behaviour of the statistics of  work and entropy production $\Sigma$ for quantum quenches~\cite{Gambassi1106,Dorner2012,Fusco2014a,PhysRevE92022108,Goold2018,PhysRevA99043603}, with emphasis in quantum phase transitions~\cite{Dorner2012,Mascarenhas1307,PhysRevA95063615,Pagnelli,PhysRevE98022107,PhysRevB93201106,Bayocboc2015,PhysRevE97052148,Nigro}.
Motivated by this, we analyze in Sec.~\ref{sec:XYmodel} the transverse field Ising model (TFIM), and discuss the behavior of~\eqref{AB_split} in the vicinity of the quantum critical point. 
This is complementary to the analysis put forth in~\cite{Varizi2020}, which studied Eq.~\eqref{CD_split}.
Then, in Sec.~\ref{sec:LMG},  we consider a macrospin of varying size and focus on the full statistics of $\sigA$ and $\sigB$, including their probability distributions and their first four cumulants. 
We finish with conclusions and future perspectives in Sec.~\ref{sec:conclusion}.



\begin{table}
\begin{center}
\caption{\label{tabABCD} Comparison between $\sigA$, $\sigB$, $\sigC$ and $\sigD$.}
    \begin{tabular}{l|c|c||c|c}
    
                            & \;$\sigA$\; & \;$\sigB$\; & \;$\sigC$\; & \;$\sigD$\; \\
    \hline
    Fluctuation Theorem     & \xmark & \cmark & \cmark & \cmark \\
    \hline
    Fluctuation Theorem when $\Delta H\to0$ & \multicolumn{2}{c||}{\cmark} & \multicolumn{2}{c}{\cmark} \\
    \hline
    Analytic when $\Delta H\to0$ and low $T$ & \multicolumn{2}{c||}{\cmark} & \multicolumn{2}{c}{\xmark} \\
    \hline

    Resource-theoretic interpretation & \multicolumn{2}{c||}{\xmark} & \multicolumn{2}{c}{\cmark} \\
    \hline    

    Vanishing for commuting protocols & \cmark & \lmark & \cmark & \lmark \\
    \hline
    Dominant for highly coherent protocols & \cmark & \lmark & \xmark & \lmark \\
    \hline
    Dominant at low temperatures & \cmark & \lmark & \xmark & \lmark \\
    \hline
    
    \end{tabular}
    \end{center}
\end{table}


\section{\label{sec:BandA}Splittings of the entropy production}

In this section we introduce our alternative splitting of the entropy production [Eq.~\eqref{AB_split}].
We focus for now at the level of averages; the corresponding stochastic formulation will be presented in Sec.~\ref{sec:StochTrajectories}.

Let $\mathcal{O}$ denote any Hermitian observable and decompose it as $\mathcal{O} = \sum_\alpha o_\alpha \Pi_\alpha$, where $\Pi_\alpha$ are projectors onto the subspaces with eigenvalues $o_\alpha$. 
We define the dephasing operation
\begin{equation}\label{DephOperation}
    \mathbb{D}_\mathcal{O}(\bullet) = \sum\limits_\alpha \Pi_\alpha \bullet \Pi_\alpha.  
\end{equation}
The rationale of the splitting Eq.~\eqref{CD_split} was to introduce an intermediate step, associated with the state $\mathbb{D}_{H_{\tau}}(\rho_{\tau})$ (Fig.~\ref{fig:abcd}(b)). This represents the final state $\rho_\tau$ dephased in the eigenbasis of the final Hamiltonian. 
If the process generates coherences, this state will differ from the actual final state $\rho_\tau$ and their entropic distance will be precisely $\sigC$ in Eq.~\eqref{C}.

For convenience, we introduce the non-equilibrium free energy, associated with the final Hamiltonian $H_\tau$
\begin{equation}
    F(\rho) = \tr\Big\{H_\tau \rho\Big\} - T S(\rho). 
\end{equation}
Non-equilibrium free energies depend on two parameters, $H$ and $\rho$. However, in this paper, we will henceforth only need free energies defined with respect to $H_\tau$, so we write it more simply as $F(\rho)$.
In terms of $F$, the entropy production~\eqref{EntProd} can be written as 
\begin{equation}\label{Sigma_free_energy}
    \Sigma = \beta\Big\{F(\rho_\tau) - F(\rho_\tau^\text{th})\Big\}, 
\end{equation}
Similarly, one can also express $\sigC$ and $\sigD$ in terms of free energy differences. 
Since $\tr\big\{ H_\tau \mathbb{D}_{H_\tau}(\rho_\tau)\big\} = \tr\big\{ H_\tau \rho_\tau\big\}$, one finds that
\begin{IEEEeqnarray}{rCl}
\label{C_free_energy}
    \sigC &=&  \beta\Big\{ F(\rho_\tau) - F\big(\mathbb{D}_{H_\tau}(\rho_\tau)\big)\Big\}, \\[0.2cm]
\label{D_free_energy}
    \sigD &=& \beta\Big\{ F\big(\mathbb{D}_{H_\tau}(\rho_\tau)\big) - F (\rho_\tau^\text{th}) \Big\},
\end{IEEEeqnarray}
which clearly add up to $\Sigma$.

The splitting~\eqref{CD_split} uses $\mathbb{D}_{H_{\tau}}(\rho_{\tau})$ as intermediate state. 
Our new splitting~\eqref{AB_split} follows a similar logic, but in reverse: Instead of working with $\rho_\tau$ dephased in the basis of $H_\tau$, we work with $H_\tau$ dephased in the basis of $\rho_\tau$. 
More precisely, we define 
\begin{equation}\label{rho_tilde}
    \tilde{\rho}_\tau^\text{th} = \frac{\exp\{-\beta~ \mathbb{D}_{\rho_\tau}(H_\tau)\}}{\tr\Big(\exp\{-\beta~ \mathbb{D}_{\rho_\tau}(H_\tau)\}\Big)},
\end{equation}
which is a thermal state based only on the incoherent part of $H_\tau$, in the basis of $\rho_\tau$ (as a consequence, $[\tilde{\rho}_\tau^\text{th},\rho_\tau] = 0$).
With this in mind, we now define
\begin{IEEEeqnarray}{rCl}
 \label{B_free_energy}
    \sigB &=& \beta \Big\{ F(\rho_\tau) - F(\tilde{\rho}_\tau^\text{th})\Big\},
    \\[0.2cm]
\label{A_free_energy}
    \sigA &=& \beta \Big\{F(\tilde{\rho}_\tau^\text{th}) - F(\rho_\tau^\text{th})\Big\}, 
\end{IEEEeqnarray}
which add up to $\Sigma$, as in Eq.~\eqref{AB_split}.
The first term, $\sigB$, compares the two \emph{commuting} states $\rho_\tau$ and $\tilde{\rho}_\tau^\text{th}$ and is hence associated with their population mismatch. 
The nonnegativity of $\sigB$ becomes evident by noting that it can also be written as 
\begin{equation}\label{B-RelativeEntropy}
    \sigB = S(\rho_\tau || \tilde{\rho}_\tau^\text{th}). 
\end{equation}
The term $\sigA$, on the other hand, compares $\rho_\tau^\text{th} \propto e^{-\beta H_\tau}$ with $\tilde{\rho}_\tau^\text{th} \propto e^{-\beta \mathbb{D}_{\rho_\tau}(H_\tau)}$. 
Unlike $\sigB$, the contribution $\sigA$ cannot be written as a relative entropy. In fact, written down explicitly, it reads
\begin{equation}\label{A-TraceLog}
    \sigA = \tr\Big\{ \rho_\tau \Big( \ln \tilde{\rho}_\tau^\text{th} - \ln \rho_\tau^\text{th}\Big) \Big\}. 
\end{equation}
Notwithstanding, as shown in Appendix~\ref{appsec:nonnegA}, it turns out that $\sigA$ is still non-negative, and zero if and only if $[H_\tau, \rho_\tau] = 0$.

Throughout this paper we will provide several additional justifications as to why the choices~\eqref{B_free_energy} and~\eqref{A_free_energy} are physically reasonable, starting in Sec.~\ref{subsec:AB-Quench}. 
But before doing so, let us briefly revisit the minimal qubit model defined above Eq.~\eqref{qubit_Sigma}.
The process is a quench ($U = 1$), so $\rho_\tau = \rho_0^\text{th}$. 
Hence, all we need to do in order to compute $\sigA$ is to dephase the final Hamiltonian $H_\tau = \omega(\sigma^z \cos\theta + \sigma^x \sin\theta)$ in the basis of $\rho_0^\text{th}$.
Or, what is equivalent, in the basis of $H_0$. The result is thus simply 
$\mathbb{D}_{\rho_\tau}(H_\tau) = \omega \cos(\theta) \sigma_z$.
Using this in~\eqref{A_free_energy}  yields Eq.~\eqref{qubit_A},
which is the result plotted in Fig.~\ref{fig:abcd}(d) and discussed in Sec.~\ref{sec:int}.



\subsection{\label{subsec:AB-Quench} Infinitesimal quenches}

The physics of the problem becomes particularly simpler in the case of infinitesimal quenches. 
We therefore now specialize the above results to this scenario. 
This will provide strong justifications in favor of the new splitting~\eqref{AB_split}. Furthermore, in this limit the splitting~\eqref{AB_split} becomes equivalent to the one recently put forward in~\cite{Scandi2019}. More precisely,  in~\cite{Scandi2019} the authors describe quasi-isothermal processes as a series of infinitesimal quenches, and in particular consider how $\Sigma$ splits into a classical and quantum contribution. Focusing on a single  infinitesimal quench, both approaches become directly comparable and, as we will show, agree with each other.  

We thus analyze what happens if we take $U = 1$, and assume that $H$ changes only by a small amount $\Delta H$ (i.e., we write $H_{\tau} = H_0 + \Delta H$).
Since $U=1$, the state of the system remains unchanged: $\rho_{\tau} = \rho^\text{th}_0$.
Therefore, dephasing $H_\tau$ in the basis of $\rho_\tau$ is equivalent to dephasing in the basis of $H_0$:
\begin{equation}
    \mathbb{D}_{\rho_{\tau}}(H_{\tau}) = \mathbb{D}_{\rho^\text{th}_0}(H_{\tau}) = \mathbb{D}_{H_0}(H_{\tau}).
\end{equation}

Let us define the dephased (incoherent) and coherent parts of the perturbation $\Delta H$, in the initial energy basis, $\Delta H^\text{d} = \mathbb{D}_{H_0}(\Delta H)$ and $\Delta H^\text{c} = H_{\tau} - \mathbb{D}_{H_0}(H_{\tau})$. 
Then, following a procedure detailed  in Appendix B of Ref.~\cite{Scandi2019}, one may show that, 
\begin{IEEEeqnarray}{rCl}
\label{rho-deph-expan}
    \tilde{\rho}^{\text{th} }_{\tau} &=& \rho^\text{th}_0 - \beta~ \mathbb{J}_{\rho^\text{th}_0} [\Delta H^\text{d} - \langle\Delta H^\text{d}\rangle_0] + \mathcal{O}(\Delta H^2),
    \\[0.2cm]
\label{rho-tau-expan}
    \rho^{\text{th}}_{\tau} &=& \rho^\text{th}_0 - \beta~ \mathbb{J}_{\rho^\text{th}_0} [\Delta H - \langle\Delta H\rangle_0] + \mathcal{O}(\Delta H^2),
\end{IEEEeqnarray}
where $\langle \ldots \rangle_0 = \tr\{\ldots 
\rho_0^\text{th}\}$ and $\mathbb{J}_{\rho}$ is a super-operator defined as
\begin{equation}
    \mathbb{J}_{\rho}[\bullet] = \int_0^1 \rho^{t}\bullet\rho^{1-t}\mathrm{d}t.
\end{equation}
We see that both $\rho_\tau^\text{th}$ and $\tilde{\rho}_\tau^\text{th}$ can be expanded essentially in a power series in $\beta \Delta H$. 
Conversely, the same is not true for the state $\mathbb{D}_{H_\tau}(\rho_0^\text{th})$ entering~\eqref{D_free_energy} and~\eqref{C_free_energy}. 
In fact, one may show that to order $\Delta H$ \footnote{This is done by noting that the dephasing $\mathbb{D}_{H}(\rho)$ can be also given by
\begin{equation*}
    \mathbb{D}_{H}(\rho) = \lim_{s\to\infty}\frac{1}{s} \int_0^s \mathrm{d}t e^{-iHt}\rho e^{iHt}.
\end{equation*}
We then use that $e^{t(H_0+\Delta H)} = e^{tH_0} + t\mathbb{J}_{e^{tH_0}}[\Delta H] + \mathcal{O}(\Delta H^2)$ and $[\rho_0^{\text{th} }, H_0]=0$. To order $\Delta H$ this gives Eq.~\eqref{Deph-state-expansion}}
\begin{equation}\label{Deph-state-expansion}
    \mathbb{D}_{H_0+\Delta H}(\rho_0^{\text{th} }) = \rho_0^{\text{th} } + \lim_{s\to\infty}\frac{i}{s}\int
    _0^{s}\mathrm{d}t \int_0^1\mathrm{d}x \, t~e^{-i x H_0 t}[\rho_0^{\text{th} },\Delta H]e^{i x H_0 t}.
\end{equation}
Even though this is an expansion in $\Delta H$, the dependence on $\beta$ enters in a highly non-trivial way. 
This explains the non-analytic behavior of $\sigD$ and $\sigC$ at low temperatures, discussed in Sec.~\ref{sec:int}.

Plugging~\eqref{rho-deph-expan}-\eqref{rho-tau-expan} in  Eqs.~\eqref{EntProd},~\eqref{B-RelativeEntropy} and~\eqref{A-TraceLog}  leads, up to second order, to \begin{IEEEeqnarray}{rCl}
\label{Sigma_Expand}
    \Sigma &=& \frac{\beta^2}{2} \tr{ \Delta H \, \mathbb{J}_{\rho^\text{th}_0} [\Delta H - \langle\Delta H\rangle_0] } = \sigB + \sigA,
    \\[0.2cm]
\label{B_Expand}
    \sigB &=& \frac{\beta^2}{2} \tr{ \Delta H^\text{d} \, \mathbb{J}_{\rho^\text{th}_0} [\Delta H^\text{d} - \langle\Delta H^\text{d}\rangle_0] },
    \\[0.2cm]
\label{A_Expand}
    \sigA &=& \frac{\beta^2}{2} \tr{ \Delta H^\text{c} \, \mathbb{J}_{\rho^\text{th}_0} [\Delta H^\text{c}] },
\end{IEEEeqnarray}
where we used the fact that $\langle \Delta H^\text{d}\rangle_0 = \langle \Delta H \rangle_0$.
The interesting aspect of these results is that, within this infinitesimal quench limit, $\sigB$ and $\sigA$ are found to be related to $\Sigma$ via the simple separation of the perturbation, $\Delta H = \Delta H^\text{d} + \Delta H^\text{c} $, into a dephased and a coherent part. 
These results also coincide with the splitting proposed in~\cite{Scandi2019}.

%

An additional justification for the splitting~\eqref{AB_split} can be given in terms of the fluctuation-dissipation relation (FDR).
As shown in Refs.~\cite{Miller2019,Scandi2019}, Eq.~\eqref{Sigma_Expand} can also be written as
\begin{equation}\label{FDRm}
    \Sigma = \frac{1}{2} \beta^2 \, \text{Var}_0[\Delta H] - \beta\mathcal{Q},
\end{equation}
where $\text{Var}_0 [\Delta H] = \langle \Delta H^2 \rangle_0 - \langle \Delta H \rangle_0^2$, is the variance of the perturbation,
and
\begin{equation}
    \mathcal{Q} = \frac{\beta}{2} \int_0^1\mathrm{d}y \, I^y(\rho^\text{th}_0,\Delta H ) \geqslant 0,
\end{equation}
is a measure of quantum coherence, associated with the so-called Wigner-Yanase-Dyson skew information~\cite{Petz2002}
\begin{equation}\label{WYD-info}
    I^y(\varrho,X) = -\frac{1}{2} \tr{ [\varrho^y,X] [\varrho^{1-y}, X] }.
\end{equation}
For incoherent processes one recovers the usual FDR $\Sigma = \frac{\beta^2}{2} \text{Var}_0[\Delta H]$ \cite{Jarzynski1997}. 
But when the process is coherent, the FDR is broken by a term $-\beta \mathcal{Q}$.
Repeating the same procedure for $\sigB$ and 
$\sigA$, one readily finds that 
\begin{equation}\label{BA_infinitesimal}
    \sigB = \frac{\beta^2}{2}  \, \text{Var}_0 [ \Delta H^\text{d} ], 
    \qquad
      \sigA = \frac{\beta^2 }{2} \, \text{Var}_0 [ \Delta H^\text{c} ] - \beta \mathcal{Q}.
\end{equation}
Whence, $\sigB$ always satisfies a standard FDR, and all violations are associated to $\sigA$. 
This provides additional justification as to why $\sigA$ is referred to as a quantum contribution.

\change{In the case of high temperatures ($\beta \to 0$), one may show that $\mathcal{Q} \sim \mathcal{O}(\beta)^3$. 
Moreover, the state entering the variances in Eq.~\eqref{BA_infinitesimal} can be replaced with the maximally mixed state $\mathbb{I}/d$. 
As a consequence, we find that to leading order in $\beta$, 
\begin{equation}\label{BA_infinitesimal2}
    \sigB = \frac{\beta^2}{2}  \, \text{Var}_{\mathbb{I}/d} [ \Delta H^\text{d} ], 
    \qquad
      \sigA = \frac{\beta^2 }{2} \, \text{Var}_{\mathbb{I}/d} [ \Delta H^\text{c} ].
\end{equation}
Both contributions are thus found to scale as $\beta^2$ in this limit, which agrees with the observations in Fig.~\ref{fig:abcd}(c) and (d).
However, their relative contribution will be determined by the variance of $\Delta H^d$ and $\Delta H^c$ in the maximally mixed state; hence, which term will be dominant will depend on the details of the process (either a commuting or a non-commuting drive). 
This is also expected to remain true for general drives.
}

%
\section{\label{sec:StochTrajectories}Stochastic trajectories}
%

We now discuss how to formulate  the splittings~\eqref{CD_split} and \eqref{AB_split} at the level of stochastic trajectories, based on a standard two-point measurement (TPM) scheme~\cite{Talkner2007}. 
Since $\Sigma = \beta(\langle W \rangle - \Delta F)$, the statistics of $\Sigma$ can be obtained  solely from measurements in the eigenbasis of the initial and final Hamiltonians. 
As first shown in~\cite{Francica2019}, a major advantage of the original splitting~\eqref{CD_split} is that this remains true when assessing the individual contributions $\sigD$ and $\sigC$; that is, no additional measurements are necessary.
As we will now show, the same is also true for $\sigB$ and $\sigA$ [Eq.~\eqref{AB_split}].
This means that both splittings can be assessed, at the stochastic level, with the same amount of information as a standard TPM. 

Irrespective of the splitting one is interested in, the protocol may therefore be described as follows. 
Initially the system is in the thermal state $\rho_0^\text{th}$, associated with the Hamiltonian $H_0=\sum_i \epsilon_i^0 | i_0 \rangle \langle i_0 |$. 
The first measurement is performed in the basis $|i_0\rangle$, which occurs with probability 
$p_i^0 =  e^{-\beta \epsilon_i^0}/Z_0$.
Conversely, the second measurement is performed at time $\tau$, after the map~\eqref{rho_tau}, and in the eigenbasis of the final Hamiltonian 
$H_{\tau} = \sum_j \epsilon_j^{\tau} | j_\tau \rangle \langle j_\tau |$.
The bases $\{|i_0\rangle\}$ and $\{|j_\tau\rangle\}$ are, in general, not compatible.

The conditional probability of finding the system in $|j_\tau\rangle$ given that it was initially in $|i_0\rangle$ is
$|\langle j_\tau | U | i_0 \rangle|^2$.
The probability associated with the forward protocol $|i_0\rangle \to |j_\tau\rangle$ is thus 
$\mathcal{P}_F[i,j] = |\langle j_\tau| U | i_0 \rangle|^2~p_i^0$. 
The dynamics is defined as being incoherent when 
$|\langle j_\tau | U | i_0 \rangle|^2 = \delta_{i,j}$, which means $U$ is not able to generate transitions between states of the initial and final Hamiltonians. 
Similarly, in the backward protocol the system starts in $\rho_\tau^\text{th}$ and one measures first in the basis of $H_\tau$, yielding $|j_\tau\rangle$ with probability 
$p_j^\tau = e^{-\beta \epsilon_j^\tau}/Z_\tau$. The time-reversed unitary $U^\dagger$ is then applied, after which  one measures in the basis $|i_0\rangle$ of $H_0$. 
This yields the backward distribution 
$\mathcal{P}_B[i,j] = |\langle i_0 | U^\dagger | j_\tau \rangle|^2 p_j^\tau$.

The entropy production associated to the trajectory $|i_0\rangle \to |j_\tau\rangle$ is now defined as usual:
\begin{equation}\label{ent_stoch_1}
    \sigma[i,j] = \ln \frac{\mathcal{P}_F[i,j]}{\mathcal{P}_B[i,j]} = \ln p_i^0/p_j^\tau. 
\end{equation}
The second equality follows from the fact that $|\langle i_0 | U^\dagger | j_\tau\rangle|^2 = |\langle j_\tau | U | i_0 \rangle|^2$.
As a consequence, $\sigma[i,j]$ depends only on the equilibrium populations $p_i^0$ and $p_j^\tau$, associated with the initial and final Hamiltonians. 
As can be readily verified, $\langle \sigma[i,j]\rangle = \sum_{i,j} \sigma[i,j] \mathcal{P}_F[i,j] = \Sigma$, returns precisely Eq.~\eqref{EntProd}.
In addition, $\sigma[i,j]$ also satisfies an integral fluctuation theorem $\langle e^{-\sigma} \rangle = 1$ (see Eq.~\eqref{CGF_sigma} for more details).

\subsection{\label{ssec:stoch_def}Stochastic definitions for the splittings~\eqref{CD_split} and~\eqref{AB_split}}

Following~\cite{Francica2019}, we now define stochastic quantities associated to $\sigD$ and $\sigC$. 
In order to do that, we first write the dephased state $\mathbb{D}_{H_{\tau}}(\rho_{\tau})$ as
$ \mathbb{D}_{H_{\tau}}(\rho_{\tau}) = \sum_j q_j^{\tau} | j_\tau \rangle \langle j_\tau |$, where
\begin{equation}\label{q_def}
    q_j^\tau = \langle j_\tau |\rho_\tau| j_\tau\rangle =  \sum\limits_i |\langle j_\tau| U | i_0 \rangle|^2 p_i^0.
\end{equation}
In passing, we note that $q_j^\tau = \sum_i \mathcal{P}_F[i,j]$, so $q_j^\tau$ can also be interpreted as the marginal distribution of the final measurement.
As shown in~\cite{Francica2019}, we may now define 
\begin{IEEEeqnarray}{rCl}
\label{d_stoch}
    \sigDij[i,j] &=& \ln q_j^{\tau}/p_j^{\tau},
    \\[0.2cm]
\label{c_stoch}    
    \sigCij[i,j] &=& \ln p_i^0/q_j^{\tau}.
\end{IEEEeqnarray}
Clearly $\sigDij[i,j]+\sigCij[i,j] =\sigma[i,j]$, which is the stochastic analog of~\eqref{CD_split}. 
Moreover, $\langle \sigDij[i,j] \rangle = \sigD$ and $\langle \sigCij[i,j] \rangle = \sigC$.

Similarly, we construct  stochastic quantities for the new quantities $\sigB$ and $\sigA$ in Eq.~\eqref{AB_split}.
The central object now is the thermal state $\tilde{\rho}_{\tau}^{\text{th}}$, defined in Eq.~\eqref{rho_tilde} and associated with the Hamiltonian $\mathbb{D}_{\rho_\tau}(H_\tau)$.
Since the system evolves unitarily,
$\rho_\tau = U \rho_0^\text{th} U^\dagger = \sum_i p_i^0 |\psi_i\rangle\langle \psi_i |$, where $|\psi_i\rangle = U |i\rangle$. 
That is, $\rho_{\tau}$ has the same populations $p_i^0$ as $\rho_0^\text{th}$, but a rotated eigenbasis. 
Based on this, we can now write Eq.~\eqref{rho_tilde} as 
\begin{equation}
    \tilde{\rho}_{\tau}^{\text{th}}= \sum_i \tilde{p}_i^{\tau} | \psi_i \rangle \langle \psi_i |,
    \qquad 
    \tilde{p}_i^{\tau} = e^{-\beta (\tilde{\epsilon}_i^{\tau} - F(\tilde{\rho}_{\tau}^{\text{th}}))},
\end{equation}
where $\tilde{\epsilon}_i^{\tau} = \langle \psi_i | H_{\tau} | \psi_i \rangle $ are the eigenvalues of the dephased Hamiltonian $ \mathbb{D}_{\rho_{\tau}}(H_{\tau}) $ and $F(\tilde{\rho}_{\tau}^{\text{th}})$ is the same free energy as that appearing in Eq.~\eqref{B_free_energy}.
We then define
\begin{IEEEeqnarray}{rCl}
    \label{b_stoch}
    \sigBij[i,j] = \ln p_i^0/\tilde{p}_i^{\tau},
    \\[0.2 cm]
    \label{a_stoch}
    \sigAij[i,j] = \ln \tilde{p}_i^{\tau}/p_j^{\tau}.
\end{IEEEeqnarray}
These quantities satisfy $\sigBij[i,j] + \sigAij[i,j] = \sigma[i,j]$, as well as $\langle \sigBij[i,j] \rangle = \sigB$ and $\langle \sigAij[i,j]\rangle = \sigA$. 


\subsection{\label{ssec:CGF}Cumulant generating functions}

For all stochastic quantities in the previous section, we can define their corresponding probability distributions or, what is more convenient, their cumulant generating functions (CGFs). 
For instance, from~\eqref{ent_stoch_1} we  define 
\begin{equation}
P(\sigma) = \sum_{i,j} \mathcal{P}_F[i,j] \delta(\sigma-\sigma[i,j]),    
\end{equation}
from which we may  compute the CGF, $K_\sigma(v) = \ln \langle e^{-v \sigma }\rangle$.
With some manipulations, this can be neatly written as~\cite{Wei2017a,Guarnieri2018}
\begin{IEEEeqnarray}{rCl}\label{CGF_sigma}
    K_\sigma(v) &=& \ln \tr \Big\{ (\rho_\tau^\text{th})^v (\rho_\tau)^{1-v}\}
    \\[0.2cm]\nonumber
    &=& (v-1) S_{v}(\rho_{\tau}^{\text{th} } || \rho_{\tau}).
\end{IEEEeqnarray}
The second equality expresses the CGF in terms of the R\'enyi divergences $S_{v}(\rho||\sigma) = (v-1)^{-1} \ln \tr{\rho^v \sigma^{1-v}}$, which may be convenient in some situations.
Setting $v = 1$ yields $K_\sigma(1) = 0$, which is  the integral fluctuation theorem~\cite{Esposito2009,Campisi2011}
\begin{equation}\label{FT_sigma}
    \big\langle e^{-\sigma} \big\rangle = 1.
\end{equation}
In addition, from the CGF we may compute any cumulant of $\sigma$ as 
\begin{equation}
    \kappa_n(\sigma)= (-1)^n\frac{\partial^n K_\sigma}{\partial v^n} \bigg|_{v = 0},
\end{equation}
with $\kappa_1(\sigma) = \Sigma$ being the mean in Eq.~\eqref{EntProd}.

We may also compute the joint CGF of $\sigDij$ and $\sigCij$, defined as $K_{\sigDij,\sigCij}(v,u) = \ln \langle e^{-v \sigDij - u\sigCij} \rangle$.
With similar manipulations, it may  be written as 
\begin{equation}\label{CGF_CD}
    K_{\sigDij,\sigCij}(v,u) = \ln \tr\Big\{
    (\rho_\tau^\text{th})^v \big[\mathbb{D}_{H_\tau}(\rho_\tau)\big]^{u-v}
    (\rho_\tau)^{1-u}    
    \Big\}.
\end{equation}
The CGF of $\sigma = \sigDij+\sigCij$,  Eq.~\eqref{CGF_sigma}, is recovered by setting $u = v$; that is $K_\sigma(v) = K_{\sigDij,\sigCij}(v,v)$.
The reduced CGFs of $\sigDij$ and $\sigCij$ are found by setting $u =0$ or $v=0$, respectively:
\begin{IEEEeqnarray}{rCl}
\label{d_CGF}
    K_{\sigDij}(v)&=& \ln \tr \Big\{  (\rho_\tau^\text{th})^v \big[\mathbb{D}_{H_\tau}(\rho_\tau)\big]^{-v}
    \rho_\tau
    \Big\},\\[0.2cm]
\label{c_CGF}
    K_{\sigCij}(u)&=& \ln \tr\Big\{
     \big[\mathbb{D}_{H_\tau}(\rho_\tau)\big]^{u}
    (\rho_\tau)^{1-u}
    \Big\}. 
\end{IEEEeqnarray}
From this one may verify that $\sigDij$ and $\sigCij$ individually satisfy fluctuation theorems 
\begin{equation}
    \langle e^{- \sigDij} \rangle = \langle e^{- \sigCij} \rangle = 1.   
\end{equation}
Note also that, except in certain particular cases, Eq.~\eqref{CGF_CD} cannot be written as a sum of two CGFs, which means $\sigDij$ and $\sigCij$ are statistically dependent. 

Similarly, we compute the joint CGF of $\sigBij$ and $\sigAij$, defined as 
$K_{\sigBij,\sigAij}(v,u) = \ln \langle e^{-v \sigBij - u\sigAij} \rangle$.
It reads
\begin{equation}
    \label{CGF_AB}
    K_{\sigBij,\sigAij}(v,u) = \ln\tr\Big\{ ( \rho^{\text{th}}_{\tau} )^{u} ( \tilde{\rho}^{\text{th}}_{\tau} )^{v-u} ( \rho_\tau )^{1-v} \Big\}.
\end{equation}
The reduced CGFs of $\sigBij$ and $\sigAij$ are again found by setting $u=0$ an $v=0$,
\begin{IEEEeqnarray}{rCl}
\label{b_CGF}
    K_{\sigBij}(v) &=& \ln \tr\Big\{ (\tilde{\rho}_\tau^\text{th})^v (\rho_\tau)^{1-v}\Big\}
    %
    \\[0.2cm]
\label{a_CGF}
    K_{\sigAij}(u) &=& \ln \tr{ ( \rho^{\text{th}}_{\tau} )^{u} ( \tilde{\rho}^{\text{th}}_{\tau} )^{-u} \rho_\tau},
\end{IEEEeqnarray}
Once again, $\sigBij$ and $\sigAij$ are, in general, statistically dependent. 
Eq.~\eqref{b_CGF}  implies that $\sigBij$ satisfies a fluctuation theorem, 
\begin{equation}
    \langle e^{-\sigBij}\rangle = 1. 
\end{equation}
But the same is not true for $\sigAij$. 
Notwithstanding, as we will show, this property is recovered in the limit of infinitesimal quenches.

\subsection{\label{ssec:stoch_infinitesimal}Infinitesimal quenches}

As before, we now specialize the above expressions to the case of infinitesimal quenches.
Since $U=1$, the path probability reduces to $\mathcal{P}_F[i,j] = |\langle j_\tau| i_0\rangle|^2 p_i^0$. 
Moreover, since $\Delta H$ is assumed to be small, $|j_\tau\rangle$ will be close to $|i_0\rangle$ and   $\epsilon_j^\tau$ will be close to $\epsilon_j^0$.
For concreteness, we assume that the spectra of $H_0$ is non-degenerate. 
Standard perturbation theory then yields, to order $\Delta H^2$, 
\begin{equation}
    \epsilon_j^{\tau} = \epsilon_j^{0} + \Delta H_{jj} + E_j^{(2)},
\end{equation}
where $\Delta H_{ij} = \langle i_0 | \Delta H | j_0\rangle$ and 
$E_j^{(2)} = \sum_{l \neq j} |\Delta H_{jl}|^2/(\epsilon_j^0 - \epsilon_l^0)$. 
Note that if we split $\Delta H = \Delta H^\text{d} + \Delta H^\text{c}$, the first non-trivial contribution of the former is $\Delta H_{jj}$, while that of the latter is $E_j^{(2)}$.
Similarly, the eigenstates $|j_\tau\rangle$ of the final Hamiltonian can be expanded as 
\begin{equation}
    |\langle j_\tau| i_0\rangle|^2 = 
     \frac{|\Delta H_{ij}|^2}{(\epsilon_j^0 - \epsilon_i^0)^2},
\end{equation}
for $i \neq j$, while $|\langle j_\tau | j_0\rangle|^2 = 1 - \sum_{\ell\neq j} |\langle j_\tau| \ell_0\rangle|^2$.

Using this, we can expand all relevant probabilities $\{\tilde{p}_j^{\tau}\}$, $\{q_j^{\tau}\}$ and $\{p_j^{\tau}\}$ entering in the stochastic definitions~\eqref{ent_stoch_1}, \eqref{d_stoch}, \eqref{c_stoch}, \eqref{b_stoch} and \eqref{a_stoch}:
\begin{IEEEeqnarray}{rCl}
\label{prob-deph-hamilt-expand}
 \tilde{p}_i^{\tau} &=& p_i^0 ( 1 - \tilde{f}_i ),
 \\[0.2cm]
 \label{prob-fim-equil-state-expand}
 p_j^{\tau} &=& p_j^0 (1 - f_j),
 \\[0.2cm]
 \label{prob-deph-state-expand}
 q_j^{\tau} &=& p_j^0(1 - s_j)
\end{IEEEeqnarray}
where
\begin{IEEEeqnarray}{rCl}
\label{tilde-f-function}
 \tilde{f}_{j} &=& \beta(\Delta H_{jj} - \langle \Delta H^{\text{d} } \rangle_0) + \beta^2\langle \Delta H^{\text{d} } \rangle_0 (\Delta H_{jj} - \langle \Delta H^{\text{d} } \rangle_0)
 \\\nonumber
 & &+ \frac{1}{2}\beta^2[\Delta H_{jj}^2 - \langle (\Delta H^\text{d})^2 \rangle_0],
 \\[0.2cm]
\label{f-function}
 f_j &=& \tilde{f}_{j} + \beta[ E_{j}^{(2)} - \langle E^{(2)}\rangle_0 ],
 \\[0.2cm]
 \label{s-function}
 s_j &=& \sum_{\ell\neq j}\frac{1-e^{-\beta (\epsilon_\ell^0 - \epsilon_j^0)}}{(\epsilon_j^0 - \epsilon_{\ell}^0)^2}|\Delta H_{\ell j}|^2,
\end{IEEEeqnarray}
and $\langle E^{(2)}\rangle_0 = \sum_i p_i^0 E_i^{(2)}$. 
Note how $\tilde{f}_j$ depends only on the diagonal part of the perturbation, $\Delta H^{\text{d} }$. 
This is in line with Eq.~\eqref{rho-deph-expan}.
Conversely, $f_j$, which is associated with the full probabilities $p_j^\tau$, also has an additional contribution from $E_j^{(2)}$, which is the term associated to coherences.

Inserting Eqs.~\eqref{prob-deph-hamilt-expand}-\eqref{prob-fim-equil-state-expand} into Eqs.~\eqref{ent_stoch_1}, \eqref{d_stoch}, \eqref{c_stoch}, \eqref{b_stoch} and \eqref{a_stoch} we obtain,
\begin{IEEEeqnarray}{rCl}
\label{sigma_stoch-expand}
    \sigma[i,j] &=& \ln p_i^0/p_j^0 - \ln (1 - f_j),
    \\[0.2cm]
\label{d_stoch_expand}
    \sigDij[i,j] &=& \ln (1 - s_j) - \ln (1 - f_j),
    \\[0.2cm]
\label{c_stoch_expand}
    \sigCij[i,j] &=& \ln{p_i^0/p_j^0} - \ln(1 - s_j),
    \\[0.2cm]
\label{b_stoch_expand}
    \sigBij[i,j] &=& -\ln(1 - \tilde{f}_i),
    \\[0.2cm]
\label{a_stoch_expand}
    \sigAij[i,j] &=& \ln p_i^0/p_j^0 + \ln(1 - \tilde{f}_i) - \ln (1 - f_j).
\end{IEEEeqnarray}
We are now in the position to discuss the analyticity of the entropy production and its splittings, at the stochastic level. A series expansion of $\ln(1-x)$ is convergent only for $|x|<1$. 
Thus, since $p_i^0/p_j^0=e^{-\beta(\epsilon_i^0-\epsilon_j^0)}$ is a well behaved function, the analyticity of $\sigma$,  $\sigBij$ and $\sigAij$ are all conditioned on having $|f_j|<1$ and $|\tilde{f}_j| < 1$, 
which is satisfied if $\beta|\Delta H_{ij}| \lesssim1$, as intuitively expected. 
Thus, the quantities of our new proposed splitting~\eqref{AB_split} behave, from an analytical point of view, similarly to the full entropy production.

On the other hand, Eqs.~\eqref{d_stoch_expand}-\eqref{c_stoch_expand} rely on $|s_j|<1$.
Each $s_j$ in~\eqref{s-function} is a weighted contribution from all energies $\epsilon_\ell^0$, with $\ell \neq j$. 
At low temperatures, those energies for which $\epsilon_\ell^0 < \epsilon_j^0$ will yield an exponentially large contribution $1- e^{-\beta(\epsilon_\ell^0 - \epsilon_j^0)}$ to the sum.
Conversely,  those with $\epsilon_\ell^0 > \epsilon_j^0$ will contribute negligibly. 
The expansion is thus not in powers of $\beta \Delta H$, which is also visible from~\eqref{Deph-state-expansion}. 
Instead, it is an expansion  in powers of $\Delta H$, with coefficients that depend exponentially in $\beta$.
Violating the condition $|s_j|<1$ is  thus exponentially easier at low temperatures. 
These results show that the shortcomings illustrated in Sec.~\ref{sec:int}, are in fact absolutely general. 

\begin{figure}
     \centering
     \includegraphics[width=0.45\textwidth]{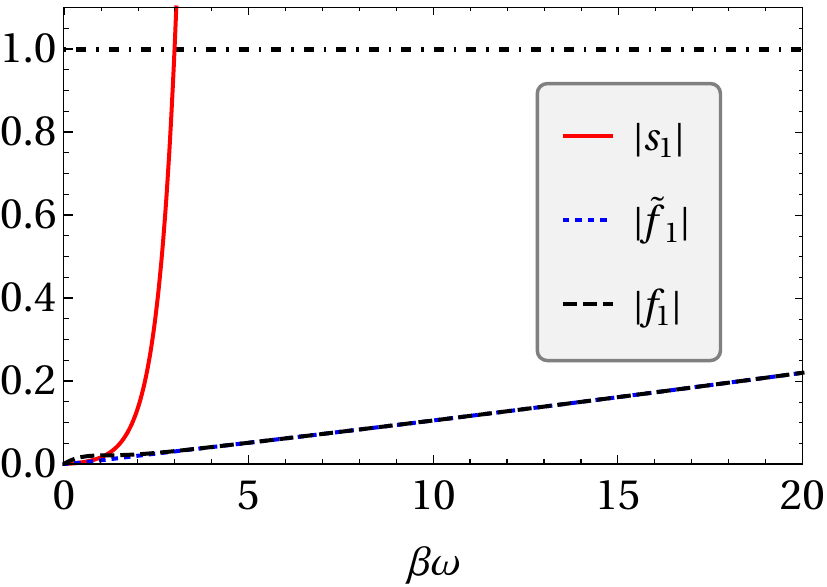}
     \caption{\label{fig:LimitsAnalyticityQubit}
     Analyticity of thermodynamic quantities at the stochastic level, for the minimal qubit model of Fig.~\ref{fig:abcd}.
    The figure compares Eqs.~\eqref{s-1-qubit} (red-solid),~\eqref{tilde-f-1-qubit} (blue-dashed) and \eqref{f-1-qubit}, as a function of $\beta\omega$ for $\theta=0.1$. The condition $|s_1|<1$, for the analyticity of $\sigDij$ and $\sigCij$, is quickly violated. For $\sigma$,  $\sigBij$ and $\sigAij$, on the other hand, the conditions $|f_1|<1$ and $|\tilde{f}_1|<1$ are satisfied for a much larger range of temperatures.
     }
 \end{figure}

To better illustrate this discussion, we revisit the minimal qubit model example treated in Sec~\ref{sec:int}. Initially the system's Hamiltonian is $H_0 = \omega\sigma^z$, and after an instantaneous quench it becomes $H_1 = \omega(\sigma^z \cos\theta + \sigma^x \sin\theta)$, where we consider $\theta$ to be small. The problematic term in this case is $s_1$ [Eq.~\eqref{s-function}], which is given by
\begin{equation}
\label{s-1-qubit}
    s_1 = \Big(1 - e^{2\beta\omega} \Big)\Bigg(\frac{\sin\theta}{2} \Bigg)^2.
\end{equation}
In comparison, we have
\begin{IEEEeqnarray}{rCl}
\label{tilde-f-1-qubit}
 \tilde{f}_1 &=& 2 \beta\omega \sin^2(\theta/2) (1+\tanh \beta\omega)\\\nonumber
 &\times& \big[ 1 + 2 \beta\omega \sin^2(\theta/2) \tanh\ \beta\omega \big]
 \\[0.2cm]
\label{f-1-qubit}
 f_1 &=& \tilde{f}_1 + \frac{1}{2} \beta\omega \sin\theta (1 - \tanh \beta\omega).
\end{IEEEeqnarray}
In Fig.~\ref{fig:LimitsAnalyticityQubit} we plot Eqs.~\eqref{s-1-qubit}-\eqref{f-1-qubit} as a function of $\beta\omega$, for $\theta=0.1$. 
The condition for analyticity of $\sigDij$ and $\sigCij$ in this case, $|s_1|<1$, is rapidly violated with increasing $\beta\omega$. For $\sigma$, $\sigBij$ and $\sigAij$, instead, $|f_1|<1$ and $|\tilde{f}_1|<1$ for a much larger range of temperatures.
It is also interesting to note that, at low temperatures, the  excited state thermal probabilities $p_1^0$, $\tilde{p}_1^{\tau}$ and $p_1^{\tau} \propto e^{-\beta\omega}$ all tend to zero exponentially as 
$e^{-\beta\omega}$. 
Conversely, $q_1^{\tau}$ tends to $(\sin\theta/2)^2$. \change{This corroborates the use of thermal states, such as~\eqref{rho_tilde}, as intermediate states for the splitting of $\Sigma$, as it ensures that the resulting functions are analytic.}

We now move on to discuss what becomes of the CGFs of Sec.~\ref{ssec:CGF} in the infinitesimal quench regime. 
We start we the CGFs of $\sigma$, $\sigBij$ and $\sigAij$ in
Eqs.~\eqref{CGF_sigma}, \eqref{CGF_AB}-\eqref{a_CGF}. Using  Eqs.~\eqref{sigma_stoch-expand}, \eqref{b_stoch_expand} and \eqref{a_stoch_expand}, together with the path probability $\mathcal{P}_{F}[i,j] = p_i^0|\langle j_{\tau}|i_0\rangle|^2$, we find to order $\Delta H^2$,
that 
\begin{equation}\label{ab_CGF-expan}
    K_{\sigBij,\sigAij} (v, u) = K_{\sigBij}(v) + K_{\sigAij}(u),
\end{equation}
where 
\begin{IEEEeqnarray}{rCl}
\label{b_CGF-expan}
    K_{\sigBij} (v) &=& \frac{ \beta^2 }{ 2 } ( v^2 - v ) \text{Var}_0 [ \Delta H^\text{d} ],
    \\[0.2cm]
\label{a_CGF-expan}
    K_{\sigAij} (u) &=& \frac{ \beta^2 }{ 2 } ( u^2 - u ) \text{Var}_0 [ \Delta H^\text{c} ] \nonumber\\
    & &+ \frac{ \beta^2 }{ 2 } \int_0^{u} \mathrm{d} x \int_x^{1-x} \mathrm{d} y I^y ( \rho^{\text{th} }_0 , \Delta H^\text{c} ).
\end{IEEEeqnarray}
These results are quite illuminating. 
Eq.~\eqref{ab_CGF-expan} implies $\sigBij$ and $\sigAij$ become \emph{statistically independent} in this limit.
Moreover, since $K_\sigma(v) = K_{\sigBij,\sigAij}(v,v)$, we now find that 
\begin{equation}\label{sigma_GGF-expan}
    K_{\sigma} (v) = K_{\sigBij} (v) + K_{\sigAij}(v).
\end{equation}
This means that all cumulants of $\sigma$ can be split as a sum of the cumulants of $\sigBij$ and $\sigAij$:  $\kappa_n(\sigma) = \kappa_n(\sigBij) + \kappa_n(\sigAij)$.
For all intents and purposes, the two channels of entropy production, $\sigB$ and $\sigA$, may thus be regarded as stemming from \emph{independent} processes: 
$\sigB$ gives the entropy production associated with a quench from $H_0 \to \mathbb{D}_{H_0}(H_{\tau})$, while $\sigA$ is associated with a second quench from $\mathbb{D}_{H_0}(H_{\tau}) \to H_{\tau}$.
We also note that, from Eq.~\eqref{sigma_GGF-expan}, it can now be seen that in this limit $\sigAij$ satisfies an integral fluctuation theorem: $\langle e^{-\sigAij} \rangle = 1$.

In contrast, for the original splitting~\eqref{CD_split}, we have
\begin{equation}
    K_{\sigDij,\sigCij}(v,u)= \ln \sum_{i,j} (p_j^0/p_i^0)^u (1-s_j)^{u-v}(1-f_j)^v p_i^0|\langle j_\tau |i_0 \rangle|^2.
\end{equation}
Once again, a series expansion of $(1-s_j)^{-x}$ is convergent only if $|s_j|<1$. 
However, if $|s_j|<1$ is satisfied, which happens for sufficiently high temperatures, one may show that, to order $\Delta H^2$, we can also split $K_{\sigDij,\sigCij} (v, u) = K_{\sigDij}(v) + K_{\sigCij}(u)$.
And, what is more important, $K_{\sigDij}$ and $K_{\sigCij}$ coincide with $K_{\sigBij}$ and $K_{\sigAij}$ respectively. 
Whence, at sufficiently high temperatures the splittings~\eqref{CD_split} and~\eqref{AB_split} coincide, even at the stochastic level.
\change{However, this is only true for infinitesimal quenches. Otherwise, the two splittings may differ, even at high temperatures.}




\section{\label{sec:XYmodel} Transverse field Ising model}

We now turn to discuss applications of our framework.
We begin with the behavior of the splitting~\eqref{AB_split} in the one-dimensional transverse field Ising model (TFIM), which is a prototypical model of a quantum critical system. 
An analysis of~\eqref{CD_split} for the same model was recently made in~\cite{Varizi2020}. 
Here we aim to contrast those results with our new 
splitting~\eqref{AB_split}. 
We thus  restrict the analysis to quench protocols, and study the problem at the level of the averages $\sigB$ and $\sigA$ (Eqs.~\eqref{B_free_energy} and~\eqref{A_free_energy}).
Non-trivial unitaries and higher order statistics will be studied in Sec~\ref{sec:LMG}, for a different model.

We begin by introducing the model and delineating the steps to compute $\sigB$ and $\sigA$.
To make the paper self-consistent, some additional details are provided in Appendices~\ref{app:xymodel} and~\ref{app:ABIsing}. 
Consider a linear chain of $N$ spins, each described by Pauli operators $\sigma_j^{x,y,z}$ and interacting via the Hamiltonian 
\begin{equation}\label{XYModel}
 H(g) = - \sum_{j=1}^{N} \bigg( J\sigma_j^x \sigma_{j+1}^x + g\sigma_j^z \bigg),
\end{equation}
where $g$ is the applied magnetic field and $J$ is the coupling strength, which we henceforth set to unity ($J=1$).
We consider periodic boundary conditions, $\sigma_{N+1}^\alpha=\sigma_1^\alpha$. 
This model presents critical points at $g = \pm 1$, where the system changes from a ferromagnetic phase, for $|g|<1$, to a paramagnetic phase, for $|g|>1$.

After a series of transformations (see Appendix~\ref{app:xymodel}) this Hamiltonian can be written as 
\begin{equation}\label{XYModel_diagonal_H}
 H(g) = \sum_{k} \epsilon_k(g) \big( 2 \eta_k^{\dagger} \eta_k - 1 \big),
\end{equation}
where $\{\eta_k\}$ are fermionic operators and \begin{equation}
    \epsilon_k(g) = \sqrt{ (g-\cos k)^2 + \sin^2 k},
\end{equation} 
are the single-particle energies. 
Here, $k=\pm(2n+1)\pi/N$, with $n = 0,1,...,N/2-1$, denotes the system's quasi-momenta.
We consider that the system initially has 
a transverse field $g_0$ and is prepared in the thermal state $\rho_0^{\text{th}} = e^{-\beta H_0}/Z_0$.
The full expression for $\rho_0^{\text{th} }$ can be found in Appendix~\ref{app:ABIsing}. 
Due to the structure of~\eqref{XYModel_diagonal_H}, it can be decomposed as a product over individual $k$ modes, which greatly facilitates the calculation of all thermodynamic quantities.



The system is then decoupled from the reservoir and undergoes an instantaneous quench, where the field is changed to $g_\tau = g_0 + \delta g$. 
Since the quench is instantaneous, the state of the system remains the same, but its Hamiltonian changes, from $H_0$ to $H_{\tau} = H_0 + \Delta H$, where
$\Delta H = -\delta g\sum_j \sigma_j^z$.
Full details on the computation of $\sigB$ and $\sigA$ are provided in Appendices~\ref{app:xymodel} and~\ref{app:ABIsing}.
The overall contributions of the diagonal~vs.~off-diagonal is described by the Bogoliubov angle $\cos\theta_k = (g_0-\cos k)/\epsilon_k^0$ and $\sin \theta_k = \sin k/\epsilon_k^0$.
And the state~\eqref{rho_tilde}, associated with the dephased final Hamiltonian, is described by the modified energies $\tilde{\epsilon}_k^{\tau} = \epsilon_k^0 + \delta g\cos\theta_k$.

We are interested in the thermodynamic limit ($N$ very large), where $k$ sums can be converted to integrals and all quantities become extensive in $N$. 
In this case, we ultimately find that 
\begin{equation}\label{B-XY-Quench}
 \sigB = N \int_0^{\pi} \frac{\mathrm{d}k}{2\pi} \, 2\Bigg\{ \ln \Bigg[ \frac{ \cosh \Big( \beta \tilde{\epsilon}_k^{\tau} \Big) }{ \cosh \Big( \beta \epsilon_k^0 \Big)} \Bigg] + \beta \Big( \epsilon_k^0 - \tilde{\epsilon}_k^{\tau} \Big) \tanh \Big( \beta \epsilon_k^0 \Big) \Bigg\},
\end{equation}
and
\begin{equation}\label{A-XY-Quench}
 \sigA = N \int_0^{\pi} \frac{\mathrm{d}k}{2\pi} \, 2 \ln \Bigg[ \frac{ \cosh \Big( \beta \epsilon_k^{\tau} \Big) }{ \cosh \Big( \beta \tilde{\epsilon}_k^{\tau} \Big)} \Bigg].
\end{equation}
Adding both contributions recovers the full  entropy production $\Sigma$, which was computed in~\cite{Dorner2012,Bayocboc2015,Varizi2020} and reads
\begin{equation}
    \Sigma = N \int_0^{\pi} \frac{\mathrm{d}k}{2\pi} \, 2\Bigg\{ \ln \Bigg[ \frac{ \cosh \Big( \beta \epsilon_k^{\tau} \Big) }{ \cosh \Big( \beta \epsilon_k^0 \Big)} \Bigg] + \beta \Big( \epsilon_k^0 - \tilde{\epsilon}_k^{\tau} \Big) \tanh \Big( \beta \epsilon_k^0 \Big) \Bigg\}.
\end{equation}
For comparison, in Appendix~\ref{app:CDIsing} we also  present the formulas for the splitting~\eqref{CD_split}, which were developed  in~\cite{Varizi2020}. 
We also mention, in passing, that Eqs.~\eqref{B-XY-Quench} and~\eqref{A-XY-Quench} are not perturbative in the quench magnitude. 
That is, they hold for arbitrary quenches $\delta g$. The only assumption is that $U = 1$. 
For completeness, their behavior in the infinitesimal case is presented in Eqs.~\eqref{B-XY-Quench-Inf} and~\eqref{A-XY-Quench-Inf}.

\begin{figure}[h]
 \includegraphics[width=0.45\textwidth]{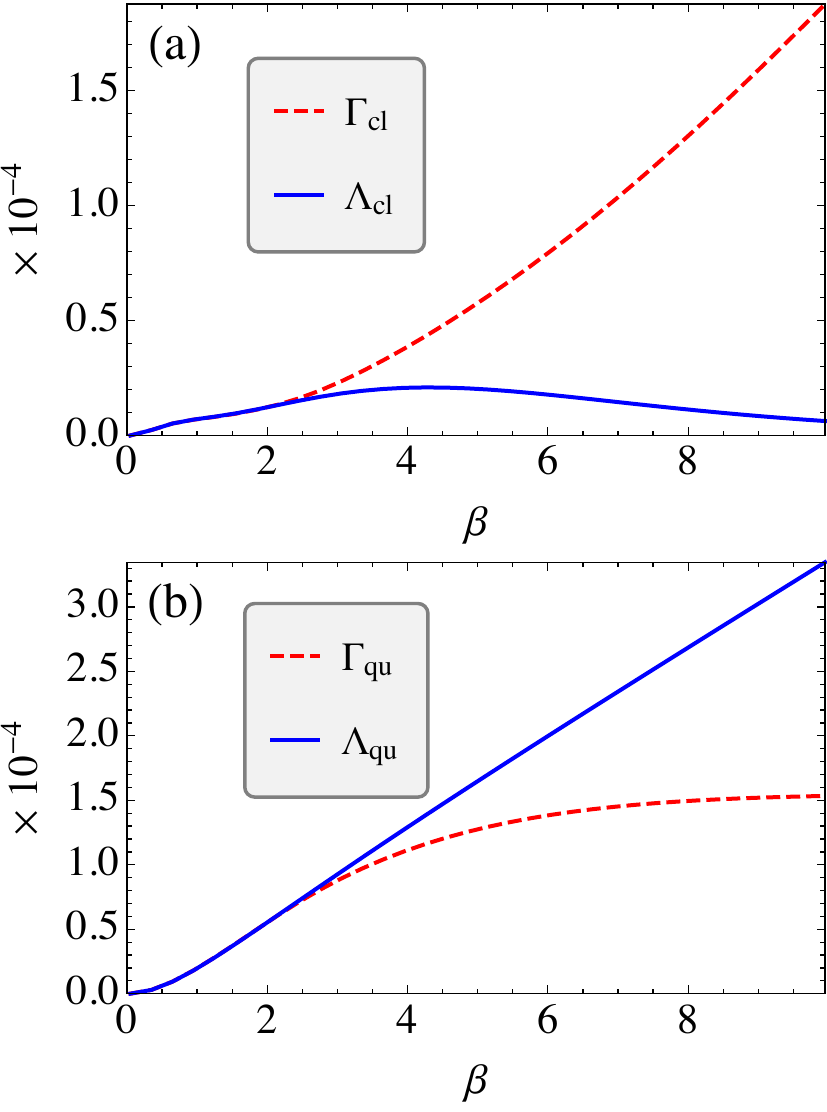}\\
\caption{\label{fig:IsingABCD_compare} Comparison between (a) $\sigD$ and $\sigB$, and (b) $\sigC$ and $\sigA$, for the TFIM, as a function of $\beta$, with $g_0 = 0.75$ and $\delta g = 0.01$. 
}
\end{figure}

Fig.~\ref{fig:IsingABCD_compare} compares the two splittings~\eqref{CD_split} and~\eqref{AB_split} as a function of $\beta$, with fixed $g_0 = 0.75$ (outside criticality) and $\delta g = 0.01$.
At high temperatures, one clearly sees how both splittings coincide.
\change{This corresponds to the region of parameters where $\sigD$ and $\sigC$ are analytic.}
But as the system is cooled, they eventually begin to differ. 
In particular, $\sigD$ tends to grow linearly with $\beta$, while $\sigB$ tends to zero. 
For the quantum components the opposite is observed: $\sigC$ tends to saturate while $\sigA$ tends to grow. Thus, at very low temperatures $\sigA$ becomes the dominant contribution in~\eqref{AB_split}, while $\sigD$ becomes the dominant one in~\eqref{CD_split}.

\begin{figure}
    \centering
    \includegraphics[width=0.45\textwidth]{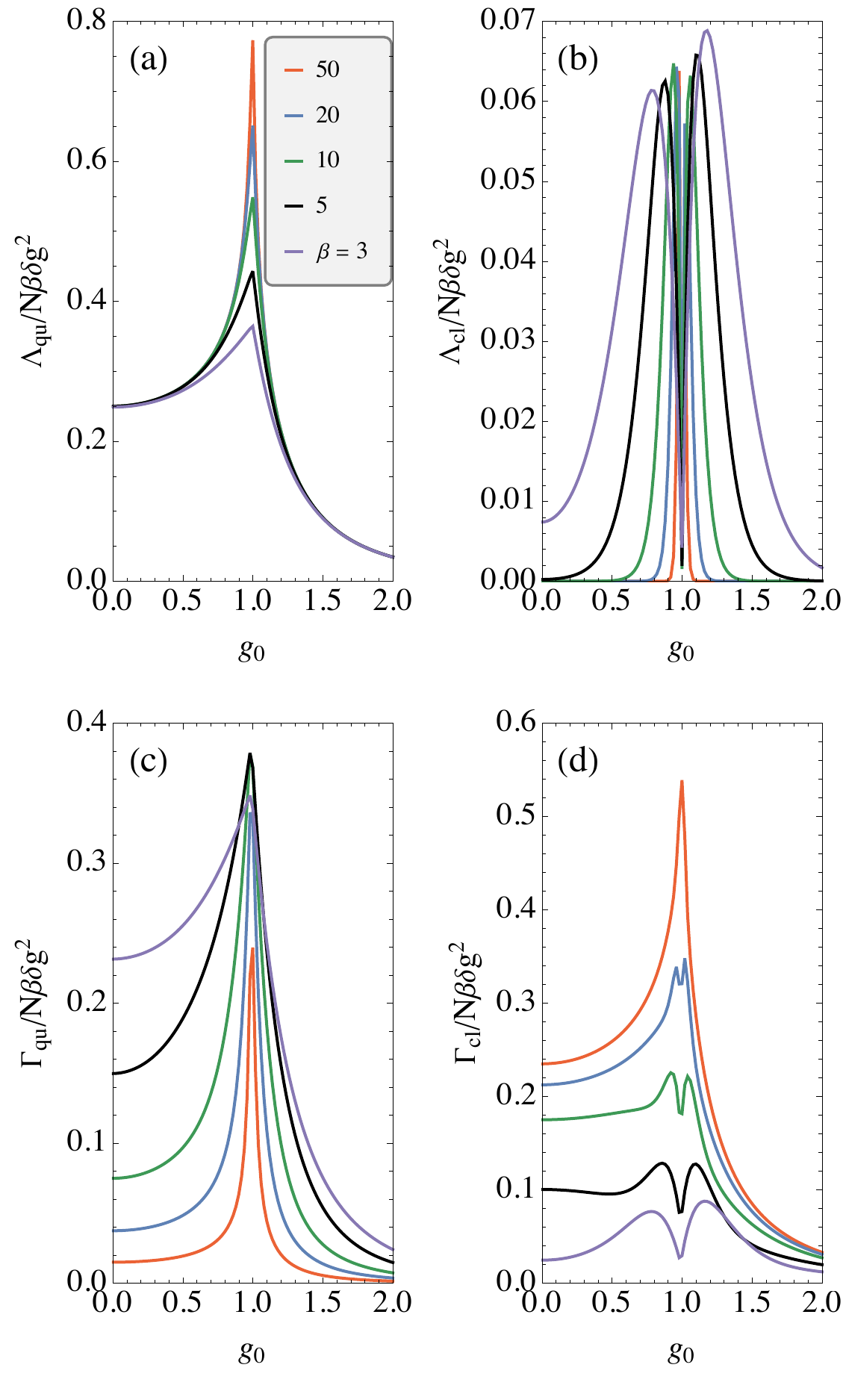}
    \caption{\label{fig:ABCDGrid} $\sigA$, $\sigB$, $\sigC$ and $\sigD$ for the TFIM as a function of $g_0$, for different values of $\beta$ (in the low temperature regime) and $\delta g= 0.01$.
    All quantities are scaled by $\beta \delta g^2$.
    }
\end{figure}

Next we turn to the behavior near criticality. 
In Fig.~\ref{fig:ABCDGrid} we plot $\sigD$, $\sigC$, $\sigB$ and $\sigA$ as a function of the initial field $g_0$, for different values of $\beta$ (focusing on low temperatures) and fixed quench magnitude of $\delta g = 0.01$.
The full entropy production $\Sigma$ behaves similarly to $\sigA$ in Fig.~\ref{fig:ABCDGrid}(a); for any finite $T$ it presents a peak at $g_0 = 1$, which eventually tends to a divergence as $\beta \to \infty$.
As is clear by comparing Figs.~\ref{fig:ABCDGrid}(a) and (b), the dominant contribution to the splitting~\eqref{AB_split} is $\sigA$. 
Moreover, $\sigA$ is found to always grow (and eventually diverge) with $\beta$ at $g_0=1$, while $\sigB$ sharply drops to zero.
Conversely, for the splitting~\eqref{CD_split}, we find in Figs.~\ref{fig:ABCDGrid}(c), (d) that the dominant contribution is instead that of the populations $\sigD$. 
Crucially, we find that in this case $\sigC$ remains finite as $\beta \to \infty$, while $\sigD$ diverges~\cite{Varizi2020}.
\change{We also call attention to the non-monotonic dependence on $\beta$, of the quantities in Fig.~\ref{fig:ABCDGrid}(c). This is  an artifact of the fact that $\Gamma_{\text{qu}}$ is scaled by  $\beta$. The quantity $\Gamma_{\text{qu}}$ itself is monotonic, but its behavior changes from $\beta^2$ at high temperatures, to $\beta^0$ in low temperatures~\cite{Varizi2020}.
}

\change{As highlighted in \cite{Varizi2020}, the entropy production in this limit
results entirely from the changes in occupations, i.e. creation/annihilation of particles, in the modes $\pm k$, when the quench is performed.
This enters in $\sigD$ as a population mismatch between the initial and final equilibrium states. 
Conversely, in the split~\eqref{AB_split}, it enters in $\sigA$ as resulting from the rotating energy basis. On the other hand, $\sigB$ only quantifies the contribution to the entropy production resulting from a change in the energy levels given by $\tilde{\epsilon}_k^{\tau} - \epsilon_k^0 = \delta g \cos\theta_k$. In the low temperature limit, only the ground and low lying excited states are relevant, and close to the critical point $g_0=1$, the latter corresponds to creating  excitations with momentum $k\to0$; but one can easily show that at $g_0=1$, $\cos\theta_k=|\sin(k/2)|$, which goes to zero when $k\to0$. This explains the drop in $\sigB$ at this point.
Overall, Fig.~\ref{fig:ABCDGrid} therefore } shows that the critical properties of these quantities depend crucially on the type of splitting used.

%

\section{\label{sec:LMG} Macrospin model}

\begin{figure*}
    \centering
    \includegraphics[width=\textwidth]{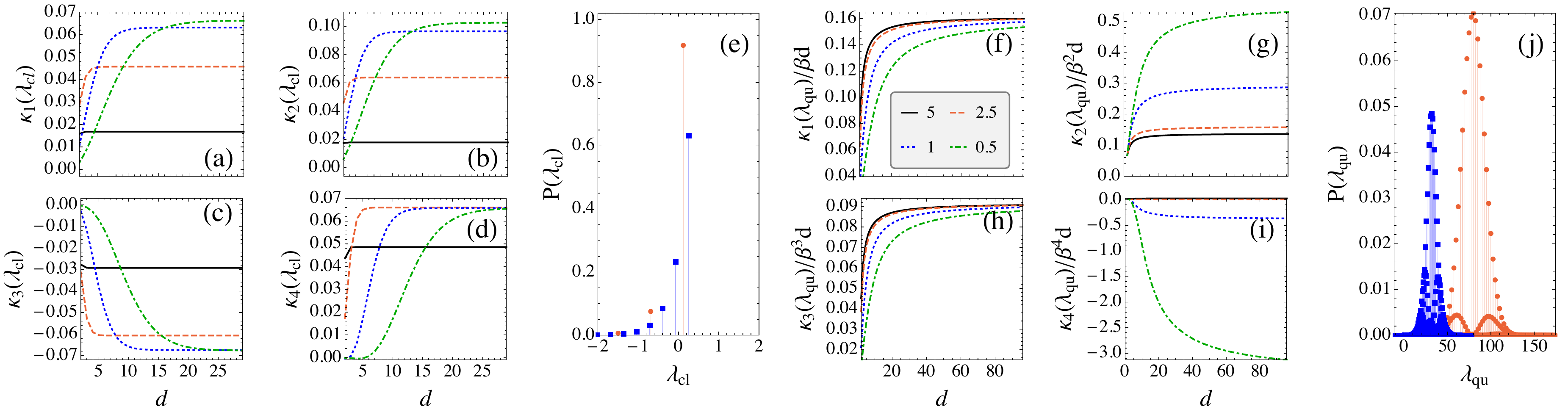}
    \caption{Statistics of the splitting~\eqref{AB_split} for the macrospin model, as a function of the Hilbert space dimension $d$ and different values of $\beta$, as shown in image (f).
    (a)-(d) First four cumulants of $\sigBij$. 
    (e) $P(\sigBij)$ for $d=200$, with $\beta = 1$ (blue) and $\beta = 2.5$ (red).
    (f)-(j) Same, but for $\sigAij$.
    Some of the cumulants are scaled by powers of $\beta$ and $d$, whenever such a scaling law exists. 
    Additional parameters: $h_z = 1$, $h_x = 0.5$ and $\tau = 2$.
    }
    \label{fig:macrospin_lambda}
\end{figure*}

\begin{figure*}
    \centering
    \includegraphics[width=\textwidth]{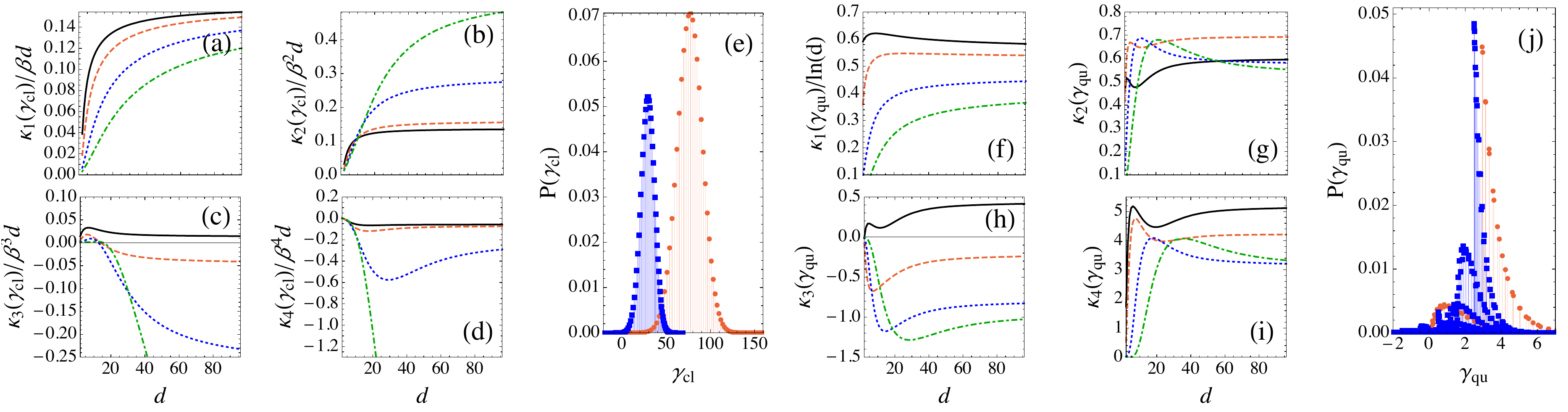}
    \caption{Same as Fig.~\ref{fig:macrospin_lambda}, but for the splitting~\eqref{CD_split}.
    }
    \label{fig:macrospin_gamma}
\end{figure*}

Finally, we analyze our framework from the stochastic perspective developed in Sec.~\ref{sec:StochTrajectories}. 
To emphasize the generality of our results, we also focus on non-quench scenarios ($U\neq 1$). 
We consider a macrospin model, with $d =2S+1$ levels and spin operators $S_x, S_y, S_z$~\cite{Sakurai2010}.
We consider a scenario similar to that of Fig.~\ref{fig:qubit2}: 
The initial and final Hamiltonians are taken to coincide, being  given by $H_0 = H_\tau = - h_z S_z$. 
And the unitary is driven by a magnetic pulse in the $x$ direction, so $U = \exp\{-i (H_0 - h_x S_x)\tau\}$.
Since the Hamiltonian is cyclic, the eigenbases $|i_0\rangle$ and $|j_\tau\rangle$ coincide. 
However, since the unitary is now non-trivial, the final state $\rho_\tau = U \rho_0^\text{th} U^\dagger$ will contain coherences in the $S_z$-basis.

A panel summarizing the results for the splitting~\eqref{AB_split} is shown in Fig.~\ref{fig:macrospin_lambda}, where we plot the first four cumulants of $\sigBij$ (images (a)-(d)) and those of $\sigAij$ (images (f)-(i)), as a function of the Hilbert space dimension $d$ and for different values of $\beta$. 
In Figs.~\ref{fig:macrospin_lambda}(e) and (j), we also show exemplary plots of the full distributions $P(\sigBij)$ and $P(\sigAij)$, for fixed $d = 200$ and two values of $\beta$. 
For comparison, a similar panel, but for the quantities in~\eqref{CD_split}, is shown in Fig.~\ref{fig:macrospin_gamma}. 
Note also that some cumulants are scaled by either $d$ or $\beta$, whenever a simple scaling rule could be found. 

From these plots the following conclusions can be drawn. 
Concerning Fig.~\ref{fig:macrospin_lambda}, all cumulants of $\sigBij$ are found to be intensive, saturating at a finite value when $d\to \infty$. 
Conversely, all cumulants of $\sigAij$ are extensive, scaling proportionally to $d$. 
The cumulants of $\sigAij$ also scale with powers of $\beta$ at low temperatures (Figs.~\ref{fig:macrospin_lambda}(f)-(i)), but for higher order cumulants this scaling only becomes good at very low temperatures. 
For the splitting~\eqref{CD_split} the situation is reversed: now the cumulants of $\sigDij$ become extensive (and quite similar to those of $\sigAij$), while those of $\sigCij$ tend to saturate. The only exception is $\kappa_1(\sigCij)$, which is found to grow logarithmically with $d$. 
Notice that this dependence on the dimensionality is fundamentally different from what was found in the TFIM (Sec.~\ref{sec:XYmodel}), where all quantities were extensive in the number of particles. 
We also note that the statistics of $\sigBij$ (Fig.~\ref{fig:macrospin_lambda}(e)) has significantly smaller support than that of $\sigAij$.
This is a consequence of the fact that, from its definition in Eq.~\eqref{b_stoch}, $\sigBij$ depends only on the initial points $|i_0\rangle$, while $\sigAij$ depends on both $|i_0\rangle$ and $|j_\tau\rangle$. 

\change{The results in Fig.~\ref{fig:macrospin_lambda}(e) indicate that even when $d \to \infty$, the distribution of $\sigBij$ will never tend to a Gaussian. 
Conversely, for $\sigDij$ in Fig.~\ref{fig:macrospin_gamma}(e), this is clearly the case. 
This is  supported by a comparison of the corresponding cumulants in images (a)-(d), which are intensive for $\sigBij$ and extensive for $\sigDij$. 
As for $\sigAij$ and $\sigCij$, even though the histograms in Figs.~\ref{fig:macrospin_lambda}(j) and~\ref{fig:macrospin_gamma}(j) do not seem to indicate a Gaussian behavior, this is expected to eventually occur for sufficiently large $d$. 
For $\sigAij$, the scaling with $d$ is similar to $\sigDij$, and hence the same argument as above applies. 
Conversely, the situation for $\sigCij$ is more delicate, since the first cumulant scales only logarithmically (and hence very slowly) with $d$. 
Extremely large sizes may thus be necessary for a Gaussian behavior to be observed.
}

\change{One might expect that in the limit $d \to \infty$ one should recover a classical spin model. 
This does not happen, however, as is evidenced by the fact that the coherent terms $\sigC$ and $\sigA$ do not vanish in this limit, but actually increase with $d$. 
The explanation for this rests essentially on a coarse-graining argument. 
Even though we take $d \to \infty$, we continue to assume we have full access to all eigenstates of the system, as appears, for instance, in the dephasing operations involved in constructing the intermediate states. 
}


\section{\label{sec:conclusion} Conclusion}

In this article, we studied how entropy production can be divided into a classical and quantum contribution, when  a system is driven out of  equilibrium. A popular choice in the literature is  given in Eq. \eqref{CD_split}, see in particular~\cite{Francica2019}. 
This splitting has several interesting properties, including individual fluctuation theorems for each term~\cite{Francica2019} and a resource-theoretic interpretation~\cite{janzing2006quantum,Lostaglio2015,Santos2019}.
However, we here noted it also has two major shortcomings. First, we showed that 
the classical contribution $\sigD$ in Eq. \eqref{CD_split} dominates for highly coherent processes and at low temperatures, in contrast with what might be expected. We observed this undesired behaviour in all considered systems, from a simple driven qubit to a many-body Ising model at criticality, and identified the divergence of the relative entropy in \eqref{D}, at low temperatures, as the underlying cause. 
Second, given a perturbation $\delta g$ of the Hamiltonian, the radius of convergence of $\sigC$ and $\sigD$ tends to zero exponentially fast as $\beta \rightarrow \infty$, making this splitting  impractical to characterise the entropy production of quenched systems at low temperatures.

In order to overcome  these shortcomings, we suggested a new splitting for the entropy production  given in Eq. \eqref{AB_split}, which was motivated by the developments of Ref.~\cite{Scandi2019} for infinitesimal quenches. The definition is valid arbitrarily out-of-equilibrium. We also provided a formulation in terms of stochastic trajectories and a physical interpretation,  highlighting how it can be obtained following a similar logic to the one behind \eqref{CD_split}. Indeed, both  \eqref{CD_split} and \eqref{AB_split} can be understood by introducing  intermediate  states for comparison. The different choices, however, turn out to have crucial consequences, especially for highly coherent processes. Indeed, in the low-temperature regime  the quantum term $\sigA$ dominates in Eq. \eqref{AB_split}, but the classical one does in \eqref{CD_split}. For high temperatures and infinitesimal quenches, both splittings coincide.
A comparison between the two approaches is summarized in Table~\ref{tabABCD}.

More generally, our considerations illustrate  that it is  non-trivial to identify the classical and quantum contributions in entropy production for an arbitrarily out-of-equilibrium process. In analogy with the definition of work for coherent processes \cite{Bumer2018,Niedenzu2019}, the splitting of $\Sigma$ in a classical and quantum term may not be unique, and will depend on the specific context into consideration. Nevertheless, there are some relevant scenarios where such a splitting is unambiguous. One is in a thermalization process described by either a Markovian master equation or as a resource-theoretic state transformation; in both cases, such a distinction seems to be very well captured by Eq.~\eqref{CD_split}~\cite{janzing2006quantum,Lostaglio2015,Santos2019}. On the other hand, when an equilibrium state is slightly moved out of equilibrium  (e.g. by  an infinitesimal quench),  the splitting \eqref{AB_split} provides a more accurate description of the quantum and classical contributions. In fact, in such a scenario, the entropy production can be decomposed into a classical and quantum contribution at all levels of the statistics, as shown in Sec.~\ref{ssec:CGF} (see also~\cite{Scandi2019,Miller2020Landauer,miller2020joint}). For general out-of-equilibrium processes, however, classical and quantum contributions become inevitably mixed. Still, our results show that the splitting \eqref{AB_split} has a more reasonable behaviour (i.e., the quantum term dominates at low temperatures and for highly coherent processes).

In a second part of the article, we applied these ideas to a transverse field Ising model, and to a macrospin undergoing finite time dynamics. For the Ising model, we found that the behavior close to criticality is fundamentally different for both splittings, with the quantum component playing a predominant role for~\eqref{AB_split} and the classical component being dominant in~\eqref{CD_split}. 
For the macrospin model, we focused not only on the average, but on the full statistics, including the first four cumulants and the corresponding probability distributions. 
We have found that different cumulants scale with the Hilbert space dimension $d$ in non-trivial ways, some being extensive, others intensive or even logarithmic. 

We hope that these results help to motivate further investigations on the non-trivial way in which populations and coherences intermix in quantum thermodynamic processes. 
We are particularly interested in further understanding how this unfolds for many-body systems in general. 
In particular, the analysis of higher order cumulants for these models has been seldom explored in the literature, even for $\Sigma$ itself. 
It would also be interesting to generalize the present results for open systems, undergoing generic interactions with a heat bath. This can be done for quasi-static processes, following the approach in~\cite{Scandi2019}. Or it can be constructed in a controllable way using collisional models~\cite{DeChiara2018}. 
Finally, these ideas could also be extended to describe quantum correlations in bipartite systems. 
For instance, instead of studying the entropy production in a work protocol, one may analyze it in the context of heat exchange between two quantum correlated systems, which are locally thermal, as studied in~\cite{Micadei2017}.

\section*{Acknowledgements}

We thank M. Scandi for  insightful discussions. We acknowledge financial support from the Brazilian agencies Conselho Nacional de Desenvolvimento Cient\'ifico e Tecnol\'ogico and Coordena\c c\~ao de Aperfei\c coamento de Pessoal de N\'ivel  Superior.
GTL acknowledges the financial support of the São Paulo Funding Agency FAPESP (Grants No. 2017/50304-7, 2017/07973-5 and No. 2018/12813-0).

\appendix

\section{\label{appsec:nonnegA}Nonnegativity of $\sigA$}

In this appendix we show that $\sigA$ defined in Eq.~\eqref{A_free_energy} is also non-negative, even though it cannot be written as a relative entropy. 
The proof is essentially based on the Bogoliubov variational theorem \cite{Callen1985}.
The first term in Eq.~\eqref{A_free_energy} reads explicitly 
$F(\tilde{\rho}_{\tau}^{\text{th} }) = \tr{\tilde{\rho}_{\tau}^{\text{th} }H_{\tau}} - TS(\tilde{\rho}_{\tau}^{\text{th}})$.
At first sight, this is not an equilibrium free energy, because the Hamiltonian $H_\tau$ is not the same as the one appearing in the exponent of $\tilde{\rho}^\text{th}_\tau$ [Eq.~\eqref{rho_tilde}]. 
However, due to the presence of the trace, we can equivalently write this as 
$F(\tilde{\rho}_{\tau}^{\text{th} }) = \tr{\tilde{\rho}_{\tau}^{\text{th} } \mathbb{D}_{\rho_\tau}(H_{\tau})} - TS(\tilde{\rho}_{\tau}^{\text{th}})$, 
which shows that it is actually an equilibrium free energy. 
Next, we note that the final Hamiltonian can be rewritten as $H_{\tau} = \mathbb{D}_{\rho_{\tau}}(H_{\tau}) + H_{\tau}^{\text{c} }$, where $H_{\tau}^{\text{c} } = H_\tau - \mathbb{D}_{\rho_{\tau}}(H_{\tau}) $.
The Bogoliubov variational theorem \cite{Callen1985} then yields
\begin{equation}\label{BogIneq}
 F(\rho_{\tau}^{ \text{th} }) \leqslant F(\tilde{\rho}_{\tau}^{ \text{th} }) + \tr{\tilde{\rho}^{\text{th}}_{\tau} H_\tau^\text{c} }.
\end{equation}
But, by construction, $ H_{\tau}^{\text{c} } = H_{\tau} - \mathbb{D}_{\rho_{\tau}} (H_{\tau})$ has only off-diagonal elements in the common eigenbasis of $\tilde{\rho}^{\text{th}}_{\tau}$ and $\rho_{\tau}$. 
Thus, the second term in Eq.~\eqref{BogIneq} vanishes. Plugging the resulting inequality back into Eq.~\eqref{A_free_energy}, we finally conclude that $\sigA\geqslant 0$.

We can also show that for any finite temperature $\sigA$ is zero if and only if $\rho_{\tau}$ is incoherent in the eigenbasis of $H_{\tau}$.
The \emph{if} part of this statement is easy: when $\rho_{\tau}$ is incoherent in the final energy eigenbasis, $\mathbb{D}_{\rho_{\tau}} (H_{\tau}) = \mathbb{D}_{H_{\tau}}(H_{\tau}) = H_{\tau}$, which leads to $\sigA(\rho_{\tau}) = 0$. 
Conversely, if we assume that $\sigA(\rho_{\tau}) = 0$ and $\beta>0$, we must have $F(\tilde{\rho}_{\tau}^{ \text{th} }) - F(\rho_{\tau}^{\text{th} }) = 0$.
This implies that
\begin{equation}
    \sum_{k=0}^{+\infty} \frac{(-\beta)^k}{k!}\tr{H_{\tau}^k - \mathbb{D}_{\rho_{\tau}}(H_{\tau})^k} = 0,
\end{equation}
which means $\tr{ H_{\tau}^k - \mathbb{D}_{\rho_{\tau}}(H_{\tau})^k }=0,\, \forall\, k \in \mathbb{N}$.
The case $k=0$ is trivial and the case $k=1$ follows directly from the definition of $\mathbb{D}_{\rho_{\tau}}(H_{\tau})$.
For the case $k=2$, we use that
\begin{equation*}\begin{aligned}
    \tr{ H_{\tau}^2 } &= \tr{ \Big( \mathbb{D}_{\rho_{\tau}}(H_{\tau}) + H_{\tau}^c \Big)^2 }\\
    &= \tr{ \mathbb{D}_{\rho_{\tau}}(H_{\tau})^2 + 2\mathbb{D}_{\rho_{\tau}}(H_{\tau}) H_{\tau}^c + (H_{\tau}^c)^2 }.
\end{aligned}\end{equation*}
Again, using the definition of $\mathbb{D}_{\rho_{\tau}}(H_{\tau})$ one may verify that $\tr{ \mathbb{D}_{\rho_{\tau}}(H_{\tau}) H_{\tau}^c } = 0$.
Therefore we are left with
\begin{equation}
     \tr{ H_{\tau}^2 - \mathbb{D}_{\rho_{\tau}}(H_{\tau})^2 } = \tr { \Big( H_{\tau} - \mathbb{D}_{\rho_{\tau}}(H_{\tau} ) \Big)^2 } = 0.
\end{equation}
But since $H_{\tau} - \mathbb{D}_{\rho_{\tau} }(H_{\tau})$ is also Hermitian, we must have $H_{\tau} - \mathbb{D}_{\rho_{\tau}}(H_{\tau}) = 0$.
Then, since $\mathbb{D}_{\rho_\tau}(H_\tau)=H_\tau$, for $k>3$, $\tr{H_\tau^k-\mathbb{D}_{\rho_\tau}^k(H_{\tau})}=0$ follows trivially, and $\rho_\tau$ must be incoherent in the eigenbasis of $H_\tau$, i.e., we must have $[\rho_\tau,H_\tau]=0$.


\section{\label{app:xymodel} Diagonalization of the transverse field Ising model}

The TFIM Hamiltonian in Eq.~\eqref{XYModel} can be diagonalized by a series of transformations, as shown in~\cite{Lieb1961}. 
Our notation follows closely that of Ref.~ \cite{Varizi2020}, which contains a self-contained derivation of these results.
The first step is the introduction of a Jordan-Wigner transformation, that maps the spin chain onto an equivalent system of spinless fermions,
\begin{equation}\label{JWtransf}
 \sigma^x_j = ( c_j^{\dagger} +  c_j) \prod_{i<j} (1 - 2 c_i^{\dagger}  c_i ),
 \quad \sigma^z_j = 1 - 2 c_j^{\dagger}  c_j,
\end{equation}
where $c_j^{\dagger}$ and $c_j$ are canonical creation and annihilation fermionic operators.
We assume $N$ is large and even. We may then ignore boundary terms \cite{Damski}, and introduce the Fourier transform
\begin{equation}\label{Fouriertransf}
  c_j = \frac{e^{-\imath\pi/4}}{\sqrt{N}} \sum_{k}  c_k e^{\imath k j},
\end{equation}
where $k = \pm (2n+1) \frac{\pi}{N}$ and $n = 0, 1, ..., N/2-1$.
Eq.~\eqref{XYModel} is then transformed to
\begin{equation}
 H(g) = \sum_{k>0} \Big[ ( g - \cos k ) (c_k^{\dagger} c_k - c_{-k} c_{-k}^{\dagger} ) + \sin k (c_k^{\dagger} c_{-k}^{\dagger} + c_{-k}c_{k} ) \Big ].
\end{equation}

Next, we introduce a new set of Fermionic operators $\eta_k$ through the Bogoliubov transformation
\begin{equation}\label{BogTrans}
 \eta_k = \cos(\theta_k/2) c_k + \sin(\theta_k/2) c_{-k}^{\dagger}.
\end{equation}
With the definitions
\begin{equation}\label{BogAngles}\begin{aligned}
  \epsilon_k (g) &= \sqrt{(g-\cos k)^2 + \sin^2 k},
  \\[0.2cm]
  (\sin \theta_k, \, \cos \theta_k) &= \bigg( \frac{\sin(k)}{\epsilon_k}, \, \frac{g-\cos(k)}{\epsilon_k} \bigg),
\end{aligned}\end{equation}
we then finally obtain
\begin{equation}
 H(g) = \sum_{k} \epsilon_k(g) \big( 2 \eta_k^{\dagger} \eta_k - 1 \big),
\end{equation}
Exploring the fact that $\epsilon_{-k}(g)=\epsilon_k(g)$, we can rewrite $H(g)$ as a sum over only positive values of $k$,
\begin{equation}
    H(g) = \sum_{k>0} 2\epsilon_k(g) (\eta_k^{\dagger}\eta_k + \eta_{-k}^{\dagger}\eta_{-k} - 1).
\end{equation}
This is useful because, as we will see, a perturbation $\delta g$ couples pairs of modes $+k$ and $-k$. Finally, if we let $|n_{-k}n_k\rangle$ be the joint eigenstates of $\eta_{-k}^\dagger \eta_{-k}$ and $\eta_k^\dagger \eta_k$, where $n_{\pm k} = 0,1$, we may also write
\begin{equation}
    H(g) = \sum_{k>0}2\epsilon_k(g) \big( -| 0_{-k} 0_k \rangle \langle 0_{-k} 0_k | + | 1_{-k} 1_k\rangle \langle 1_{-k} 1_k | \big).
\end{equation}

If we consider now a perturbation $\delta g$ in the field, we have
\begin{equation}
    \Delta H = - \delta g \sum_{j=1}^N \sigma_j^{z}= \delta g \sum_{j=1}^N (2c_j^{\dagger}c_j - 1) = \delta g \sum_k (2c_k^{\dagger}c_k - 1),
\end{equation}
where we used Eqs.~\eqref{JWtransf} and~\eqref{Fouriertransf}. Finally, using Eqs.~\eqref{BogTrans} and~\eqref{BogAngles} we obtain
\begin{equation}
    \Delta H = 2\delta g \sum_{k>0} \big[ \cos\theta_k (\eta_k^{\dagger}\eta_k + \eta_{-k}^{\dagger}\eta_{-k} - 1) + \sin\theta_k(\eta_{-k}^{\dagger}\eta_k^{\dagger} - \eta_{-k}\eta_k)\big],
\end{equation}
where the coupling between $+k$ and $-k$ modes is clear from the second term. Alternatively, this can be written as $\Delta H = \Delta H^\text{d} + \Delta H^\text{c}$, with
\begin{IEEEeqnarray}{rCl}
 \label{XY-Deph-Perturb}
 \Delta H^\text{d} &=& 2 \delta g \sum_{k>0}
 \cos\theta_k \big( -| 0_{-k} 0_k \rangle \langle 0_{-k} 0_k | + | 1_{-k} 1_k\rangle \langle 1_{-k} 1_k | \big),\IEEEeqnarraynumspace
 \\[0.2cm]
 \label{XY-Cohe-Perturb}
 \Delta H^\text{c} &=& 2\delta g \sum_{k>0}
 \sin\theta_k \Big( | 0_{-k} 0_k \rangle \langle 1_{-k} 1_k | + | 1_{-k} 1_k \rangle \langle 0_{-k} 0_k | \Big),
\end{IEEEeqnarray}
where $\Delta H^{\text{d} }$ and $\Delta H^{\text{c} }$
are the dephased and coherent parts of the perturbation, respectively.
\section{\label{app:ABIsing} $\sigB$ and $\sigA$ for the TFIM}

Using the results from~\eqref{app:xymodel}, we now show how to compute $\sigB$ and $\sigA$ using Eqs.~\eqref{B_free_energy} and~\eqref{A_free_energy}.
Since we consider the initial field to be $g_0$, the initial Hamiltonian is given by
\begin{equation}\begin{aligned}
    H_0 &= \sum_k \epsilon_k^0 (2\eta_k^{\dagger}\eta_k - 1),
    \\[0.2cm]
    &= \sum_{k>0} 2\epsilon_k^0 \big( -| 0_{-k} 0_k \rangle \langle 0_{-k} 0_k | + | 1_{-k} 1_k\rangle \langle 1_{-k} 1_k | \big)
\end{aligned}\end{equation}
where $\epsilon_k^0=\epsilon_k(g_0)$ and $|n_{-k}n_k\rangle$ are the joint eigenstates of $\eta_{k}^{\dagger}\eta_k$ and $\eta_{-k}^{\dagger}\eta_{-k}$. Thus, the initial state $\rho_0^{\text{th}}=e^{-\beta H_0}/Z_0$ can be written as
\begin{subequations}\label{Gibbs0}
 \begin{align}
 \rho_0^{\text{th}} &= \prod\limits_{k>0} \rho_{0\, |\pm k}^{\text{th}},
 \\[0.2cm]
 \rho_{0\, |\pm k}^{\text{th}} &= \sum_{n_{k} = 0,1,~n_{-k} = 0,1} \frac{e^{ 2 \beta \epsilon_k^0 (1 - n_k - n_{-k})} }{4\cosh^2(\beta\epsilon_k^0)}|  n_{-k} n_k\rangle \langle n_{-k} n_k |.
 \end{align}\end{subequations}

After the instantaneous quench, which changes the field to its final value $g_{\tau}= g_0 + \delta g$, we have the final Hamiltonian,
\begin{equation}\begin{aligned}
    H_{\tau} &= \sum_{k} \epsilon_k^{\tau} (2 \xi^{\dagger}_k \xi_k - 1),
    \\[0.2cm]
    &= \sum_{k>0} 2 \epsilon_k^{\tau} \big( - | 0_{-k}^{\tau}0_k^{\tau} \rangle \langle 0_{-k}^{\tau}0_{k}^{\tau} | + | 1_{-k}^{\tau} 1_k^{\tau} \rangle \langle 1_{-k}^{\tau} 1_k^{\tau} | \big)
\end{aligned}\end{equation}
where $\epsilon_k^{\tau} = \epsilon_k(g_{\tau})$ and $|n_{-k}^{\tau} n_k^{\tau}\rangle$ are the joint eigenstates of the post-quench fermionic operators $\xi_k^{\dagger}\xi_k$ and $\xi_{-k}^{\dagger}\xi_{-k}$, which are related to the pre-quench operators $\{\eta_k\}$ according to \cite{Dorner2012}
\begin{equation}\label{operatorsmix}
 \xi_k = \cos(\Delta_k/2)\eta_k + \sin(\Delta_k/2)\eta_{-k}^{\dagger},
\end{equation}
where 
$\sin\Delta_k = -\delta g \sin(k)/ \epsilon_k^\tau \epsilon_k^0$. As discussed in Appendix~\ref{app:xymodel}, Eq.~\eqref{operatorsmix} shows us that the perturbation couples pairs of modes $+k$ and $-k$. This is why it is more convenient to write all quantities as  $\rho_0^\text{th} = \prod_{k>0} \rho_{0|\pm k}^\text{th}$ and $H_0 = \sum_{k>0} H_{0|\pm k}$, instead of a product/sum  over the negative values of $k$.

The corresponding final equilibrium state $\rho_{\tau}^{\text{th}} = e^{-\beta H_{\tau}}/Z_{\tau}$ is given by
\begin{subequations}\label{Gibbstau}
 \begin{align}
 \rho_{\tau}^{\text{th}} &= \prod\limits_{k>0} \rho_{\tau\, |\pm k}^{\text{th}},
 \\[0.2cm]
 \rho_{\tau\, |\pm k}^{\text{th}} &= \sum_{n_{k}^{\tau} = 0,1,~n_{-k}^{\tau} = 0,1} \frac{e^{ 2 \beta \epsilon_k^{\tau} (1 - n_k^{\tau} - n_{-k}^{\tau})} }{4\cosh^2(\beta\epsilon_k^{\tau})} |  n_{-k}^{\tau} n_k^{\tau}\rangle \langle n_{-k}^{\tau} n_k^{\tau} |.
 \end{align}\end{subequations}
We can proceed now to calculate $\tilde{\rho}_\tau^\text{th}$ in Eq.~\eqref{rho_tilde}. We first compute the dephased Hamiltonian $H_0 + \Delta H^\text{d}$, where $\Delta H^\text{d}$ is given in Eq.~\eqref{XY-Deph-Perturb}. That is,
\begin{equation}\begin{aligned}
   \mathbb{D}_{\rho_0^\text{th}}(H_{\tau}) 
   &= \sum_{k>0} 2\tilde{\epsilon}_k^{\tau}
   \big( -| 0_{-k} 0_k \rangle \langle 0_{-k} 0_k | + | 1_{-k} 1_k\rangle \langle 1_{-k} 1_k | \big),
\end{aligned}\end{equation}
where $\tilde{\epsilon}_k^{\tau} = \epsilon_k^{\tau} \cos \Delta_k = \epsilon_k^0 + \delta g \cos\theta_k $.
From this, one then finds the associated thermal state [Eq.~\eqref{rho_tilde}]
\begin{subequations}\label{GibbsDeph0}
 \begin{align}
 \tilde{\rho}^{\text{th}}_{\tau} &= \prod\limits_{k>0} \tilde{\rho}_{\tau\, |\pm k}^{\text{th}},
 \\[0.2cm]
 \tilde{\rho}_{\tau\, |\pm k}^{\text{th}} &= \sum_{n_{k} = 0,1,~n_{-k} = 0,1} \frac{e^{ 2 \beta \tilde{\epsilon}_k^{\tau} (1 - n_k - n_{-k})} }{4\cosh^2(\beta\tilde{\epsilon}_k^{\tau})}|  n_{-k} n_k\rangle \langle n_{-k} n_k |.
 \end{align}\end{subequations}

We have all we need to compute $\sigB$ and $\sigA$ now. We just have to plug Eqs.~\eqref{Gibbs0}, ~\eqref{Gibbstau} and~\eqref{GibbsDeph0} into~\eqref{B_free_energy} and~\eqref{A_free_energy}.
Because all states $\rho_0^{\text{th} }$, $\tilde{\rho}_{\tau}^{\text{th} }$ and $\rho_{\tau}^{\text{th} }$ are separable in terms of $\pm k$ modes, $\sigB$ and $\sigA$ will be given as sums over $k$. Hence, we find
\begin{equation}\label{B-XY-Quench-sum}
 \sigB = \sum_{k>0} 2\Bigg\{ \ln \Bigg[ \frac{ \cosh \Big( \beta \tilde{\epsilon}_k^{\tau} \Big) }{ \cosh \Big( \beta \epsilon_k^0 \Big)} \Bigg] + \beta \Big( \epsilon_k^0 - \tilde{\epsilon}_k^{\tau} \Big) \tanh \Big( \beta \epsilon_k^0 \Big) \Bigg\},
\end{equation}
and
\begin{equation}\label{A-XY-Quench-sum}
 \sigA = \sum_{k>0} 2 \ln \Bigg[ \frac{ \cosh \Big( \beta \epsilon_k^{\tau} \Big) }{ \cosh \Big( \beta \tilde{\epsilon}_k^{\tau} \Big)} \Bigg].
\end{equation}
Finally, in the limit of very large $N$, all $k$-sums can be converted to integrals and all quantities become extensive in $N$. In particular, we can substitute $\sum_{k>0}\to N \int_0^{\pi}\frac{\mathrm{d}k}{2\pi}$ in Eqs.~\eqref{B-XY-Quench-sum} and~\eqref{A-XY-Quench-sum} to obtain Eqs.~\eqref{B-XY-Quench} and~\eqref{A-XY-Quench}.

We note that Eqs.~\eqref{B-XY-Quench} and~\eqref{A-XY-Quench} do not assume that the quench is infinitesimal. 
All they assume is that $U = 1$. 
If, in particular, we are interested in infinitesimal quenches, then we may series expand these expressions in powers of $\delta g$, leading to 
\begin{IEEEeqnarray}{rCl}
 \label{B-XY-Quench-Inf}
    \sigB &=& N 
    \beta^2 \delta g^2 \int_0^{\pi} \frac{\mathrm{d}k}{2\pi} \, 
    \sech^2( \beta \epsilon_k^0 ) \cos^2 \theta_k,
    \\[0.25cm]
\label{A-XY-Quench-Inf}
    \sigA &=& N 
    \beta^2 \delta g^2 \int_0^{\pi} \frac{\mathrm{d}k}{2\pi} \, 
    \frac{ \tanh( \beta \epsilon_k^0 )}{\beta \epsilon_k^0} \sin^2 \theta_k,
\end{IEEEeqnarray}
where it is clear the relation of $\sigB$ and $\sigA$ with the dephased and coherent parts of the perturbation in Eqs.~\eqref{XY-Deph-Perturb} and~\eqref{XY-Cohe-Perturb}. Furthermore, it is easy to check that they satisfy Eq.~\eqref{BA_infinitesimal}.

\begin{widetext}

\section{\label{app:CDIsing} $\sigD$ and $\sigC$ for the TFIM}

For completeness, in this appendix we write down the expressions for $\sigD$ and $\sigC$ for the TFIM, computed in \cite{Varizi2020}:
\begin{IEEEeqnarray}{rCl}
 \label{C-XY-Quench}
 \sigC &=& N \int_0^{\pi} \frac{\mathrm{d}k}{2\pi}\, \Bigg\{ \frac{1}{2} \tanh \big( \beta \epsilon_k^0 \big) \Bigg[ \ln \bigg[ \frac{1 + \tanh \big( 2 \beta \epsilon_k^0 \big) }{1 - \tanh \big( 2 \beta \epsilon_k^0 \big) } \bigg] - \cos (\Delta_k) \ln \bigg[ \frac{1 + \tanh \big( 2 \beta \epsilon_k^0 \big) \cos(\Delta_k) }{1 - \tanh \big( 2 \beta \epsilon_k^0 \big) \cos(\Delta_k) } \bigg] \Bigg] - \frac{ \cosh \big( 2 \beta \epsilon_k^0 \big) }{ 4 \cosh^2 \big( \beta \epsilon_k^0 \big) } \nonumber \\ &\qquad& \times \ln \Big[ 1 + \sinh^2 \big( 2 \beta \epsilon_k^0 \big) \sin^2 (\Delta_k) \Big] \Bigg \},
 \\[0.2cm]
 \label{D-XY-Quench}
 \sigD &=& N \int_0^{\pi} \frac{\mathrm{d}k}{2\pi}\, \Bigg\{   2 \ln \bigg[ \frac{ \cosh \big( \beta \epsilon_k^\tau \big) }{ \cosh \big( \beta \epsilon_k^0 \big) } \bigg] - \frac{1}{2} \tanh \big( \beta \epsilon_k^0 \big) \cos (\Delta_k) \Bigg[ \ln \bigg[ \frac{1 + \tanh \big(2 \beta \epsilon_k^{\tau} \big)}{1 - \tanh \big( 2 \beta \epsilon_k^{\tau} \big)} \bigg] - \ln \bigg[ \frac{ 1 + \tanh \big( 2 \beta \epsilon_k^0 \big) \cos (\Delta_k) }{ 1 - \tanh \big( 2 \beta \epsilon_k^0 \big) \cos (\Delta_k)} \bigg] \Bigg] \nonumber \\
 &\qquad& + \frac{ \cosh \big ( 2 \beta \epsilon_k^0 \big)}{ 4 \cosh^2 \big( \beta \epsilon_k^0 \big) } \ln \Big[ 1 + \sinh^2 \big( 2 \beta \epsilon_k^0 \big) \sin^2 (\Delta_k) \Big] \Bigg\}.
\end{IEEEeqnarray}
 These expressions were used in plotting Figs.~\ref{fig:ABCDGrid}(c) and (d). 
The problem in the analyticities of these quantities stem from the last term in both integrals: In order to series expand them we need to satisfy the condition $\sinh^2(2\beta\epsilon_k^0)\sin^2(\Delta_k)<1$. This is prohibitive at low temperatures, since this function scales exponentially with $\beta$, through $\sinh^2(2\beta\epsilon_k^0)$, but only polynomially with the perturbation, through $\sin^2(\Delta_k)$.

\end{widetext}


\bibliography{library,bibl}

\begin{thebibliography}{95}%
\makeatletter
\providecommand \@ifxundefined [1]{%
 \@ifx{#1\undefined}
}%
\providecommand \@ifnum [1]{%
 \ifnum #1\expandafter \@firstoftwo
 \else \expandafter \@secondoftwo
 \fi
}%
\providecommand \@ifx [1]{%
 \ifx #1\expandafter \@firstoftwo
 \else \expandafter \@secondoftwo
 \fi
}%
\providecommand \natexlab [1]{#1}%
\providecommand \enquote  [1]{``#1''}%
\providecommand \bibnamefont  [1]{#1}%
\providecommand \bibfnamefont [1]{#1}%
\providecommand \citenamefont [1]{#1}%
\providecommand \href@noop [0]{\@secondoftwo}%
\providecommand \href [0]{\begingroup \@sanitize@url \@href}%
\providecommand \@href[1]{\@@startlink{#1}\@@href}%
\providecommand \@@href[1]{\endgroup#1\@@endlink}%
\providecommand \@sanitize@url [0]{\catcode `\\12\catcode `\$12\catcode
  `\&12\catcode `\#12\catcode `\^12\catcode `\_12\catcode `\%12\relax}%
\providecommand \@@startlink[1]{}%
\providecommand \@@endlink[0]{}%
\providecommand \url  [0]{\begingroup\@sanitize@url \@url }%
\providecommand \@url [1]{\endgroup\@href {#1}{\urlprefix }}%
\providecommand \urlprefix  [0]{URL }%
\providecommand \Eprint [0]{\href }%
\providecommand \doibase [0]{http://dx.doi.org/}%
\providecommand \selectlanguage [0]{\@gobble}%
\providecommand \bibinfo  [0]{\@secondoftwo}%
\providecommand \bibfield  [0]{\@secondoftwo}%
\providecommand \translation [1]{[#1]}%
\providecommand \BibitemOpen [0]{}%
\providecommand \bibitemStop [0]{}%
\providecommand \bibitemNoStop [0]{.\EOS\space}%
\providecommand \EOS [0]{\spacefactor3000\relax}%
\providecommand \BibitemShut  [1]{\csname bibitem#1\endcsname}%
\let\auto@bib@innerbib\@empty
\bibitem [{\citenamefont {Goold}\ \emph {et~al.}(2015)\citenamefont {Goold},
  \citenamefont {Huber}, \citenamefont {Riera}, \citenamefont {del Rio},\ and\
  \citenamefont {Skrzypczyk}}]{Goold2016}%
  \BibitemOpen
  \bibfield  {author} {\bibinfo {author} {\bibfnamefont {J.}~\bibnamefont
  {Goold}}, \bibinfo {author} {\bibfnamefont {M.}~\bibnamefont {Huber}},
  \bibinfo {author} {\bibfnamefont {A.}~\bibnamefont {Riera}}, \bibinfo
  {author} {\bibfnamefont {L.}~\bibnamefont {del Rio}}, \ and\ \bibinfo
  {author} {\bibfnamefont {P.}~\bibnamefont {Skrzypczyk}},\ }\href {\doibase
  10.1088/1751-8113/49/14/143001} {\bibfield  {journal} {\bibinfo  {journal}
  {1505.07835}\ }\textbf {\bibinfo {volume} {143001}},\ \bibinfo {pages} {31}
  (\bibinfo {year} {2015})},\ \Eprint {http://arxiv.org/abs/1505.07835}
  {arXiv:1505.07835} \BibitemShut {NoStop}%
\bibitem [{\citenamefont {Vinjanampathy}\ and\ \citenamefont
  {Anders}(2016)}]{Vinjanampathy2016}%
  \BibitemOpen
  \bibfield  {author} {\bibinfo {author} {\bibfnamefont {S.}~\bibnamefont
  {Vinjanampathy}}\ and\ \bibinfo {author} {\bibfnamefont {J.}~\bibnamefont
  {Anders}},\ }\href@noop {} {\bibfield  {journal} {\bibinfo  {journal}
  {Contemporary Physics}\ }\textbf {\bibinfo {volume} {57}},\ \bibinfo {pages}
  {545} (\bibinfo {year} {2016})}\BibitemShut {NoStop}%
\bibitem [{\citenamefont {Allahverdyan}\ \emph {et~al.}(2004)\citenamefont
  {Allahverdyan}, \citenamefont {Balian},\ and\ \citenamefont
  {Nieuwenhuizen}}]{Allahverdyan_2004}%
  \BibitemOpen
  \bibfield  {author} {\bibinfo {author} {\bibfnamefont {A.~E.}\ \bibnamefont
  {Allahverdyan}}, \bibinfo {author} {\bibfnamefont {R.}~\bibnamefont
  {Balian}}, \ and\ \bibinfo {author} {\bibfnamefont {T.~M.}\ \bibnamefont
  {Nieuwenhuizen}},\ }\href {\doibase 10.1209/epl/i2004-10101-2} {\bibfield
  {journal} {\bibinfo  {journal} {Europhysics Letters ({EPL})}\ }\textbf
  {\bibinfo {volume} {67}},\ \bibinfo {pages} {565} (\bibinfo {year}
  {2004})}\BibitemShut {NoStop}%
\bibitem [{\citenamefont {Scully}\ \emph {et~al.}(2007)\citenamefont {Scully},
  \citenamefont {Zubairy}, \citenamefont {Agarwal},\ and\ \citenamefont
  {Walther}}]{Scully2007}%
  \BibitemOpen
  \bibfield  {author} {\bibinfo {author} {\bibfnamefont {M.~O.}\ \bibnamefont
  {Scully}}, \bibinfo {author} {\bibfnamefont {M.~S.}\ \bibnamefont {Zubairy}},
  \bibinfo {author} {\bibfnamefont {G.~S.}\ \bibnamefont {Agarwal}}, \ and\
  \bibinfo {author} {\bibfnamefont {H.}~\bibnamefont {Walther}},\ }\href
  {\doibase 10.1126/science.1078955} {\bibfield  {journal} {\bibinfo  {journal}
  {Science}\ }\textbf {\bibinfo {volume} {299}},\ \bibinfo {pages} {862}
  (\bibinfo {year} {2007})}\BibitemShut {NoStop}%
\bibitem [{\citenamefont {Korzekwa}\ \emph {et~al.}(2016)\citenamefont
  {Korzekwa}, \citenamefont {Lostaglio}, \citenamefont {Oppenheim},\ and\
  \citenamefont {Jennings}}]{Korzekwa2016}%
  \BibitemOpen
  \bibfield  {author} {\bibinfo {author} {\bibfnamefont {K.}~\bibnamefont
  {Korzekwa}}, \bibinfo {author} {\bibfnamefont {M.}~\bibnamefont {Lostaglio}},
  \bibinfo {author} {\bibfnamefont {J.}~\bibnamefont {Oppenheim}}, \ and\
  \bibinfo {author} {\bibfnamefont {D.}~\bibnamefont {Jennings}},\ }\href@noop
  {} {\bibfield  {journal} {\bibinfo  {journal} {New Journal of Physics}\
  }\textbf {\bibinfo {volume} {18}},\ \bibinfo {pages} {023045} (\bibinfo
  {year} {2016})}\BibitemShut {NoStop}%
\bibitem [{\citenamefont {Manzano}\ \emph {et~al.}(2018)\citenamefont
  {Manzano}, \citenamefont {Plastina},\ and\ \citenamefont
  {Zambrini}}]{Manzano2018}%
  \BibitemOpen
  \bibfield  {author} {\bibinfo {author} {\bibfnamefont {G.}~\bibnamefont
  {Manzano}}, \bibinfo {author} {\bibfnamefont {F.}~\bibnamefont {Plastina}}, \
  and\ \bibinfo {author} {\bibfnamefont {R.}~\bibnamefont {Zambrini}},\ }\href
  {\doibase 10.1103/PhysRevLett.121.120602} {\bibfield  {journal} {\bibinfo
  {journal} {Physical Review Letters}\ }\textbf {\bibinfo {volume} {121}},\
  \bibinfo {pages} {120602} (\bibinfo {year} {2018})},\ \Eprint
  {http://arxiv.org/abs/1805.08184} {arXiv:1805.08184} \BibitemShut {NoStop}%
\bibitem [{\citenamefont {L\"{o}rch}\ \emph {et~al.}(2018)\citenamefont
  {L\"{o}rch}, \citenamefont {Bruder}, \citenamefont {Brunner},\ and\
  \citenamefont {Hofer}}]{Lrch2018}%
  \BibitemOpen
  \bibfield  {author} {\bibinfo {author} {\bibfnamefont {N.}~\bibnamefont
  {L\"{o}rch}}, \bibinfo {author} {\bibfnamefont {C.}~\bibnamefont {Bruder}},
  \bibinfo {author} {\bibfnamefont {N.}~\bibnamefont {Brunner}}, \ and\
  \bibinfo {author} {\bibfnamefont {P.~P.}\ \bibnamefont {Hofer}},\ }\href
  {\doibase 10.1088/2058-9565/aacbf3} {\bibfield  {journal} {\bibinfo
  {journal} {Quantum Science and Technology}\ }\textbf {\bibinfo {volume}
  {3}},\ \bibinfo {pages} {035014} (\bibinfo {year} {2018})}\BibitemShut
  {NoStop}%
\bibitem [{\citenamefont {Rodrigues}\ \emph {et~al.}(2019)\citenamefont
  {Rodrigues}, \citenamefont {{De Chiara}}, \citenamefont {Paternostro},\ and\
  \citenamefont {Landi}}]{Rodrigues2019}%
  \BibitemOpen
  \bibfield  {author} {\bibinfo {author} {\bibfnamefont {F.~L.~S.}\
  \bibnamefont {Rodrigues}}, \bibinfo {author} {\bibfnamefont {G.}~\bibnamefont
  {{De Chiara}}}, \bibinfo {author} {\bibfnamefont {M.}~\bibnamefont
  {Paternostro}}, \ and\ \bibinfo {author} {\bibfnamefont {G.~T.}\ \bibnamefont
  {Landi}},\ }\href {\doibase 10.1103/PhysRevLett.123.140601} {\bibfield
  {journal} {\bibinfo  {journal} {Physical Review Letters}\ }\textbf {\bibinfo
  {volume} {123}},\ \bibinfo {pages} {140601} (\bibinfo {year} {2019})},\
  \Eprint {http://arxiv.org/abs/1906.08203} {arXiv:1906.08203} \BibitemShut
  {NoStop}%
\bibitem [{\citenamefont {Francica}\ \emph {et~al.}(2020)\citenamefont
  {Francica}, \citenamefont {Binder}, \citenamefont {Guarnieri}, \citenamefont
  {Mitchison}, \citenamefont {Goold},\ and\ \citenamefont
  {Plastina}}]{Francica2020}%
  \BibitemOpen
  \bibfield  {author} {\bibinfo {author} {\bibfnamefont {G.}~\bibnamefont
  {Francica}}, \bibinfo {author} {\bibfnamefont {F.~C.}\ \bibnamefont
  {Binder}}, \bibinfo {author} {\bibfnamefont {G.}~\bibnamefont {Guarnieri}},
  \bibinfo {author} {\bibfnamefont {M.~T.}\ \bibnamefont {Mitchison}}, \bibinfo
  {author} {\bibfnamefont {J.}~\bibnamefont {Goold}}, \ and\ \bibinfo {author}
  {\bibfnamefont {F.}~\bibnamefont {Plastina}},\ }\href {\doibase
  10.1103/PhysRevLett.125.180603} {\bibfield  {journal} {\bibinfo  {journal}
  {Physical Review Letters}\ }\textbf {\bibinfo {volume} {125}},\ \bibinfo
  {pages} {180603} (\bibinfo {year} {2020})},\ \Eprint
  {http://arxiv.org/abs/2006.05424} {arXiv:2006.05424} \BibitemShut {NoStop}%
\bibitem [{\citenamefont {Hovhannisyan}\ \emph {et~al.}(2013)\citenamefont
  {Hovhannisyan}, \citenamefont {Perarnau-Llobet}, \citenamefont {Huber},\ and\
  \citenamefont {Ac{\'{\i}}n}}]{Hovhannisyan2013}%
  \BibitemOpen
  \bibfield  {author} {\bibinfo {author} {\bibfnamefont {K.~V.}\ \bibnamefont
  {Hovhannisyan}}, \bibinfo {author} {\bibfnamefont {M.}~\bibnamefont
  {Perarnau-Llobet}}, \bibinfo {author} {\bibfnamefont {M.}~\bibnamefont
  {Huber}}, \ and\ \bibinfo {author} {\bibfnamefont {A.}~\bibnamefont
  {Ac{\'{\i}}n}},\ }\href {\doibase 10.1103/physrevlett.111.240401} {\bibfield
  {journal} {\bibinfo  {journal} {Physical Review Letters}\ }\textbf {\bibinfo
  {volume} {111}} (\bibinfo {year} {2013}),\
  10.1103/physrevlett.111.240401}\BibitemShut {NoStop}%
\bibitem [{\citenamefont {Campaioli}\ \emph {et~al.}(2017)\citenamefont
  {Campaioli}, \citenamefont {Pollock}, \citenamefont {Binder}, \citenamefont
  {C{\'{e}}leri}, \citenamefont {Goold}, \citenamefont {Vinjanampathy},\ and\
  \citenamefont {Modi}}]{Campaioli2017}%
  \BibitemOpen
  \bibfield  {author} {\bibinfo {author} {\bibfnamefont {F.}~\bibnamefont
  {Campaioli}}, \bibinfo {author} {\bibfnamefont {F.~A.}\ \bibnamefont
  {Pollock}}, \bibinfo {author} {\bibfnamefont {F.~C.}\ \bibnamefont {Binder}},
  \bibinfo {author} {\bibfnamefont {L.}~\bibnamefont {C{\'{e}}leri}}, \bibinfo
  {author} {\bibfnamefont {J.}~\bibnamefont {Goold}}, \bibinfo {author}
  {\bibfnamefont {S.}~\bibnamefont {Vinjanampathy}}, \ and\ \bibinfo {author}
  {\bibfnamefont {K.}~\bibnamefont {Modi}},\ }\href {\doibase
  10.1103/PhysRevLett.118.150601} {\bibfield  {journal} {\bibinfo  {journal}
  {Physical Review Letters}\ }\textbf {\bibinfo {volume} {118}},\ \bibinfo
  {pages} {150601} (\bibinfo {year} {2017})},\ \Eprint
  {http://arxiv.org/abs/1612.04991} {arXiv:1612.04991} \BibitemShut {NoStop}%
\bibitem [{\citenamefont {Campaioli}\ \emph {et~al.}(2018)\citenamefont
  {Campaioli}, \citenamefont {Pollock},\ and\ \citenamefont
  {Vinjanampathy}}]{campaioli2018quantum}%
  \BibitemOpen
  \bibfield  {author} {\bibinfo {author} {\bibfnamefont {F.}~\bibnamefont
  {Campaioli}}, \bibinfo {author} {\bibfnamefont {F.~A.}\ \bibnamefont
  {Pollock}}, \ and\ \bibinfo {author} {\bibfnamefont {S.}~\bibnamefont
  {Vinjanampathy}},\ }in\ \href {\doibase 10.1007/978-3-319-99046-0_8} {\emph
  {\bibinfo {booktitle} {Thermodynamics in the Quantum Regime}}}\ (\bibinfo
  {publisher} {Springer},\ \bibinfo {year} {2018})\ pp.\ \bibinfo {pages}
  {207--225}\BibitemShut {NoStop}%
\bibitem [{\citenamefont {Juli\`a-Farr\'e}\ \emph {et~al.}(2020)\citenamefont
  {Juli\`a-Farr\'e}, \citenamefont {Salamon}, \citenamefont {Riera},
  \citenamefont {Bera},\ and\ \citenamefont {Lewenstein}}]{Julia2020}%
  \BibitemOpen
  \bibfield  {author} {\bibinfo {author} {\bibfnamefont {S.}~\bibnamefont
  {Juli\`a-Farr\'e}}, \bibinfo {author} {\bibfnamefont {T.}~\bibnamefont
  {Salamon}}, \bibinfo {author} {\bibfnamefont {A.}~\bibnamefont {Riera}},
  \bibinfo {author} {\bibfnamefont {M.~N.}\ \bibnamefont {Bera}}, \ and\
  \bibinfo {author} {\bibfnamefont {M.}~\bibnamefont {Lewenstein}},\ }\href
  {\doibase 10.1103/PhysRevResearch.2.023113} {\bibfield  {journal} {\bibinfo
  {journal} {Phys. Rev. Research}\ }\textbf {\bibinfo {volume} {2}},\ \bibinfo
  {pages} {023113} (\bibinfo {year} {2020})}\BibitemShut {NoStop}%
\bibitem [{\citenamefont {Correa}\ \emph {et~al.}(2014)\citenamefont {Correa},
  \citenamefont {Palao}, \citenamefont {Alonso},\ and\ \citenamefont
  {Adesso}}]{Correa2014}%
  \BibitemOpen
  \bibfield  {author} {\bibinfo {author} {\bibfnamefont {L.~A.}\ \bibnamefont
  {Correa}}, \bibinfo {author} {\bibfnamefont {J.~P.}\ \bibnamefont {Palao}},
  \bibinfo {author} {\bibfnamefont {D.}~\bibnamefont {Alonso}}, \ and\ \bibinfo
  {author} {\bibfnamefont {G.}~\bibnamefont {Adesso}},\ }\href {\doibase
  10.1038/srep03949} {\bibfield  {journal} {\bibinfo  {journal} {Scientific
  reports}\ }\textbf {\bibinfo {volume} {4}},\ \bibinfo {pages} {3949}
  (\bibinfo {year} {2014})},\ \Eprint {http://arxiv.org/abs/arXiv:1308.4174v1}
  {arXiv:arXiv:1308.4174v1} \BibitemShut {NoStop}%
\bibitem [{\citenamefont {Rossnagel}\ \emph {et~al.}(2014)\citenamefont
  {Rossnagel}, \citenamefont {Abah}, \citenamefont {Schmidt-Kaler},
  \citenamefont {Singer},\ and\ \citenamefont {Lutz}}]{Ronagel2014}%
  \BibitemOpen
  \bibfield  {author} {\bibinfo {author} {\bibfnamefont {J.}~\bibnamefont
  {Rossnagel}}, \bibinfo {author} {\bibfnamefont {O.}~\bibnamefont {Abah}},
  \bibinfo {author} {\bibfnamefont {F.}~\bibnamefont {Schmidt-Kaler}}, \bibinfo
  {author} {\bibfnamefont {K.}~\bibnamefont {Singer}}, \ and\ \bibinfo {author}
  {\bibfnamefont {E.}~\bibnamefont {Lutz}},\ }\href {\doibase
  10.1103/PhysRevLett.112.030602} {\bibfield  {journal} {\bibinfo  {journal}
  {Physical Review Letters}\ }\textbf {\bibinfo {volume} {112}},\ \bibinfo
  {pages} {030602} (\bibinfo {year} {2014})},\ \Eprint
  {http://arxiv.org/abs/1308.5935} {arXiv:1308.5935} \BibitemShut {NoStop}%
\bibitem [{\citenamefont {Brunner}\ \emph {et~al.}(2014)\citenamefont
  {Brunner}, \citenamefont {Cavalcanti}, \citenamefont {Pironio}, \citenamefont
  {Scarani},\ and\ \citenamefont {Wehner}}]{Brunner2014}%
  \BibitemOpen
  \bibfield  {author} {\bibinfo {author} {\bibfnamefont {N.}~\bibnamefont
  {Brunner}}, \bibinfo {author} {\bibfnamefont {D.}~\bibnamefont {Cavalcanti}},
  \bibinfo {author} {\bibfnamefont {S.}~\bibnamefont {Pironio}}, \bibinfo
  {author} {\bibfnamefont {V.}~\bibnamefont {Scarani}}, \ and\ \bibinfo
  {author} {\bibfnamefont {S.}~\bibnamefont {Wehner}},\ }\href {\doibase
  10.1103/RevModPhys.86.419} {\bibfield  {journal} {\bibinfo  {journal}
  {Reviews of Modern Physics}\ }\textbf {\bibinfo {volume} {86}},\ \bibinfo
  {pages} {419} (\bibinfo {year} {2014})},\ \Eprint
  {http://arxiv.org/abs/1303.2849} {arXiv:1303.2849} \BibitemShut {NoStop}%
\bibitem [{\citenamefont {Uzdin}\ \emph {et~al.}(2015)\citenamefont {Uzdin},
  \citenamefont {Levy},\ and\ \citenamefont {Kosloff}}]{Uzdin2015}%
  \BibitemOpen
  \bibfield  {author} {\bibinfo {author} {\bibfnamefont {R.}~\bibnamefont
  {Uzdin}}, \bibinfo {author} {\bibfnamefont {A.}~\bibnamefont {Levy}}, \ and\
  \bibinfo {author} {\bibfnamefont {R.}~\bibnamefont {Kosloff}},\ }\href
  {\doibase 10.1103/PhysRevX.5.031044} {\bibfield  {journal} {\bibinfo
  {journal} {Physical Review X}\ }\textbf {\bibinfo {volume} {5}},\ \bibinfo
  {pages} {031044} (\bibinfo {year} {2015})}\BibitemShut {NoStop}%
\bibitem [{\citenamefont {Manzano}\ \emph {et~al.}(2016)\citenamefont
  {Manzano}, \citenamefont {Galve}, \citenamefont {Zambrini},\ and\
  \citenamefont {Parrondo}}]{Manzano2016}%
  \BibitemOpen
  \bibfield  {author} {\bibinfo {author} {\bibfnamefont {G.}~\bibnamefont
  {Manzano}}, \bibinfo {author} {\bibfnamefont {F.}~\bibnamefont {Galve}},
  \bibinfo {author} {\bibfnamefont {R.}~\bibnamefont {Zambrini}}, \ and\
  \bibinfo {author} {\bibfnamefont {J.~M.~R.}\ \bibnamefont {Parrondo}},\
  }\href {\doibase 10.1103/PhysRevE.93.052120} {\bibfield  {journal} {\bibinfo
  {journal} {Physical Review E}\ }\textbf {\bibinfo {volume} {93}},\ \bibinfo
  {pages} {052120} (\bibinfo {year} {2016})}\BibitemShut {NoStop}%
\bibitem [{\citenamefont {Hammam}\ \emph {et~al.}(2021)\citenamefont {Hammam},
  \citenamefont {Hassouni}, \citenamefont {Fazio},\ and\ \citenamefont
  {Manzano}}]{hammam2021optimizing}%
  \BibitemOpen
  \bibfield  {author} {\bibinfo {author} {\bibfnamefont {K.}~\bibnamefont
  {Hammam}}, \bibinfo {author} {\bibfnamefont {Y.}~\bibnamefont {Hassouni}},
  \bibinfo {author} {\bibfnamefont {R.}~\bibnamefont {Fazio}}, \ and\ \bibinfo
  {author} {\bibfnamefont {G.}~\bibnamefont {Manzano}},\ }\href@noop {}
  {\bibfield  {journal} {\bibinfo  {journal} {arXiv preprint arXiv:2101.11572}\
  } (\bibinfo {year} {2021})}\BibitemShut {NoStop}%
\bibitem [{\citenamefont {Janzing}(2006)}]{janzing2006quantum}%
  \BibitemOpen
  \bibfield  {author} {\bibinfo {author} {\bibfnamefont {D.}~\bibnamefont
  {Janzing}},\ }\href {\doibase 10.1007/s10955-006-9220-x} {\bibfield
  {journal} {\bibinfo  {journal} {Journal of statistical physics}\ }\textbf
  {\bibinfo {volume} {125}},\ \bibinfo {pages} {761} (\bibinfo {year}
  {2006})}\BibitemShut {NoStop}%
\bibitem [{\citenamefont {Lostaglio}\ \emph {et~al.}(2015)\citenamefont
  {Lostaglio}, \citenamefont {Jennings},\ and\ \citenamefont
  {Rudolph}}]{Lostaglio2015}%
  \BibitemOpen
  \bibfield  {author} {\bibinfo {author} {\bibfnamefont {M.}~\bibnamefont
  {Lostaglio}}, \bibinfo {author} {\bibfnamefont {D.}~\bibnamefont {Jennings}},
  \ and\ \bibinfo {author} {\bibfnamefont {T.}~\bibnamefont {Rudolph}},\ }\href
  {\doibase 10.1038/ncomms7383} {\bibfield  {journal} {\bibinfo  {journal}
  {Nature communications}\ }\textbf {\bibinfo {volume} {6}},\ \bibinfo {pages}
  {6383} (\bibinfo {year} {2015})},\ \Eprint {http://arxiv.org/abs/1405.2188}
  {arXiv:1405.2188} \BibitemShut {NoStop}%
\bibitem [{\citenamefont {Cwiklinski}\ \emph {et~al.}(2015)\citenamefont
  {Cwiklinski}, \citenamefont {Studzinski}, \citenamefont {Horodecki},\ and\
  \citenamefont {Oppenheim}}]{Cwikli2015}%
  \BibitemOpen
  \bibfield  {author} {\bibinfo {author} {\bibfnamefont {P.}~\bibnamefont
  {Cwiklinski}}, \bibinfo {author} {\bibfnamefont {M.}~\bibnamefont
  {Studzinski}}, \bibinfo {author} {\bibfnamefont {M.}~\bibnamefont
  {Horodecki}}, \ and\ \bibinfo {author} {\bibfnamefont {J.}~\bibnamefont
  {Oppenheim}},\ }\href {\doibase 10.1103/PhysRevLett.115.210403} {\bibfield
  {journal} {\bibinfo  {journal} {Physical Review Letters}\ }\textbf {\bibinfo
  {volume} {115}},\ \bibinfo {pages} {210403} (\bibinfo {year} {2015})},\
  \Eprint {http://arxiv.org/abs/1405.5029} {arXiv:1405.5029} \BibitemShut
  {NoStop}%
\bibitem [{\citenamefont {Allahverdyan}(2014)}]{Allahverdyan2014}%
  \BibitemOpen
  \bibfield  {author} {\bibinfo {author} {\bibfnamefont {A.~E.}\ \bibnamefont
  {Allahverdyan}},\ }\href {\doibase 10.1103/physreve.90.032137} {\bibfield
  {journal} {\bibinfo  {journal} {Physical Review E}\ }\textbf {\bibinfo
  {volume} {90}} (\bibinfo {year} {2014}),\
  10.1103/physreve.90.032137}\BibitemShut {NoStop}%
\bibitem [{\citenamefont {Talkner}\ and\ \citenamefont
  {H{\"{a}}nggi}(2016)}]{Talkner2016}%
  \BibitemOpen
  \bibfield  {author} {\bibinfo {author} {\bibfnamefont {P.}~\bibnamefont
  {Talkner}}\ and\ \bibinfo {author} {\bibfnamefont {P.}~\bibnamefont
  {H{\"{a}}nggi}},\ }\href {\doibase 10.1103/PhysRevE.93.022131} {\bibfield
  {journal} {\bibinfo  {journal} {Physical Review E}\ }\textbf {\bibinfo
  {volume} {93}},\ \bibinfo {pages} {022131} (\bibinfo {year} {2016})},\
  \Eprint {http://arxiv.org/abs/1512.02516} {arXiv:1512.02516} \BibitemShut
  {NoStop}%
\bibitem [{\citenamefont {Hofer}\ \emph {et~al.}(2017)\citenamefont {Hofer},
  \citenamefont {Brask}, \citenamefont {Perarnau-Llobet},\ and\ \citenamefont
  {Brunner}}]{Hofer2017}%
  \BibitemOpen
  \bibfield  {author} {\bibinfo {author} {\bibfnamefont {P.~P.}\ \bibnamefont
  {Hofer}}, \bibinfo {author} {\bibfnamefont {J.~B.}\ \bibnamefont {Brask}},
  \bibinfo {author} {\bibfnamefont {M.}~\bibnamefont {Perarnau-Llobet}}, \ and\
  \bibinfo {author} {\bibfnamefont {N.}~\bibnamefont {Brunner}},\ }\href@noop
  {} {\ ,\ \bibinfo {pages} {23} (\bibinfo {year} {2017})},\ \Eprint
  {http://arxiv.org/abs/1703.03719} {arXiv:1703.03719} \BibitemShut {NoStop}%
\bibitem [{\citenamefont {B\"{a}umer}\ \emph {et~al.}(2018)\citenamefont
  {B\"{a}umer}, \citenamefont {Lostaglio}, \citenamefont {Perarnau-Llobet},\
  and\ \citenamefont {Sampaio}}]{Bumer2018}%
  \BibitemOpen
  \bibfield  {author} {\bibinfo {author} {\bibfnamefont {E.}~\bibnamefont
  {B\"{a}umer}}, \bibinfo {author} {\bibfnamefont {M.}~\bibnamefont
  {Lostaglio}}, \bibinfo {author} {\bibfnamefont {M.}~\bibnamefont
  {Perarnau-Llobet}}, \ and\ \bibinfo {author} {\bibfnamefont {R.}~\bibnamefont
  {Sampaio}},\ }in\ \href {\doibase 10.1007/978-3-319-99046-0_11} {\emph
  {\bibinfo {booktitle} {Fundamental Theories of Physics}}}\ (\bibinfo
  {publisher} {Springer International Publishing},\ \bibinfo {year} {2018})\
  pp.\ \bibinfo {pages} {275--300}\BibitemShut {NoStop}%
\bibitem [{\citenamefont {Levy}\ and\ \citenamefont
  {Lostaglio}(2020)}]{Levy2020}%
  \BibitemOpen
  \bibfield  {author} {\bibinfo {author} {\bibfnamefont {A.}~\bibnamefont
  {Levy}}\ and\ \bibinfo {author} {\bibfnamefont {M.}~\bibnamefont
  {Lostaglio}},\ }\href {\doibase 10.1103/PRXQuantum.1.010309} {\bibfield
  {journal} {\bibinfo  {journal} {PRX Quantum}\ }\textbf {\bibinfo {volume}
  {1}},\ \bibinfo {pages} {010309} (\bibinfo {year} {2020})}\BibitemShut
  {NoStop}%
\bibitem [{\citenamefont {Micadei}\ \emph {et~al.}(2020)\citenamefont
  {Micadei}, \citenamefont {Landi},\ and\ \citenamefont {Lutz}}]{Micadei2020}%
  \BibitemOpen
  \bibfield  {author} {\bibinfo {author} {\bibfnamefont {K.}~\bibnamefont
  {Micadei}}, \bibinfo {author} {\bibfnamefont {G.~T.}\ \bibnamefont {Landi}},
  \ and\ \bibinfo {author} {\bibfnamefont {E.}~\bibnamefont {Lutz}},\ }\href
  {\doibase 10.1103/PhysRevLett.124.090602} {\bibfield  {journal} {\bibinfo
  {journal} {Phys. Rev. Lett.}\ }\textbf {\bibinfo {volume} {124}},\ \bibinfo
  {pages} {090602} (\bibinfo {year} {2020})}\BibitemShut {NoStop}%
\bibitem [{\citenamefont {Miller}\ \emph {et~al.}(2019)\citenamefont {Miller},
  \citenamefont {Scandi}, \citenamefont {Anders},\ and\ \citenamefont
  {Perarnau-Llobet}}]{Miller2019}%
  \BibitemOpen
  \bibfield  {author} {\bibinfo {author} {\bibfnamefont {H.~J.~D.}\
  \bibnamefont {Miller}}, \bibinfo {author} {\bibfnamefont {M.}~\bibnamefont
  {Scandi}}, \bibinfo {author} {\bibfnamefont {J.}~\bibnamefont {Anders}}, \
  and\ \bibinfo {author} {\bibfnamefont {M.}~\bibnamefont {Perarnau-Llobet}},\
  }\href {\doibase 10.1103/PhysRevLett.123.230603} {\bibfield  {journal}
  {\bibinfo  {journal} {Physical Review Letters}\ }\textbf {\bibinfo {volume}
  {123}},\ \bibinfo {pages} {230603} (\bibinfo {year} {2019})},\ \Eprint
  {http://arxiv.org/abs/1905.07328} {arXiv:1905.07328} \BibitemShut {NoStop}%
\bibitem [{\citenamefont {Scandi}\ \emph {et~al.}(2020)\citenamefont {Scandi},
  \citenamefont {Miller}, \citenamefont {Anders},\ and\ \citenamefont
  {Perarnau-Llobet}}]{Scandi2019}%
  \BibitemOpen
  \bibfield  {author} {\bibinfo {author} {\bibfnamefont {M.}~\bibnamefont
  {Scandi}}, \bibinfo {author} {\bibfnamefont {H.~J.~D.}\ \bibnamefont
  {Miller}}, \bibinfo {author} {\bibfnamefont {J.}~\bibnamefont {Anders}}, \
  and\ \bibinfo {author} {\bibfnamefont {M.}~\bibnamefont {Perarnau-Llobet}},\
  }\href {\doibase 10.1103/PhysRevResearch.2.023377} {\bibfield  {journal}
  {\bibinfo  {journal} {Physical Review Research}\ }\textbf {\bibinfo {volume}
  {2}},\ \bibinfo {pages} {023377} (\bibinfo {year} {2020})},\ \Eprint
  {http://arxiv.org/abs/1911.04306} {arXiv:1911.04306} \BibitemShut {NoStop}%
\bibitem [{\citenamefont {Miller}\ \emph
  {et~al.}(2020{\natexlab{a}})\citenamefont {Miller}, \citenamefont
  {Mohammady}, \citenamefont {Perarnau-Llobet},\ and\ \citenamefont
  {Guarnieri}}]{miller2020joint}%
  \BibitemOpen
  \bibfield  {author} {\bibinfo {author} {\bibfnamefont {H.~J.}\ \bibnamefont
  {Miller}}, \bibinfo {author} {\bibfnamefont {M.~H.}\ \bibnamefont
  {Mohammady}}, \bibinfo {author} {\bibfnamefont {M.}~\bibnamefont
  {Perarnau-Llobet}}, \ and\ \bibinfo {author} {\bibfnamefont {G.}~\bibnamefont
  {Guarnieri}},\ }\href@noop {} {\bibfield  {journal} {\bibinfo  {journal}
  {arXiv preprint arXiv:2011.11589}\ } (\bibinfo {year}
  {2020}{\natexlab{a}})}\BibitemShut {NoStop}%
\bibitem [{\citenamefont {Lloyd}(1989)}]{Lloyd1989}%
  \BibitemOpen
  \bibfield  {author} {\bibinfo {author} {\bibfnamefont {S.}~\bibnamefont
  {Lloyd}},\ }\href {\doibase 10.1103/PhysRevA.39.5378} {\bibfield  {journal}
  {\bibinfo  {journal} {Physical Review A}\ }\textbf {\bibinfo {volume} {39}},\
  \bibinfo {pages} {5378} (\bibinfo {year} {1989})}\BibitemShut {NoStop}%
\bibitem [{\citenamefont {Jennings}\ and\ \citenamefont
  {Rudolph}(2010)}]{Jennings2010}%
  \BibitemOpen
  \bibfield  {author} {\bibinfo {author} {\bibfnamefont {D.}~\bibnamefont
  {Jennings}}\ and\ \bibinfo {author} {\bibfnamefont {T.}~\bibnamefont
  {Rudolph}},\ }\href {\doibase 10.1103/PhysRevE.81.061130} {\bibfield
  {journal} {\bibinfo  {journal} {Physical Review E - Statistical, Nonlinear,
  and Soft Matter Physics}\ }\textbf {\bibinfo {volume} {81}},\ \bibinfo
  {pages} {061130} (\bibinfo {year} {2010})},\ \Eprint
  {http://arxiv.org/abs/1002.0314} {arXiv:1002.0314} \BibitemShut {NoStop}%
\bibitem [{\citenamefont {Jevtic}\ \emph {et~al.}(2015)\citenamefont {Jevtic},
  \citenamefont {Rudolph}, \citenamefont {Jennings}, \citenamefont {Hirono},
  \citenamefont {Nakayama},\ and\ \citenamefont {Murao}}]{Jevtic2015a}%
  \BibitemOpen
  \bibfield  {author} {\bibinfo {author} {\bibfnamefont {S.}~\bibnamefont
  {Jevtic}}, \bibinfo {author} {\bibfnamefont {T.}~\bibnamefont {Rudolph}},
  \bibinfo {author} {\bibfnamefont {D.}~\bibnamefont {Jennings}}, \bibinfo
  {author} {\bibfnamefont {Y.}~\bibnamefont {Hirono}}, \bibinfo {author}
  {\bibfnamefont {S.}~\bibnamefont {Nakayama}}, \ and\ \bibinfo {author}
  {\bibfnamefont {M.}~\bibnamefont {Murao}},\ }\href {\doibase
  10.1103/PhysRevE.92.042113} {\bibfield  {journal} {\bibinfo  {journal}
  {Physical Review E}\ }\textbf {\bibinfo {volume} {92}},\ \bibinfo {pages}
  {042113} (\bibinfo {year} {2015})},\ \Eprint {http://arxiv.org/abs/1204.3571}
  {arXiv:1204.3571} \BibitemShut {NoStop}%
\bibitem [{\citenamefont {Micadei}\ \emph {et~al.}(2019)\citenamefont
  {Micadei}, \citenamefont {Peterson}, \citenamefont {Souza}, \citenamefont
  {Sarthour}, \citenamefont {Oliveira}, \citenamefont {Landi}, \citenamefont
  {Batalh{\~{a}}o}, \citenamefont {Serra},\ and\ \citenamefont
  {Lutz}}]{Micadei2017}%
  \BibitemOpen
  \bibfield  {author} {\bibinfo {author} {\bibfnamefont {K.}~\bibnamefont
  {Micadei}}, \bibinfo {author} {\bibfnamefont {J.~P.~S.}\ \bibnamefont
  {Peterson}}, \bibinfo {author} {\bibfnamefont {A.~M.}\ \bibnamefont {Souza}},
  \bibinfo {author} {\bibfnamefont {R.~S.}\ \bibnamefont {Sarthour}}, \bibinfo
  {author} {\bibfnamefont {I.~S.}\ \bibnamefont {Oliveira}}, \bibinfo {author}
  {\bibfnamefont {G.~T.}\ \bibnamefont {Landi}}, \bibinfo {author}
  {\bibfnamefont {T.~B.}\ \bibnamefont {Batalh{\~{a}}o}}, \bibinfo {author}
  {\bibfnamefont {R.~M.}\ \bibnamefont {Serra}}, \ and\ \bibinfo {author}
  {\bibfnamefont {E.}~\bibnamefont {Lutz}},\ }\href {\doibase
  10.1038/s41467-019-10333-7} {\bibfield  {journal} {\bibinfo  {journal}
  {Nature Communications}\ }\textbf {\bibinfo {volume} {10}},\ \bibinfo {pages}
  {2456} (\bibinfo {year} {2019})},\ \Eprint {http://arxiv.org/abs/1711.03323}
  {arXiv:1711.03323} \BibitemShut {NoStop}%
\bibitem [{\citenamefont {Santos}\ \emph {et~al.}(2019)\citenamefont {Santos},
  \citenamefont {C{\'{e}}leri}, \citenamefont {Landi},\ and\ \citenamefont
  {Paternostro}}]{Santos2019}%
  \BibitemOpen
  \bibfield  {author} {\bibinfo {author} {\bibfnamefont {J.~P.}\ \bibnamefont
  {Santos}}, \bibinfo {author} {\bibfnamefont {L.~C.}\ \bibnamefont
  {C{\'{e}}leri}}, \bibinfo {author} {\bibfnamefont {G.~T.}\ \bibnamefont
  {Landi}}, \ and\ \bibinfo {author} {\bibfnamefont {M.}~\bibnamefont
  {Paternostro}},\ }\href {\doibase https://doi.org/10.1038/s41534-019-0138-y}
  {\bibfield  {journal} {\bibinfo  {journal} {npj Quantum Information}\
  }\textbf {\bibinfo {volume} {5}},\ \bibinfo {pages} {23} (\bibinfo {year}
  {2019})},\ \Eprint {http://arxiv.org/abs/1707.08946} {arXiv:1707.08946}
  \BibitemShut {NoStop}%
\bibitem [{\citenamefont {Mohammady}\ \emph {et~al.}(2020)\citenamefont
  {Mohammady}, \citenamefont {Auff{\`{e}}ves},\ and\ \citenamefont
  {Anders}}]{Mohammady2020}%
  \BibitemOpen
  \bibfield  {author} {\bibinfo {author} {\bibfnamefont {M.~H.}\ \bibnamefont
  {Mohammady}}, \bibinfo {author} {\bibfnamefont {A.}~\bibnamefont
  {Auff{\`{e}}ves}}, \ and\ \bibinfo {author} {\bibfnamefont {J.}~\bibnamefont
  {Anders}},\ }\href {\doibase 10.1038/s42005-020-0356-9} {\bibfield  {journal}
  {\bibinfo  {journal} {Communications Physics}\ }\textbf {\bibinfo {volume}
  {3}},\ \bibinfo {pages} {89} (\bibinfo {year} {2020})},\ \Eprint
  {http://arxiv.org/abs/1907.06559} {arXiv:1907.06559} \BibitemShut {NoStop}%
\bibitem [{\citenamefont {Francica}\ \emph {et~al.}(2019)\citenamefont
  {Francica}, \citenamefont {Goold},\ and\ \citenamefont
  {Plastina}}]{Francica2019}%
  \BibitemOpen
  \bibfield  {author} {\bibinfo {author} {\bibfnamefont {G.}~\bibnamefont
  {Francica}}, \bibinfo {author} {\bibfnamefont {J.}~\bibnamefont {Goold}}, \
  and\ \bibinfo {author} {\bibfnamefont {F.}~\bibnamefont {Plastina}},\ }\href
  {\doibase 10.1103/PhysRevE.99.042105} {\bibfield  {journal} {\bibinfo
  {journal} {Physical Review E}\ }\textbf {\bibinfo {volume} {99}},\ \bibinfo
  {pages} {042105} (\bibinfo {year} {2019})},\ \Eprint
  {http://arxiv.org/abs/1707.06950} {arXiv:1707.06950} \BibitemShut {NoStop}%
\bibitem [{\citenamefont {Varizi}\ \emph {et~al.}(2020)\citenamefont {Varizi},
  \citenamefont {Vieira}, \citenamefont {Cormick}, \citenamefont {Drumond},\
  and\ \citenamefont {Landi}}]{Varizi2020}%
  \BibitemOpen
  \bibfield  {author} {\bibinfo {author} {\bibfnamefont {A.~D.}\ \bibnamefont
  {Varizi}}, \bibinfo {author} {\bibfnamefont {A.~P.}\ \bibnamefont {Vieira}},
  \bibinfo {author} {\bibfnamefont {C.}~\bibnamefont {Cormick}}, \bibinfo
  {author} {\bibfnamefont {R.~C.}\ \bibnamefont {Drumond}}, \ and\ \bibinfo
  {author} {\bibfnamefont {G.~T.}\ \bibnamefont {Landi}},\ }\href {\doibase
  10.1103/PhysRevResearch.2.033279} {\bibfield  {journal} {\bibinfo  {journal}
  {Physical Review Research}\ }\textbf {\bibinfo {volume} {2}},\ \bibinfo
  {pages} {033279} (\bibinfo {year} {2020})},\ \Eprint
  {http://arxiv.org/abs/2004.00616} {arXiv:2004.00616} \BibitemShut {NoStop}%
\bibitem [{\citenamefont {Esposito}\ \emph {et~al.}(2009)\citenamefont
  {Esposito}, \citenamefont {Harbola},\ and\ \citenamefont
  {Mukamel}}]{Esposito2009}%
  \BibitemOpen
  \bibfield  {author} {\bibinfo {author} {\bibfnamefont {M.}~\bibnamefont
  {Esposito}}, \bibinfo {author} {\bibfnamefont {U.}~\bibnamefont {Harbola}}, \
  and\ \bibinfo {author} {\bibfnamefont {S.}~\bibnamefont {Mukamel}},\ }\href
  {\doibase 10.1103/RevModPhys.81.1665} {\bibfield  {journal} {\bibinfo
  {journal} {Reviews of Modern Physics}\ }\textbf {\bibinfo {volume} {81}},\
  \bibinfo {pages} {1665} (\bibinfo {year} {2009})}\BibitemShut {NoStop}%
\bibitem [{\citenamefont {Campisi}\ \emph {et~al.}(2011)\citenamefont
  {Campisi}, \citenamefont {H{\"{a}}nggi},\ and\ \citenamefont
  {Talkner}}]{Campisi2011}%
  \BibitemOpen
  \bibfield  {author} {\bibinfo {author} {\bibfnamefont {M.}~\bibnamefont
  {Campisi}}, \bibinfo {author} {\bibfnamefont {P.}~\bibnamefont
  {H{\"{a}}nggi}}, \ and\ \bibinfo {author} {\bibfnamefont {P.}~\bibnamefont
  {Talkner}},\ }\href {\doibase 10.1103/RevModPhys.83.771} {\bibfield
  {journal} {\bibinfo  {journal} {Reviews of Modern Physics}\ }\textbf
  {\bibinfo {volume} {83}},\ \bibinfo {pages} {771} (\bibinfo {year}
  {2011})}\BibitemShut {NoStop}%
\bibitem [{\citenamefont {Kawai}\ \emph {et~al.}(2007)\citenamefont {Kawai},
  \citenamefont {Parrondo},\ and\ \citenamefont {{Van Den
  Broeck}}}]{Kawai2007}%
  \BibitemOpen
  \bibfield  {author} {\bibinfo {author} {\bibfnamefont {R.}~\bibnamefont
  {Kawai}}, \bibinfo {author} {\bibfnamefont {J.~M.}\ \bibnamefont {Parrondo}},
  \ and\ \bibinfo {author} {\bibfnamefont {C.}~\bibnamefont {{Van Den
  Broeck}}},\ }\href {\doibase 10.1103/PhysRevLett.98.080602} {\bibfield
  {journal} {\bibinfo  {journal} {Physical Review Letters}\ }\textbf {\bibinfo
  {volume} {98}},\ \bibinfo {pages} {080602} (\bibinfo {year} {2007})},\
  \Eprint {http://arxiv.org/abs/0701397} {arXiv:0701397 [cond-mat]}
  \BibitemShut {NoStop}%
\bibitem [{\citenamefont {Vaikuntanathan}\ and\ \citenamefont
  {Jarzynski}(2009)}]{Vaikuntanathan2009}%
  \BibitemOpen
  \bibfield  {author} {\bibinfo {author} {\bibfnamefont {S.}~\bibnamefont
  {Vaikuntanathan}}\ and\ \bibinfo {author} {\bibfnamefont {C.}~\bibnamefont
  {Jarzynski}},\ }\href {\doibase 10.1209/0295-5075/87/60005} {\bibfield
  {journal} {\bibinfo  {journal} {EPL (Europhysics Letters)}\ }\textbf
  {\bibinfo {volume} {87}},\ \bibinfo {pages} {60005} (\bibinfo {year}
  {2009})}\BibitemShut {NoStop}%
\bibitem [{\citenamefont {Parrondo}\ \emph {et~al.}(2009)\citenamefont
  {Parrondo}, \citenamefont {{Van Den Broeck}},\ and\ \citenamefont
  {Kawai}}]{Parrondo2009}%
  \BibitemOpen
  \bibfield  {author} {\bibinfo {author} {\bibfnamefont {J.~M.}\ \bibnamefont
  {Parrondo}}, \bibinfo {author} {\bibfnamefont {C.}~\bibnamefont {{Van Den
  Broeck}}}, \ and\ \bibinfo {author} {\bibfnamefont {R.}~\bibnamefont
  {Kawai}},\ }\href {\doibase 10.1088/1367-2630/11/7/073008} {\bibfield
  {journal} {\bibinfo  {journal} {New Journal of Physics}\ }\textbf {\bibinfo
  {volume} {11}},\ \bibinfo {pages} {073008} (\bibinfo {year} {2009})},\
  \Eprint {http://arxiv.org/abs/0904.1573} {arXiv:0904.1573} \BibitemShut
  {NoStop}%
\bibitem [{\citenamefont {Deffner}\ and\ \citenamefont
  {Lutz}(2010)}]{Deffner2010}%
  \BibitemOpen
  \bibfield  {author} {\bibinfo {author} {\bibfnamefont {S.}~\bibnamefont
  {Deffner}}\ and\ \bibinfo {author} {\bibfnamefont {E.}~\bibnamefont {Lutz}},\
  }\href {\doibase 10.1103/PhysRevLett.105.170402} {\bibfield  {journal}
  {\bibinfo  {journal} {Physical Review Letters}\ }\textbf {\bibinfo {volume}
  {105}},\ \bibinfo {pages} {170402} (\bibinfo {year} {2010})},\ \Eprint
  {http://arxiv.org/abs/1005.4495} {arXiv:1005.4495} \BibitemShut {NoStop}%
\bibitem [{\citenamefont {Jarzynski}(1997)}]{Jarzynski1997}%
  \BibitemOpen
  \bibfield  {author} {\bibinfo {author} {\bibfnamefont {C.}~\bibnamefont
  {Jarzynski}},\ }\href {\doibase 10.1103/PhysRevLett.78.2690} {\bibfield
  {journal} {\bibinfo  {journal} {Physical Review Letters}\ }\textbf {\bibinfo
  {volume} {78}},\ \bibinfo {pages} {2690} (\bibinfo {year}
  {1997})}\BibitemShut {NoStop}%
\bibitem [{\citenamefont {Kurchan}(1998)}]{Kurchan1998}%
  \BibitemOpen
  \bibfield  {author} {\bibinfo {author} {\bibfnamefont {J.}~\bibnamefont
  {Kurchan}},\ }\href {\doibase 10.1088/0305-4470/31/16/003} {\bibfield
  {journal} {\bibinfo  {journal} {Journal of Physics A: Mathematical and
  General}\ }\textbf {\bibinfo {volume} {31}},\ \bibinfo {pages} {3719}
  (\bibinfo {year} {1998})}\BibitemShut {NoStop}%
\bibitem [{\citenamefont {Talkner}\ \emph {et~al.}(2007)\citenamefont
  {Talkner}, \citenamefont {Lutz},\ and\ \citenamefont
  {H{\"{a}}nggi}}]{Talkner2007}%
  \BibitemOpen
  \bibfield  {author} {\bibinfo {author} {\bibfnamefont {P.}~\bibnamefont
  {Talkner}}, \bibinfo {author} {\bibfnamefont {E.}~\bibnamefont {Lutz}}, \
  and\ \bibinfo {author} {\bibfnamefont {P.}~\bibnamefont {H{\"{a}}nggi}},\
  }\href {\doibase 10.1103/PhysRevE.75.050102} {\bibfield  {journal} {\bibinfo
  {journal} {Physical Review E}\ }\textbf {\bibinfo {volume} {75}},\ \bibinfo
  {pages} {050102} (\bibinfo {year} {2007})}\BibitemShut {NoStop}%
\bibitem [{\citenamefont {Derrida}\ and\ \citenamefont
  {Lebowitz}(1998)}]{Derrida1998}%
  \BibitemOpen
  \bibfield  {author} {\bibinfo {author} {\bibfnamefont {B.}~\bibnamefont
  {Derrida}}\ and\ \bibinfo {author} {\bibfnamefont {J.~L.}\ \bibnamefont
  {Lebowitz}},\ }\href {\doibase 10.1103/PhysRevLett.80.209} {\bibfield
  {journal} {\bibinfo  {journal} {Physical Review Letters}\ }\textbf {\bibinfo
  {volume} {80}},\ \bibinfo {pages} {209} (\bibinfo {year} {1998})},\ \Eprint
  {http://arxiv.org/abs/9809044} {arXiv:9809044 [cond-mat]} \BibitemShut
  {NoStop}%
\bibitem [{\citenamefont {Crooks}(1998)}]{Crooks1998}%
  \BibitemOpen
  \bibfield  {author} {\bibinfo {author} {\bibfnamefont {G.~E.}\ \bibnamefont
  {Crooks}},\ }\href {http://link.springer.com/article/10.1023/A:1023208217925}
  {\bibfield  {journal} {\bibinfo  {journal} {Journal of Statistical Physics}\
  }\textbf {\bibinfo {volume} {90}},\ \bibinfo {pages} {1481} (\bibinfo {year}
  {1998})}\BibitemShut {NoStop}%
\bibitem [{\citenamefont {Lebowitz}\ and\ \citenamefont
  {Spohn}(1999)}]{Lebowitz1999}%
  \BibitemOpen
  \bibfield  {author} {\bibinfo {author} {\bibfnamefont {J.}~\bibnamefont
  {Lebowitz}}\ and\ \bibinfo {author} {\bibfnamefont {H.}~\bibnamefont
  {Spohn}},\ }\href {http://link.springer.com/article/10.1023/A:1004589714161}
  {\bibfield  {journal} {\bibinfo  {journal} {Journal of Statistical Physics}\
  }\textbf {\bibinfo {volume} {95}},\ \bibinfo {pages} {333} (\bibinfo {year}
  {1999})}\BibitemShut {NoStop}%
\bibitem [{\citenamefont {Mukamel}(2003)}]{Mukamel2003}%
  \BibitemOpen
  \bibfield  {author} {\bibinfo {author} {\bibfnamefont {S.}~\bibnamefont
  {Mukamel}},\ }\href {\doibase 10.1103/PhysRevLett.90.170604} {\bibfield
  {journal} {\bibinfo  {journal} {Physical Review Letters}\ }\textbf {\bibinfo
  {volume} {90}},\ \bibinfo {pages} {170604} (\bibinfo {year} {2003})},\
  \Eprint {http://arxiv.org/abs/0302190} {arXiv:0302190 [cond-mat]}
  \BibitemShut {NoStop}%
\bibitem [{\citenamefont {Guarnieri}\ \emph {et~al.}(2018)\citenamefont
  {Guarnieri}, \citenamefont {Ng}, \citenamefont {Modi}, \citenamefont
  {Eisert}, \citenamefont {Paternostro},\ and\ \citenamefont
  {Goold}}]{Guarnieri2018}%
  \BibitemOpen
  \bibfield  {author} {\bibinfo {author} {\bibfnamefont {G.}~\bibnamefont
  {Guarnieri}}, \bibinfo {author} {\bibfnamefont {N.~H.~Y.}\ \bibnamefont
  {Ng}}, \bibinfo {author} {\bibfnamefont {K.}~\bibnamefont {Modi}}, \bibinfo
  {author} {\bibfnamefont {J.}~\bibnamefont {Eisert}}, \bibinfo {author}
  {\bibfnamefont {M.}~\bibnamefont {Paternostro}}, \ and\ \bibinfo {author}
  {\bibfnamefont {J.}~\bibnamefont {Goold}},\ }\href {\doibase
  10.1103/PhysRevE.99.050101} {\bibfield  {journal} {\bibinfo  {journal}
  {Physical Review E}\ }\textbf {\bibinfo {volume} {99}},\ \bibinfo {pages}
  {050101} (\bibinfo {year} {2018})},\ \Eprint
  {http://arxiv.org/abs/1804.09962} {arXiv:1804.09962} \BibitemShut {NoStop}%
\bibitem [{\citenamefont {Liphardt}\ \emph {et~al.}(2002)\citenamefont
  {Liphardt}, \citenamefont {Dumont}, \citenamefont {Smith}, \citenamefont
  {Tinoco},\ and\ \citenamefont {Bustamante}}]{Liphardt2002}%
  \BibitemOpen
  \bibfield  {author} {\bibinfo {author} {\bibfnamefont {J.}~\bibnamefont
  {Liphardt}}, \bibinfo {author} {\bibfnamefont {S.}~\bibnamefont {Dumont}},
  \bibinfo {author} {\bibfnamefont {S.~B.}\ \bibnamefont {Smith}}, \bibinfo
  {author} {\bibfnamefont {I.}~\bibnamefont {Tinoco}}, \ and\ \bibinfo {author}
  {\bibfnamefont {C.}~\bibnamefont {Bustamante}},\ }\href {\doibase
  10.1126/science.1071152} {\bibfield  {journal} {\bibinfo  {journal} {Science
  (New York, N.Y.)}\ }\textbf {\bibinfo {volume} {296}},\ \bibinfo {pages}
  {1832} (\bibinfo {year} {2002})}\BibitemShut {NoStop}%
\bibitem [{\citenamefont {Douarche}\ \emph {et~al.}(2005)\citenamefont
  {Douarche}, \citenamefont {Ciliberto}, \citenamefont {Petrosyan},\ and\
  \citenamefont {Rabbiosi}}]{Douarche2005}%
  \BibitemOpen
  \bibfield  {author} {\bibinfo {author} {\bibfnamefont {F.}~\bibnamefont
  {Douarche}}, \bibinfo {author} {\bibfnamefont {S.}~\bibnamefont {Ciliberto}},
  \bibinfo {author} {\bibfnamefont {a.}~\bibnamefont {Petrosyan}}, \ and\
  \bibinfo {author} {\bibfnamefont {I.}~\bibnamefont {Rabbiosi}},\ }\href
  {\doibase 10.1209/epl/i2005-10024-4} {\bibfield  {journal} {\bibinfo
  {journal} {Europhysics Letters (EPL)}\ }\textbf {\bibinfo {volume} {70}},\
  \bibinfo {pages} {593} (\bibinfo {year} {2005})}\BibitemShut {NoStop}%
\bibitem [{\citenamefont {Collin}\ \emph {et~al.}(2005)\citenamefont {Collin},
  \citenamefont {Ritort}, \citenamefont {Jarzynski}, \citenamefont {Smith},
  \citenamefont {Tinoco},\ and\ \citenamefont {Bustamante}}]{Collin2005}%
  \BibitemOpen
  \bibfield  {author} {\bibinfo {author} {\bibfnamefont {D.}~\bibnamefont
  {Collin}}, \bibinfo {author} {\bibfnamefont {F.}~\bibnamefont {Ritort}},
  \bibinfo {author} {\bibfnamefont {C.}~\bibnamefont {Jarzynski}}, \bibinfo
  {author} {\bibfnamefont {S.~B.}\ \bibnamefont {Smith}}, \bibinfo {author}
  {\bibfnamefont {I.}~\bibnamefont {Tinoco}}, \ and\ \bibinfo {author}
  {\bibfnamefont {C.}~\bibnamefont {Bustamante}},\ }\href {\doibase
  10.1038/nature04061} {\bibfield  {journal} {\bibinfo  {journal} {Nature}\
  }\textbf {\bibinfo {volume} {437}},\ \bibinfo {pages} {231} (\bibinfo {year}
  {2005})}\BibitemShut {NoStop}%
\bibitem [{\citenamefont {Speck}\ \emph {et~al.}(2007)\citenamefont {Speck},
  \citenamefont {Blickle}, \citenamefont {Bechinger},\ and\ \citenamefont
  {Seifert}}]{Speck2007}%
  \BibitemOpen
  \bibfield  {author} {\bibinfo {author} {\bibfnamefont {T.}~\bibnamefont
  {Speck}}, \bibinfo {author} {\bibfnamefont {V.}~\bibnamefont {Blickle}},
  \bibinfo {author} {\bibfnamefont {C.}~\bibnamefont {Bechinger}}, \ and\
  \bibinfo {author} {\bibfnamefont {U.}~\bibnamefont {Seifert}},\ }\href
  {\doibase 10.1209/0295-5075/79/30002} {\bibfield  {journal} {\bibinfo
  {journal} {Europhysics Letters (EPL)}\ }\textbf {\bibinfo {volume} {79}},\
  \bibinfo {pages} {30002} (\bibinfo {year} {2007})}\BibitemShut {NoStop}%
\bibitem [{\citenamefont {Saira}\ \emph {et~al.}(2012)\citenamefont {Saira},
  \citenamefont {Yoon}, \citenamefont {Tanttu}, \citenamefont
  {M{\"{o}}tt{\"{o}}nen}, \citenamefont {Averin},\ and\ \citenamefont
  {Pekola}}]{Saira2012}%
  \BibitemOpen
  \bibfield  {author} {\bibinfo {author} {\bibfnamefont {O.~P.}\ \bibnamefont
  {Saira}}, \bibinfo {author} {\bibfnamefont {Y.}~\bibnamefont {Yoon}},
  \bibinfo {author} {\bibfnamefont {T.}~\bibnamefont {Tanttu}}, \bibinfo
  {author} {\bibfnamefont {M.}~\bibnamefont {M{\"{o}}tt{\"{o}}nen}}, \bibinfo
  {author} {\bibfnamefont {D.~V.}\ \bibnamefont {Averin}}, \ and\ \bibinfo
  {author} {\bibfnamefont {J.~P.}\ \bibnamefont {Pekola}},\ }\href {\doibase
  10.1103/PhysRevLett.109.180601} {\bibfield  {journal} {\bibinfo  {journal}
  {Physical Review Letters}\ }\textbf {\bibinfo {volume} {109}},\ \bibinfo
  {pages} {180601} (\bibinfo {year} {2012})},\ \Eprint
  {http://arxiv.org/abs/1206.7049} {arXiv:1206.7049} \BibitemShut {NoStop}%
\bibitem [{\citenamefont {Koski}\ \emph {et~al.}(2013)\citenamefont {Koski},
  \citenamefont {Sagawa}, \citenamefont {Saira}, \citenamefont {Yoon},
  \citenamefont {Kutvonen}, \citenamefont {Solinas}, \citenamefont
  {M{\"{o}}tt{\"{o}}nen}, \citenamefont {Ala-Nissila},\ and\ \citenamefont
  {Pekola}}]{Koski2013}%
  \BibitemOpen
  \bibfield  {author} {\bibinfo {author} {\bibfnamefont {J.~V.}\ \bibnamefont
  {Koski}}, \bibinfo {author} {\bibfnamefont {T.}~\bibnamefont {Sagawa}},
  \bibinfo {author} {\bibfnamefont {O.~P.}\ \bibnamefont {Saira}}, \bibinfo
  {author} {\bibfnamefont {Y.}~\bibnamefont {Yoon}}, \bibinfo {author}
  {\bibfnamefont {A.}~\bibnamefont {Kutvonen}}, \bibinfo {author}
  {\bibfnamefont {P.}~\bibnamefont {Solinas}}, \bibinfo {author} {\bibfnamefont
  {M.}~\bibnamefont {M{\"{o}}tt{\"{o}}nen}}, \bibinfo {author} {\bibfnamefont
  {T.}~\bibnamefont {Ala-Nissila}}, \ and\ \bibinfo {author} {\bibfnamefont
  {J.~P.}\ \bibnamefont {Pekola}},\ }\href {\doibase 10.1038/nphys2711}
  {\bibfield  {journal} {\bibinfo  {journal} {Nature Physics}\ }\textbf
  {\bibinfo {volume} {9}},\ \bibinfo {pages} {644} (\bibinfo {year} {2013})},\
  \Eprint {http://arxiv.org/abs/1303.6405} {arXiv:1303.6405} \BibitemShut
  {NoStop}%
\bibitem [{\citenamefont {Batalh{\~{a}}o}\ \emph {et~al.}(2014)\citenamefont
  {Batalh{\~{a}}o}, \citenamefont {Souza}, \citenamefont {Mazzola},
  \citenamefont {Auccaise}, \citenamefont {Sarthour}, \citenamefont {Oliveira},
  \citenamefont {Goold}, \citenamefont {{De Chiara}}, \citenamefont
  {Paternostro},\ and\ \citenamefont {Serra}}]{Batalhao2014}%
  \BibitemOpen
  \bibfield  {author} {\bibinfo {author} {\bibfnamefont {T.~B.}\ \bibnamefont
  {Batalh{\~{a}}o}}, \bibinfo {author} {\bibfnamefont {A.~M.}\ \bibnamefont
  {Souza}}, \bibinfo {author} {\bibfnamefont {L.}~\bibnamefont {Mazzola}},
  \bibinfo {author} {\bibfnamefont {R.}~\bibnamefont {Auccaise}}, \bibinfo
  {author} {\bibfnamefont {R.~S.}\ \bibnamefont {Sarthour}}, \bibinfo {author}
  {\bibfnamefont {I.~S.}\ \bibnamefont {Oliveira}}, \bibinfo {author}
  {\bibfnamefont {J.}~\bibnamefont {Goold}}, \bibinfo {author} {\bibfnamefont
  {G.}~\bibnamefont {{De Chiara}}}, \bibinfo {author} {\bibfnamefont
  {M.}~\bibnamefont {Paternostro}}, \ and\ \bibinfo {author} {\bibfnamefont
  {R.~M.}\ \bibnamefont {Serra}},\ }\href {\doibase
  10.1103/PhysRevLett.113.140601} {\bibfield  {journal} {\bibinfo  {journal}
  {Physical Review Letters}\ }\textbf {\bibinfo {volume} {113}},\ \bibinfo
  {pages} {140601} (\bibinfo {year} {2014})}\BibitemShut {NoStop}%
\bibitem [{\citenamefont {An}\ \emph {et~al.}(2014)\citenamefont {An},
  \citenamefont {Zhang}, \citenamefont {Um}, \citenamefont {Lv}, \citenamefont
  {Lu}, \citenamefont {Zhang}, \citenamefont {Yin}, \citenamefont {Quan},\ and\
  \citenamefont {Kim}}]{An2014}%
  \BibitemOpen
  \bibfield  {author} {\bibinfo {author} {\bibfnamefont {S.}~\bibnamefont
  {An}}, \bibinfo {author} {\bibfnamefont {J.-N.}\ \bibnamefont {Zhang}},
  \bibinfo {author} {\bibfnamefont {M.}~\bibnamefont {Um}}, \bibinfo {author}
  {\bibfnamefont {D.}~\bibnamefont {Lv}}, \bibinfo {author} {\bibfnamefont
  {Y.}~\bibnamefont {Lu}}, \bibinfo {author} {\bibfnamefont {J.}~\bibnamefont
  {Zhang}}, \bibinfo {author} {\bibfnamefont {Z.-Q.}\ \bibnamefont {Yin}},
  \bibinfo {author} {\bibfnamefont {H.~T.}\ \bibnamefont {Quan}}, \ and\
  \bibinfo {author} {\bibfnamefont {K.}~\bibnamefont {Kim}},\ }\href {\doibase
  10.1038/nphys3197} {\bibfield  {journal} {\bibinfo  {journal} {Nature
  Physics}\ }\textbf {\bibinfo {volume} {11}},\ \bibinfo {pages} {193}
  (\bibinfo {year} {2014})},\ \Eprint {http://arxiv.org/abs/1409.4485}
  {arXiv:1409.4485} \BibitemShut {NoStop}%
\bibitem [{\citenamefont {Batalh{\~{a}}o}\ \emph {et~al.}(2015)\citenamefont
  {Batalh{\~{a}}o}, \citenamefont {Souza}, \citenamefont {Sarthour},
  \citenamefont {Oliveira}, \citenamefont {Paternostro}, \citenamefont {Lutz},\
  and\ \citenamefont {Serra}}]{Batalhao2015}%
  \BibitemOpen
  \bibfield  {author} {\bibinfo {author} {\bibfnamefont {T.~B.}\ \bibnamefont
  {Batalh{\~{a}}o}}, \bibinfo {author} {\bibfnamefont {A.~M.}\ \bibnamefont
  {Souza}}, \bibinfo {author} {\bibfnamefont {R.~S.}\ \bibnamefont {Sarthour}},
  \bibinfo {author} {\bibfnamefont {I.~S.}\ \bibnamefont {Oliveira}}, \bibinfo
  {author} {\bibfnamefont {M.}~\bibnamefont {Paternostro}}, \bibinfo {author}
  {\bibfnamefont {E.}~\bibnamefont {Lutz}}, \ and\ \bibinfo {author}
  {\bibfnamefont {R.~M.}\ \bibnamefont {Serra}},\ }\href {\doibase
  10.1103/PhysRevLett.115.190601} {\bibfield  {journal} {\bibinfo  {journal}
  {Physical Review Letters}\ }\textbf {\bibinfo {volume} {115}},\ \bibinfo
  {pages} {190601} (\bibinfo {year} {2015})},\ \Eprint
  {http://arxiv.org/abs/1502.06704v1} {arXiv:1502.06704v1} \BibitemShut
  {NoStop}%
\bibitem [{\citenamefont {Talarico}\ \emph {et~al.}(2016)\citenamefont
  {Talarico}, \citenamefont {Monteiro}, \citenamefont {Mattei}, \citenamefont
  {Duzzioni}, \citenamefont {{Souto Ribeiro}},\ and\ \citenamefont
  {C{\'{e}}leri}}]{Talarico2016}%
  \BibitemOpen
  \bibfield  {author} {\bibinfo {author} {\bibfnamefont {M.~A.}\ \bibnamefont
  {Talarico}}, \bibinfo {author} {\bibfnamefont {P.~B.}\ \bibnamefont
  {Monteiro}}, \bibinfo {author} {\bibfnamefont {E.~C.}\ \bibnamefont
  {Mattei}}, \bibinfo {author} {\bibfnamefont {E.~I.}\ \bibnamefont
  {Duzzioni}}, \bibinfo {author} {\bibfnamefont {P.~H.}\ \bibnamefont {{Souto
  Ribeiro}}}, \ and\ \bibinfo {author} {\bibfnamefont {L.~C.}\ \bibnamefont
  {C{\'{e}}leri}},\ }\href {\doibase 10.1103/PhysRevA.94.042305} {\bibfield
  {journal} {\bibinfo  {journal} {Physical Review A}\ }\textbf {\bibinfo
  {volume} {94}},\ \bibinfo {pages} {042305} (\bibinfo {year} {2016})},\
  \Eprint {http://arxiv.org/abs/1604.07237} {arXiv:1604.07237} \BibitemShut
  {NoStop}%
\bibitem [{\citenamefont {Zhang}\ \emph {et~al.}(2018)\citenamefont {Zhang},
  \citenamefont {Wang}, \citenamefont {Xiang}, \citenamefont {Jia},
  \citenamefont {Duan}, \citenamefont {Cai}, \citenamefont {Zhan},
  \citenamefont {Zong}, \citenamefont {Wu}, \citenamefont {Sun}, \citenamefont
  {Yin},\ and\ \citenamefont {Guo}}]{Zhang2018a}%
  \BibitemOpen
  \bibfield  {author} {\bibinfo {author} {\bibfnamefont {Z.}~\bibnamefont
  {Zhang}}, \bibinfo {author} {\bibfnamefont {T.}~\bibnamefont {Wang}},
  \bibinfo {author} {\bibfnamefont {L.}~\bibnamefont {Xiang}}, \bibinfo
  {author} {\bibfnamefont {Z.}~\bibnamefont {Jia}}, \bibinfo {author}
  {\bibfnamefont {P.}~\bibnamefont {Duan}}, \bibinfo {author} {\bibfnamefont
  {W.}~\bibnamefont {Cai}}, \bibinfo {author} {\bibfnamefont {Z.}~\bibnamefont
  {Zhan}}, \bibinfo {author} {\bibfnamefont {Z.}~\bibnamefont {Zong}}, \bibinfo
  {author} {\bibfnamefont {J.}~\bibnamefont {Wu}}, \bibinfo {author}
  {\bibfnamefont {L.}~\bibnamefont {Sun}}, \bibinfo {author} {\bibfnamefont
  {Y.}~\bibnamefont {Yin}}, \ and\ \bibinfo {author} {\bibfnamefont
  {G.}~\bibnamefont {Guo}},\ }\href {\doibase 10.1088/1367-2630/aad4e7}
  {\bibfield  {journal} {\bibinfo  {journal} {New Journal of Physics}\ }\textbf
  {\bibinfo {volume} {20}},\ \bibinfo {pages} {085001} (\bibinfo {year}
  {2018})},\ \Eprint {http://arxiv.org/abs/1805.10879} {arXiv:1805.10879}
  \BibitemShut {NoStop}%
\bibitem [{\citenamefont {Smith}\ \emph {et~al.}(2018)\citenamefont {Smith},
  \citenamefont {Lu}, \citenamefont {An}, \citenamefont {Zhang}, \citenamefont
  {Zhang}, \citenamefont {Gong}, \citenamefont {Quan}, \citenamefont
  {Jarzynski},\ and\ \citenamefont {Kim}}]{Smith2017}%
  \BibitemOpen
  \bibfield  {author} {\bibinfo {author} {\bibfnamefont {A.}~\bibnamefont
  {Smith}}, \bibinfo {author} {\bibfnamefont {Y.}~\bibnamefont {Lu}}, \bibinfo
  {author} {\bibfnamefont {S.}~\bibnamefont {An}}, \bibinfo {author}
  {\bibfnamefont {X.}~\bibnamefont {Zhang}}, \bibinfo {author} {\bibfnamefont
  {J.-N.}\ \bibnamefont {Zhang}}, \bibinfo {author} {\bibfnamefont
  {Z.}~\bibnamefont {Gong}}, \bibinfo {author} {\bibfnamefont {H.~T.}\
  \bibnamefont {Quan}}, \bibinfo {author} {\bibfnamefont {C.}~\bibnamefont
  {Jarzynski}}, \ and\ \bibinfo {author} {\bibfnamefont {K.}~\bibnamefont
  {Kim}},\ }\href {http://arxiv.org/abs/1708.01495} {\bibfield  {journal}
  {\bibinfo  {journal} {New Journal of Physics}\ }\textbf {\bibinfo {volume}
  {20}},\ \bibinfo {pages} {013008} (\bibinfo {year} {2018})},\ \Eprint
  {http://arxiv.org/abs/1708.01495} {arXiv:1708.01495} \BibitemShut {NoStop}%
\bibitem [{\citenamefont {Elouard}\ \emph {et~al.}(2020)\citenamefont
  {Elouard}, \citenamefont {Herrera-Mart{\'{i}}}, \citenamefont {Esposito},\
  and\ \citenamefont {Auff{\`{e}}ves}}]{Elouard2020}%
  \BibitemOpen
  \bibfield  {author} {\bibinfo {author} {\bibfnamefont {C.}~\bibnamefont
  {Elouard}}, \bibinfo {author} {\bibfnamefont {D.}~\bibnamefont
  {Herrera-Mart{\'{i}}}}, \bibinfo {author} {\bibfnamefont {M.}~\bibnamefont
  {Esposito}}, \ and\ \bibinfo {author} {\bibfnamefont {A.}~\bibnamefont
  {Auff{\`{e}}ves}},\ }\href {\doibase 10.1088/1367-2630/abbd6e} {\bibfield
  {journal} {\bibinfo  {journal} {New Journal of Physics}\ }\textbf {\bibinfo
  {volume} {22}},\ \bibinfo {pages} {103039} (\bibinfo {year} {2020})},\
  \Eprint {http://arxiv.org/abs/2001.08033} {arXiv:2001.08033} \BibitemShut
  {NoStop}%
\bibitem [{\citenamefont {Brand{\~{a}}o}\ \emph {et~al.}(2013)\citenamefont
  {Brand{\~{a}}o}, \citenamefont {Horodecki}, \citenamefont {Oppenheim},
  \citenamefont {Renes},\ and\ \citenamefont {Spekkens}}]{Brandao2013}%
  \BibitemOpen
  \bibfield  {author} {\bibinfo {author} {\bibfnamefont {F.~G. S.~L.}\
  \bibnamefont {Brand{\~{a}}o}}, \bibinfo {author} {\bibfnamefont
  {M.}~\bibnamefont {Horodecki}}, \bibinfo {author} {\bibfnamefont
  {J.}~\bibnamefont {Oppenheim}}, \bibinfo {author} {\bibfnamefont {J.~M.}\
  \bibnamefont {Renes}}, \ and\ \bibinfo {author} {\bibfnamefont {R.~W.}\
  \bibnamefont {Spekkens}},\ }\href {\doibase 10.1103/PhysRevLett.111.250404}
  {\bibfield  {journal} {\bibinfo  {journal} {Physical Review Letters}\
  }\textbf {\bibinfo {volume} {111}},\ \bibinfo {pages} {250404} (\bibinfo
  {year} {2013})},\ \Eprint {http://arxiv.org/abs/1111.3882} {arXiv:1111.3882}
  \BibitemShut {NoStop}%
\bibitem [{\citenamefont {Horodecki}\ and\ \citenamefont
  {Oppenheim}(2013)}]{Horodecki2013}%
  \BibitemOpen
  \bibfield  {author} {\bibinfo {author} {\bibfnamefont {M.}~\bibnamefont
  {Horodecki}}\ and\ \bibinfo {author} {\bibfnamefont {J.}~\bibnamefont
  {Oppenheim}},\ }\href {\doibase 10.1038/ncomms3059} {\bibfield  {journal}
  {\bibinfo  {journal} {Nature communications}\ }\textbf {\bibinfo {volume}
  {4}},\ \bibinfo {pages} {2059} (\bibinfo {year} {2013})},\ \Eprint
  {http://arxiv.org/abs/arXiv:1111.3834v1} {arXiv:arXiv:1111.3834v1}
  \BibitemShut {NoStop}%
\bibitem [{\citenamefont {Baumgratz}\ \emph {et~al.}(2014)\citenamefont
  {Baumgratz}, \citenamefont {Cramer},\ and\ \citenamefont
  {Plenio}}]{Baumgratz2014}%
  \BibitemOpen
  \bibfield  {author} {\bibinfo {author} {\bibfnamefont {T.}~\bibnamefont
  {Baumgratz}}, \bibinfo {author} {\bibfnamefont {M.}~\bibnamefont {Cramer}}, \
  and\ \bibinfo {author} {\bibfnamefont {M.~B.}\ \bibnamefont {Plenio}},\
  }\href {\doibase 10.1103/PhysRevLett.113.140401} {\bibfield  {journal}
  {\bibinfo  {journal} {Physical Review Letters}\ }\textbf {\bibinfo {volume}
  {113}},\ \bibinfo {pages} {140401} (\bibinfo {year} {2014})},\ \Eprint
  {http://arxiv.org/abs/1311.0275} {arXiv:1311.0275} \BibitemShut {NoStop}%
\bibitem [{\citenamefont {Streltsov}\ \emph {et~al.}(2017)\citenamefont
  {Streltsov}, \citenamefont {Adesso},\ and\ \citenamefont
  {Plenio}}]{Streltsov2016a}%
  \BibitemOpen
  \bibfield  {author} {\bibinfo {author} {\bibfnamefont {A.}~\bibnamefont
  {Streltsov}}, \bibinfo {author} {\bibfnamefont {G.}~\bibnamefont {Adesso}}, \
  and\ \bibinfo {author} {\bibfnamefont {M.~B.}\ \bibnamefont {Plenio}},\
  }\href {\doibase 10.1103/RevModPhys.89.041003} {\bibfield  {journal}
  {\bibinfo  {journal} {Reviews of Modern Physics}\ }\textbf {\bibinfo {volume}
  {89}},\ \bibinfo {pages} {041003} (\bibinfo {year} {2017})},\ \Eprint
  {http://arxiv.org/abs/1609.02439} {arXiv:1609.02439} \BibitemShut {NoStop}%
\bibitem [{\citenamefont {Nielsen}\ and\ \citenamefont
  {Chuang}(2000)}]{Nielsen}%
  \BibitemOpen
  \bibfield  {author} {\bibinfo {author} {\bibfnamefont {M.~A.}\ \bibnamefont
  {Nielsen}}\ and\ \bibinfo {author} {\bibfnamefont {I.~L.}\ \bibnamefont
  {Chuang}},\ }\href@noop {} {\emph {\bibinfo {title} {{Quantum Computation and
  Quantum Information}}}}\ (\bibinfo  {publisher} {Cambridge University
  Press},\ \bibinfo {year} {2000})\BibitemShut {NoStop}%
\bibitem [{\citenamefont {Gambassi}\ and\ \citenamefont
  {Silva}(2011)}]{Gambassi1106}%
  \BibitemOpen
  \bibfield  {author} {\bibinfo {author} {\bibfnamefont {A.}~\bibnamefont
  {Gambassi}}\ and\ \bibinfo {author} {\bibfnamefont {A.}~\bibnamefont
  {Silva}},\ }\href@noop {} {\enquote {\bibinfo {title} {Statistics of the work
  in quantum quenches, universality and the critical casimir effect},}\ }
  (\bibinfo {year} {2011}),\ \Eprint {http://arxiv.org/abs/1106.2671}
  {arXiv:1106.2671 [cond-mat.stat-mech]} \BibitemShut {NoStop}%
\bibitem [{\citenamefont {Dorner}\ \emph {et~al.}(2012)\citenamefont {Dorner},
  \citenamefont {Goold}, \citenamefont {Cormick}, \citenamefont {Paternostro},\
  and\ \citenamefont {Vedral}}]{Dorner2012}%
  \BibitemOpen
  \bibfield  {author} {\bibinfo {author} {\bibfnamefont {R.}~\bibnamefont
  {Dorner}}, \bibinfo {author} {\bibfnamefont {J.}~\bibnamefont {Goold}},
  \bibinfo {author} {\bibfnamefont {C.}~\bibnamefont {Cormick}}, \bibinfo
  {author} {\bibfnamefont {M.}~\bibnamefont {Paternostro}}, \ and\ \bibinfo
  {author} {\bibfnamefont {V.}~\bibnamefont {Vedral}},\ }\href {\doibase
  10.1103/PhysRevLett.109.160601} {\bibfield  {journal} {\bibinfo  {journal}
  {Physical Review Letters}\ }\textbf {\bibinfo {volume} {109}},\ \bibinfo
  {pages} {160601} (\bibinfo {year} {2012})}\BibitemShut {NoStop}%
\bibitem [{\citenamefont {Fusco}\ \emph {et~al.}(2014)\citenamefont {Fusco},
  \citenamefont {Pigeon}, \citenamefont {Apollaro}, \citenamefont {Xuereb},
  \citenamefont {Mazzola}, \citenamefont {Campisi}, \citenamefont {Ferraro},
  \citenamefont {Paternostro},\ and\ \citenamefont {{De Chiara}}}]{Fusco2014a}%
  \BibitemOpen
  \bibfield  {author} {\bibinfo {author} {\bibfnamefont {L.}~\bibnamefont
  {Fusco}}, \bibinfo {author} {\bibfnamefont {S.}~\bibnamefont {Pigeon}},
  \bibinfo {author} {\bibfnamefont {T.~J.~G.}\ \bibnamefont {Apollaro}},
  \bibinfo {author} {\bibfnamefont {A.}~\bibnamefont {Xuereb}}, \bibinfo
  {author} {\bibfnamefont {L.}~\bibnamefont {Mazzola}}, \bibinfo {author}
  {\bibfnamefont {M.}~\bibnamefont {Campisi}}, \bibinfo {author} {\bibfnamefont
  {A.}~\bibnamefont {Ferraro}}, \bibinfo {author} {\bibfnamefont
  {M.}~\bibnamefont {Paternostro}}, \ and\ \bibinfo {author} {\bibfnamefont
  {G.}~\bibnamefont {{De Chiara}}},\ }\href {\doibase
  10.1103/PhysRevX.4.031029} {\bibfield  {journal} {\bibinfo  {journal}
  {Physical Review X}\ }\textbf {\bibinfo {volume} {4}},\ \bibinfo {pages}
  {031029} (\bibinfo {year} {2014})}\BibitemShut {NoStop}%
\bibitem [{\citenamefont {Goold}\ \emph {et~al.}(2018)\citenamefont {Goold},
  \citenamefont {Plastina}, \citenamefont {Gambassi},\ and\ \citenamefont
  {Silva}}]{Goold2018}%
  \BibitemOpen
  \bibfield  {author} {\bibinfo {author} {\bibfnamefont {J.}~\bibnamefont
  {Goold}}, \bibinfo {author} {\bibfnamefont {F.}~\bibnamefont {Plastina}},
  \bibinfo {author} {\bibfnamefont {A.}~\bibnamefont {Gambassi}}, \ and\
  \bibinfo {author} {\bibfnamefont {A.}~\bibnamefont {Silva}},\ }in\ \href
  {\doibase 10.1007/978-3-319-99046-0_13} {\emph {\bibinfo {booktitle}
  {Thermodynamics in the quantum regime - Recent Progress and Outlook}}},\
  \bibinfo {editor} {edited by\ \bibinfo {editor} {\bibfnamefont
  {F.}~\bibnamefont {Binder}}, \bibinfo {editor} {\bibfnamefont {L.~A.}\
  \bibnamefont {Correa}}, \bibinfo {editor} {\bibfnamefont {G.}~\bibnamefont
  {C}}, \bibinfo {editor} {\bibfnamefont {J.}~\bibnamefont {Anders}}, \ and\
  \bibinfo {editor} {\bibfnamefont {G.}~\bibnamefont {Adesso}}}\ (\bibinfo
  {publisher} {Springer},\ \bibinfo {year} {2018})\ pp.\ \bibinfo {pages}
  {317--336},\ \Eprint {http://arxiv.org/abs/1804.02805} {arXiv:1804.02805}
  \BibitemShut {NoStop}%
\bibitem [{\citenamefont {Sharma}\ and\ \citenamefont
  {Dutta}(2015)}]{PhysRevE92022108}%
  \BibitemOpen
  \bibfield  {author} {\bibinfo {author} {\bibfnamefont {S.}~\bibnamefont
  {Sharma}}\ and\ \bibinfo {author} {\bibfnamefont {A.}~\bibnamefont {Dutta}},\
  }\href {\doibase 10.1103/PhysRevE.92.022108} {\bibfield  {journal} {\bibinfo
  {journal} {Phys. Rev. E}\ }\textbf {\bibinfo {volume} {92}},\ \bibinfo
  {pages} {022108} (\bibinfo {year} {2015})}\BibitemShut {NoStop}%
\bibitem [{\citenamefont {Vicari}(2019)}]{PhysRevA99043603}%
  \BibitemOpen
  \bibfield  {author} {\bibinfo {author} {\bibfnamefont {E.}~\bibnamefont
  {Vicari}},\ }\href {\doibase 10.1103/PhysRevA.99.043603} {\bibfield
  {journal} {\bibinfo  {journal} {Phys. Rev. A}\ }\textbf {\bibinfo {volume}
  {99}},\ \bibinfo {pages} {043603} (\bibinfo {year} {2019})}\BibitemShut
  {NoStop}%
\bibitem [{\citenamefont {Mascarenhas}\ \emph {et~al.}(2014)\citenamefont
  {Mascarenhas}, \citenamefont {Bragan\ifmmode~\mbox{\c{c}}\else \c{c}\fi{}a},
  \citenamefont {Dorner}, \citenamefont {Fran\ifmmode
  \mbox{\c{c}}\else~\c{c}\fi{}a Santos}, \citenamefont {Vedral}, \citenamefont
  {Modi},\ and\ \citenamefont {Goold}}]{Mascarenhas1307}%
  \BibitemOpen
  \bibfield  {author} {\bibinfo {author} {\bibfnamefont {E.}~\bibnamefont
  {Mascarenhas}}, \bibinfo {author} {\bibfnamefont {H.}~\bibnamefont
  {Bragan\ifmmode~\mbox{\c{c}}\else \c{c}\fi{}a}}, \bibinfo {author}
  {\bibfnamefont {R.}~\bibnamefont {Dorner}}, \bibinfo {author} {\bibfnamefont
  {M.}~\bibnamefont {Fran\ifmmode \mbox{\c{c}}\else~\c{c}\fi{}a Santos}},
  \bibinfo {author} {\bibfnamefont {V.}~\bibnamefont {Vedral}}, \bibinfo
  {author} {\bibfnamefont {K.}~\bibnamefont {Modi}}, \ and\ \bibinfo {author}
  {\bibfnamefont {J.}~\bibnamefont {Goold}},\ }\href {\doibase
  10.1103/PhysRevE.89.062103} {\bibfield  {journal} {\bibinfo  {journal} {Phys.
  Rev. E}\ }\textbf {\bibinfo {volume} {89}},\ \bibinfo {pages} {062103}
  (\bibinfo {year} {2014})}\BibitemShut {NoStop}%
\bibitem [{\citenamefont {Cosco}\ \emph {et~al.}(2017)\citenamefont {Cosco},
  \citenamefont {Borrelli}, \citenamefont {Silvi}, \citenamefont {Maniscalco},\
  and\ \citenamefont {De~Chiara}}]{PhysRevA95063615}%
  \BibitemOpen
  \bibfield  {author} {\bibinfo {author} {\bibfnamefont {F.}~\bibnamefont
  {Cosco}}, \bibinfo {author} {\bibfnamefont {M.}~\bibnamefont {Borrelli}},
  \bibinfo {author} {\bibfnamefont {P.}~\bibnamefont {Silvi}}, \bibinfo
  {author} {\bibfnamefont {S.}~\bibnamefont {Maniscalco}}, \ and\ \bibinfo
  {author} {\bibfnamefont {G.}~\bibnamefont {De~Chiara}},\ }\href {\doibase
  10.1103/PhysRevA.95.063615} {\bibfield  {journal} {\bibinfo  {journal} {Phys.
  Rev. A}\ }\textbf {\bibinfo {volume} {95}},\ \bibinfo {pages} {063615}
  (\bibinfo {year} {2017})}\BibitemShut {NoStop}%
\bibitem [{\citenamefont {Paganelli}\ and\ \citenamefont
  {Apollaro}(2017)}]{Pagnelli}%
  \BibitemOpen
  \bibfield  {author} {\bibinfo {author} {\bibfnamefont {S.}~\bibnamefont
  {Paganelli}}\ and\ \bibinfo {author} {\bibfnamefont {T.~J.~G.}\ \bibnamefont
  {Apollaro}},\ }\href {\doibase 10.1142/S0217979217500655} {\bibfield
  {journal} {\bibinfo  {journal} {International Journal of Modern Physics B}\
  }\textbf {\bibinfo {volume} {31}},\ \bibinfo {pages} {1750065} (\bibinfo
  {year} {2017})},\ \Eprint
  {http://arxiv.org/abs/https://doi.org/10.1142/S0217979217500655}
  {https://doi.org/10.1142/S0217979217500655} \BibitemShut {NoStop}%
\bibitem [{\citenamefont {Wang}\ \emph {et~al.}(2018)\citenamefont {Wang},
  \citenamefont {Cao},\ and\ \citenamefont {Quan}}]{PhysRevE98022107}%
  \BibitemOpen
  \bibfield  {author} {\bibinfo {author} {\bibfnamefont {Q.}~\bibnamefont
  {Wang}}, \bibinfo {author} {\bibfnamefont {D.}~\bibnamefont {Cao}}, \ and\
  \bibinfo {author} {\bibfnamefont {H.~T.}\ \bibnamefont {Quan}},\ }\href
  {\doibase 10.1103/PhysRevE.98.022107} {\bibfield  {journal} {\bibinfo
  {journal} {Phys. Rev. E}\ }\textbf {\bibinfo {volume} {98}},\ \bibinfo
  {pages} {022107} (\bibinfo {year} {2018})}\BibitemShut {NoStop}%
\bibitem [{\citenamefont {Bayat}\ \emph {et~al.}(2016)\citenamefont {Bayat},
  \citenamefont {Apollaro}, \citenamefont {Paganelli}, \citenamefont
  {De~Chiara}, \citenamefont {Johannesson}, \citenamefont {Bose},\ and\
  \citenamefont {Sodano}}]{PhysRevB93201106}%
  \BibitemOpen
  \bibfield  {author} {\bibinfo {author} {\bibfnamefont {A.}~\bibnamefont
  {Bayat}}, \bibinfo {author} {\bibfnamefont {T.~J.~G.}\ \bibnamefont
  {Apollaro}}, \bibinfo {author} {\bibfnamefont {S.}~\bibnamefont {Paganelli}},
  \bibinfo {author} {\bibfnamefont {G.}~\bibnamefont {De~Chiara}}, \bibinfo
  {author} {\bibfnamefont {H.}~\bibnamefont {Johannesson}}, \bibinfo {author}
  {\bibfnamefont {S.}~\bibnamefont {Bose}}, \ and\ \bibinfo {author}
  {\bibfnamefont {P.}~\bibnamefont {Sodano}},\ }\href {\doibase
  10.1103/PhysRevB.93.201106} {\bibfield  {journal} {\bibinfo  {journal} {Phys.
  Rev. B}\ }\textbf {\bibinfo {volume} {93}},\ \bibinfo {pages} {201106}
  (\bibinfo {year} {2016})}\BibitemShut {NoStop}%
\bibitem [{\citenamefont {Bayocboc}\ and\ \citenamefont
  {Paraan}(2015)}]{Bayocboc2015}%
  \BibitemOpen
  \bibfield  {author} {\bibinfo {author} {\bibfnamefont {F.~A.}\ \bibnamefont
  {Bayocboc}}\ and\ \bibinfo {author} {\bibfnamefont {P.~N.~C.}\ \bibnamefont
  {Paraan}},\ }\href {\doibase 10.1103/PhysRevE.92.032142} {\bibfield
  {journal} {\bibinfo  {journal} {Physical Review E}\ }\textbf {\bibinfo
  {volume} {92}},\ \bibinfo {pages} {032142} (\bibinfo {year}
  {2015})}\BibitemShut {NoStop}%
\bibitem [{\citenamefont {Pelissetto}\ \emph {et~al.}(2018)\citenamefont
  {Pelissetto}, \citenamefont {Rossini},\ and\ \citenamefont
  {Vicari}}]{PhysRevE97052148}%
  \BibitemOpen
  \bibfield  {author} {\bibinfo {author} {\bibfnamefont {A.}~\bibnamefont
  {Pelissetto}}, \bibinfo {author} {\bibfnamefont {D.}~\bibnamefont {Rossini}},
  \ and\ \bibinfo {author} {\bibfnamefont {E.}~\bibnamefont {Vicari}},\ }\href
  {\doibase 10.1103/PhysRevE.97.052148} {\bibfield  {journal} {\bibinfo
  {journal} {Phys. Rev. E}\ }\textbf {\bibinfo {volume} {97}},\ \bibinfo
  {pages} {052148} (\bibinfo {year} {2018})}\BibitemShut {NoStop}%
\bibitem [{\citenamefont {Nigro}\ \emph {et~al.}(2019)\citenamefont {Nigro},
  \citenamefont {Rossini},\ and\ \citenamefont {Vicari}}]{Nigro}%
  \BibitemOpen
  \bibfield  {author} {\bibinfo {author} {\bibfnamefont {D.}~\bibnamefont
  {Nigro}}, \bibinfo {author} {\bibfnamefont {D.}~\bibnamefont {Rossini}}, \
  and\ \bibinfo {author} {\bibfnamefont {E.}~\bibnamefont {Vicari}},\ }\href
  {\doibase 10.1088/1742-5468/ab00e2} {\bibfield  {journal} {\bibinfo
  {journal} {Journal of Statistical Mechanics: Theory and Experiment}\ }\textbf
  {\bibinfo {volume} {2019}},\ \bibinfo {pages} {023104} (\bibinfo {year}
  {2019})}\BibitemShut {NoStop}%
\bibitem [{Note1()}]{Note1}%
  \BibitemOpen
  \bibinfo {note} {This is done by noting that the dephasing $\protect \mathbb
  {D}_{H}(\rho )$ can be also given by \begin {equation*} \protect \mathbb
  {D}_{H}(\rho ) = \protect \qopname \relax m{lim}_{s\to \infty }\protect \frac
  {1}{s} \DOTSI \intop \ilimits@ _0^s \protect \mathrm {d}t e^{-iHt}\rho
  e^{iHt}. \end {equation*} We then use that $e^{t(H_0+\Delta H)} = e^{tH_0} +
  t\protect \mathbb {J}_{e^{tH_0}}[\Delta H] + \protect \mathcal {O}(\Delta
  H^2)$ and $[\rho _0^{\protect \text {th} }, H_0]=0$. To order $\Delta H$ this
  gives Eq.~\protect \textup {\hbox {\mathsurround \z@ \protect \normalfont
  (\ignorespaces \ref {Deph-state-expansion}\unskip \@@italiccorr
  )}}}\BibitemShut {NoStop}%
\bibitem [{\citenamefont {Petz}(2002)}]{Petz2002}%
  \BibitemOpen
  \bibfield  {author} {\bibinfo {author} {\bibfnamefont {D.}~\bibnamefont
  {Petz}},\ }\href {\doibase 10.1088/0305-4470/35/4/305} {\bibfield  {journal}
  {\bibinfo  {journal} {Journal of Physics A: Mathematical and General}\
  }\textbf {\bibinfo {volume} {35}},\ \bibinfo {pages} {929} (\bibinfo {year}
  {2002})},\ \Eprint {http://arxiv.org/abs/0106125} {arXiv:0106125 [quant-ph]}
  \BibitemShut {NoStop}%
\bibitem [{\citenamefont {Wei}\ and\ \citenamefont {Plenio}(2017)}]{Wei2017a}%
  \BibitemOpen
  \bibfield  {author} {\bibinfo {author} {\bibfnamefont {B.~B.}\ \bibnamefont
  {Wei}}\ and\ \bibinfo {author} {\bibfnamefont {M.~B.}\ \bibnamefont
  {Plenio}},\ }\href {\doibase 10.1088/1367-2630/aa579e} {\bibfield  {journal}
  {\bibinfo  {journal} {New Journal of Physics}\ }\textbf {\bibinfo {volume}
  {19}},\ \bibinfo {pages} {023002} (\bibinfo {year} {2017})},\ \Eprint
  {http://arxiv.org/abs/1509.07043} {arXiv:1509.07043} \BibitemShut {NoStop}%
\bibitem [{\citenamefont {Sakurai}\ and\ \citenamefont
  {Napolitano}(2010)}]{Sakurai2010}%
  \BibitemOpen
  \bibfield  {author} {\bibinfo {author} {\bibfnamefont {J.~J.}\ \bibnamefont
  {Sakurai}}\ and\ \bibinfo {author} {\bibfnamefont {J.~J.}\ \bibnamefont
  {Napolitano}},\ }\href@noop {} {\emph {\bibinfo {title} {{Modern Quantum
  Mechanics}}}},\ \bibinfo {edition} {2nd}\ ed.\ (\bibinfo  {publisher}
  {Addison-Wesley},\ \bibinfo {year} {2010})\ p.\ \bibinfo {pages}
  {550}\BibitemShut {NoStop}%
\bibitem [{\citenamefont {Niedenzu}\ \emph {et~al.}(2019)\citenamefont
  {Niedenzu}, \citenamefont {Huber},\ and\ \citenamefont
  {Boukobza}}]{Niedenzu2019}%
  \BibitemOpen
  \bibfield  {author} {\bibinfo {author} {\bibfnamefont {W.}~\bibnamefont
  {Niedenzu}}, \bibinfo {author} {\bibfnamefont {M.}~\bibnamefont {Huber}}, \
  and\ \bibinfo {author} {\bibfnamefont {E.}~\bibnamefont {Boukobza}},\ }\href
  {http://arxiv.org/abs/1907.01353} {\ ,\ \bibinfo {pages} {1} (\bibinfo {year}
  {2019})},\ \Eprint {http://arxiv.org/abs/1907.01353} {arXiv:1907.01353}
  \BibitemShut {NoStop}%
\bibitem [{\citenamefont {Miller}\ \emph
  {et~al.}(2020{\natexlab{b}})\citenamefont {Miller}, \citenamefont
  {Guarnieri}, \citenamefont {Mitchison},\ and\ \citenamefont
  {Goold}}]{Miller2020Landauer}%
  \BibitemOpen
  \bibfield  {author} {\bibinfo {author} {\bibfnamefont {H.~J.~D.}\
  \bibnamefont {Miller}}, \bibinfo {author} {\bibfnamefont {G.}~\bibnamefont
  {Guarnieri}}, \bibinfo {author} {\bibfnamefont {M.~T.}\ \bibnamefont
  {Mitchison}}, \ and\ \bibinfo {author} {\bibfnamefont {J.}~\bibnamefont
  {Goold}},\ }\href {\doibase 10.1103/PhysRevLett.125.160602} {\bibfield
  {journal} {\bibinfo  {journal} {Phys. Rev. Lett.}\ }\textbf {\bibinfo
  {volume} {125}},\ \bibinfo {pages} {160602} (\bibinfo {year}
  {2020}{\natexlab{b}})}\BibitemShut {NoStop}%
\bibitem [{\citenamefont {{De Chiara}}\ \emph {et~al.}(2018)\citenamefont {{De
  Chiara}}, \citenamefont {Landi}, \citenamefont {Hewgill}, \citenamefont
  {Reid}, \citenamefont {Ferraro}, \citenamefont {Roncaglia},\ and\
  \citenamefont {Antezza}}]{DeChiara2018}%
  \BibitemOpen
  \bibfield  {author} {\bibinfo {author} {\bibfnamefont {G.}~\bibnamefont {{De
  Chiara}}}, \bibinfo {author} {\bibfnamefont {G.}~\bibnamefont {Landi}},
  \bibinfo {author} {\bibfnamefont {A.}~\bibnamefont {Hewgill}}, \bibinfo
  {author} {\bibfnamefont {B.}~\bibnamefont {Reid}}, \bibinfo {author}
  {\bibfnamefont {A.}~\bibnamefont {Ferraro}}, \bibinfo {author} {\bibfnamefont
  {A.~J.}\ \bibnamefont {Roncaglia}}, \ and\ \bibinfo {author} {\bibfnamefont
  {M.}~\bibnamefont {Antezza}},\ }\href {\doibase
  https://doi.org/10.1088/1367-2630/aaecee} {\bibfield  {journal} {\bibinfo
  {journal} {New Journal of Physics}\ }\textbf {\bibinfo {volume} {20}},\
  \bibinfo {pages} {113024} (\bibinfo {year} {2018})},\ \Eprint
  {http://arxiv.org/abs/1808.10450} {arXiv:1808.10450} \BibitemShut {NoStop}%
\bibitem [{\citenamefont {Callen}(1985)}]{Callen1985}%
  \BibitemOpen
  \bibfield  {author} {\bibinfo {author} {\bibfnamefont {H.~B.}\ \bibnamefont
  {Callen}},\ }\href@noop {} {\emph {\bibinfo {title} {{Thermodynamics and an
  introduction to Thermostatistics}}}},\ \bibinfo {edition} {2nd}\ ed.\
  (\bibinfo  {publisher} {Wiley},\ \bibinfo {year} {1985})\ p.\ \bibinfo
  {pages} {493}\BibitemShut {NoStop}%
\bibitem [{\citenamefont {Lieb}\ \emph {et~al.}(1961)\citenamefont {Lieb},
  \citenamefont {Schultz},\ and\ \citenamefont {Mattis}}]{Lieb1961}%
  \BibitemOpen
  \bibfield  {author} {\bibinfo {author} {\bibfnamefont {E.~H.}\ \bibnamefont
  {Lieb}}, \bibinfo {author} {\bibfnamefont {T.}~\bibnamefont {Schultz}}, \
  and\ \bibinfo {author} {\bibfnamefont {D.}~\bibnamefont {Mattis}},\ }\href
  {http://www.sciencedirect.com/science/article/pii/0003491661901154}
  {\bibfield  {journal} {\bibinfo  {journal} {Annals of Physics}\ }\textbf
  {\bibinfo {volume} {16}},\ \bibinfo {pages} {407} (\bibinfo {year}
  {1961})}\BibitemShut {NoStop}%
\bibitem [{\citenamefont {Damski}\ and\ \citenamefont {Rams}(2014)}]{Damski}%
  \BibitemOpen
  \bibfield  {author} {\bibinfo {author} {\bibfnamefont {B.}~\bibnamefont
  {Damski}}\ and\ \bibinfo {author} {\bibfnamefont {M.~M.}\ \bibnamefont
  {Rams}},\ }\href {\doibase 10.1088/1751-8113/47/2/025303} {\bibfield
  {journal} {\bibinfo  {journal} {Journal of Physics A: Mathematical and
  Theoretical}\ }\textbf {\bibinfo {volume} {47}},\ \bibinfo {pages} {025303}
  (\bibinfo {year} {2014})}\BibitemShut {NoStop}%
\end{thebibliography}%
\end{document}